\documentclass[twoside,12pt]{article}
\usepackage{color}
\usepackage{epsfig}
\usepackage{multirow}
\usepackage{varwidth}

\newcommand{\be}{\begin{equation}}
\newcommand{\ee}{\end{equation}}
\newcommand{\bea}{\begin{eqnarray}}
\newcommand{\eea}{\end{eqnarray}}

\newcommand{\pA}    {p+A}
\newcommand{\dA}    {d+A}
\newcommand{\pAu}   {p+Au}
\newcommand{\dAu}   {d+Au}
\newcommand{\AuAu}  {Au+Au}
\newcommand{\pPb}   {p+Pb}
\newcommand{\PbPb}  {Pb+Pb}
\newcommand{\Ru}    {^{96}_{44}\rm{Ru}}
\newcommand{\Zr}    {^{96}_{40}\rm{Zr}}
\newcommand{\RuRu}  {^{96}_{44}\rm{Ru} + ^{96}_{44}\rm{Ru}}
\newcommand{\ZrZr}  {^{96}_{40}\rm{Zr} + ^{96}_{40}\rm{Zr}}
\newcommand{\CP}    {$\mathcal{CP}$}
\newcommand{\etal}  {{\it et al.}}
\newcommand{\snn}   {\sqrt{s_{_{\rm NN}}}}
\newcommand{\gevc}  {GeV/$c$}
\newcommand{\gevcc} {GeV/$c^2$ }
\newcommand{\mevc}  {MeV/$c$}
\newcommand{\nth}   {$n$th}
\newcommand{\Npart} {$N_{\rm part}$}
\newcommand{\Ndomain}{N_{\rm domain}}
\newcommand{\Noff}  {N^{\rm offline}_{\rm trk}}
\newcommand{\Ach}   {A_{\rm ch}}
\newcommand{\dNdeta}{dN_{\rm ch}/d\eta}
\newcommand{\pt}    {p_{T}}
\newcommand{\minv}  {m_{\rm inv}}
\newcommand{\Bsq}   {B_{\rm sq}}
\newcommand{\Bvec}  {\vec{B}}
\newcommand{\pip}   {\pi^+}
\newcommand{\pim}   {\pi^-}
\newcommand{\mpi}   {m_{\pi}}
\newcommand{\clust} {{\rm clust.}}
\newcommand{\deta}  {\Delta\eta}
\newcommand{\dphi}  {\Delta\phi}
\newcommand{\psiPP} {\psi_{\rm PP}}
\newcommand{\psiRP} {\psi_{\rm RP}}
\newcommand{\psiEP} {\psi_{\rm EP}}
\newcommand{\nEP}   {{n,\rm EP}}
\newcommand{\psiB}  {\psi_{B}}
\newcommand{\phia}  {\phi_{\alpha}}
\newcommand{\phib}  {\phi_{\beta}}

\newcommand{\phiclust}{\phi_{\clust}}
\newcommand{\gSS}   {\gamma_{\rm SS}}
\newcommand{\gOS}   {\gamma_{\rm OS}}
\newcommand{\dg}    {\Delta\gamma}
\newcommand{\dginc} {\dg_{\rm inc}}
\newcommand{\dgscale}{\dg_{\rm scaled}}
\newcommand{\dd}    {\Delta\delta}
\newcommand{\dSS}   {\delta_{\rm SS}}
\newcommand{\dOS}   {\delta_{\rm OS}}
\newcommand{\vc}    {v_{2,c}}
\newcommand{\vct}   {v_{2,c}\{2\}}
\newcommand{\vnt}   {v_n\{2\}}
\newcommand{\vnf}   {v_n\{4\}}
\newcommand{\vclust}{v_{2,\clust}}
\newcommand{\vexe}  {v_{2,{\rm ebye}}}
\newcommand{\vnexe} {v_{n,{\rm ebye}}}
\newcommand{\cme}   {{\rm CME}}
\newcommand{\bkg}   {{\rm Bkg}}
\newcommand{\fcme}  {f_{\cme}}
\newcommand{\anorm} {a_{\rm norm}}
\newcommand{\bnorm} {b_{\rm norm}}
\newcommand{\mean}[1] {\langle #1\rangle}

\begin{document}
\title{Experimental searches for the chiral magnetic effect in heavy-ion collisions}
\author{Jie Zhao,$^{1}$ Fuqiang Wang$^{1,2}$\\
\\
$^1$Department of Physics and Astronomy, Purdue University,\\
West Lafayette, Indiana 47907, USA\\
$^2$School of Science, Huzhou University, Huzhou, Zhejiang 313000, China}

\maketitle

\begin{abstract}
  The chiral magnetic effect (CME) in quantum chromodynamics (QCD) refers to a charge separation (an electric current) of chirality imbalanced quarks generated along an external strong magnetic field. The chirality imbalance results from interactions of quarks, under the approximate chiral symmetry restoration, with metastable local domains of gluon fields of non-zero topological charges out of QCD vacuum fluctuations. Those local domains violate the $\mathcal{P}$ and $\mathcal{CP}$ invariance, potentially offering a solution to the strong $\mathcal{CP}$ problem in explaining the magnitude of the matter-antimatter asymmetry in today's universe. Relativistic heavy-ion collisions, with the likely creation of the high energy density quark-gluon plasma and restoration of the approximate chiral symmetry, and the possibly long-lived strong magnetic field, provide a unique opportunity to detect the CME. Early measurements of the CME-induced charge separation in heavy-ion collisions are dominated by physics backgrounds. Major efforts have been devoted to eliminate or reduce those backgrounds. We review those efforts, with a somewhat historical perspective, and focus on the recent innovative experimental undertakings in the search for the CME in heavy-ion collisions.
\end{abstract}

\eject
\tableofcontents
\eject

\section{Introduction}

Our universe started from the Big Bang singularity~\cite{Alpher:1948ve} with equal amounts of matter and antimatter, but is today dominated by only matter. No significant concentration of antimatter has ever been found in the observable universe~\cite{Canetti:2012zc}.
This matter-antimatter asymmetry is caused by \CP\ (charge-conjugation parity) violation, a slight difference in the physics governing matter and antimatter~\cite{Sakharov:1967dj,RevModPhys.76.1}, 
as in e.g.~electroweak baryogenesis~\cite{Kuzmin:1985mm,Shaposhnikov:1987tw}.
\CP\ is violated in the weak interaction but the magnitude of the CKM quark-sector \CP\ violation~\cite{Cabibbo:1963yz,Kobayashi:1973fv} is too small to explain the present universe matter-antimatter asymmetry~\cite{Mannel:2007zz}. It is unclear whether the lepton-sector \CP\ violation through leptogenesis~\cite{Sakharov:1967dj,Davidson:2008bu} is large enough to account for the matter-antimatter asymmetry. \CP\ violation in the strong interaction in the early universe may be needed. \CP\ violation is not prohibited in the strong interaction~\cite{Kim:2008hd} but none has been experimentally observed~\cite{Baker:2006ts,Afach:2015sja}. This is called the strong \CP\ problem~\cite{Kim:2008hd,Peccei:1977hh}. 
To solve the strong \CP\ problem, Peccei and Quinn~\cite{Peccei:1977hh,Peccei:1977ur} proposed to extend the QCD (quantum chromodynamics) Lagrangian by a \CP-violating $\theta$ term, first introduced by 't~Hooft~\cite{tHooft:1976snw,tHooft:1986ooh} in resolving the axial $U(1)$ problem~\cite{Weinberg:1975ui}. It predicts the existence of a new particle called the axion. If axions exist, they would not only offer a solution to the strong \CP\ problem, but could also be a dark matter candidate~\cite{Duffy:2009ig}. On the other hand, the Peccei-Quinn mechanism would remove the large, flavour diagonal \CP\ violation, precluding a solution to the strong \CP\ problem to arise from QCD. However, axions have not been detected after four decades of search since its conception~\cite{Rosenberg:2000wb,Graham:2015ouw}. 

Here, we concentrate on another possible solution to the strong \CP\ problem, namely, \CP\ violation in local metastable domains of QCD vacuum, manifested via the chiral magnetic effect (CME) under strong magnetic fields. We review the experimental searches for the CME in relativistic heavy-ion collisions.

\subsection{The chiral magnetic effect}
\CP\ invariance is not a requirement by QCD, the theory governing the strong interaction among quarks and gluons~\cite{Marciano:1977su}. However, \CP\ appears to be conserved in the strong interaction. This may be accidental such that the simplicity and renormalizability of QCD require \CP\ conservation even though the theory itself does not require it~\cite{Kim:2008hd}. Because of vacuum fluctuations in QCD, metastable local domains of gluon fields can form with nonzero topological charges (Chern-Simons winding numbers)~\cite{Lee:1973iz,Lee:1974ma,Morley:1983wr,Kharzeev:1998kz}. 
The topological charge, $Q_W$, is proportional to the integral of the scalar product of the gluonic (color) electric and magnetic fields, and is zero in the physical vacuum~\cite{Kharzeev:1998kz,Kharzeev:2007jp}.
Transitions between QCD vacuum states of gluonic configurations can be described by instantons/sphalerons mechanisms~\cite{Kharzeev:1998kz,Kharzeev:2004ey}.
Under the approximate chiral symmetry restoration, quark interactions with those topological gluon fields would change the helicities of the quarks, thereby causing an chirality imbalance between left- and right-handed quarks, $Q_W=N_L-N_R\neq0$.
Such a quantum chiral anomaly would lead to a local parity ($\mathcal{P}$) and \CP\ violations in those metastable domains~\cite{Kharzeev:1998kz,Kharzeev:2007jp,Kharzeev:2004ey,Kharzeev:1999cz,Kharzeev:2009fn}. 
Local \CP\ violations could have happened in the early universe when the temperature and energy density were high and the universe was in the deconfined state of the quark-gluon plasma (QGP) under the approximate chiral symmetry~\cite{Shuryak:2008eq}.
Those local $\mathcal{CP}$ violations in the strong interaction could offer a solution to the strong \CP\ problem and possibly explain the magnitude of the matter-antimatter asymmetry in the present universe~\cite{RevModPhys.76.1,Kharzeev:2004ey,Fukushima:2008xe}.

\begin{figure*}[!htb]
  \centerline{\includegraphics[width=0.6\hsize]{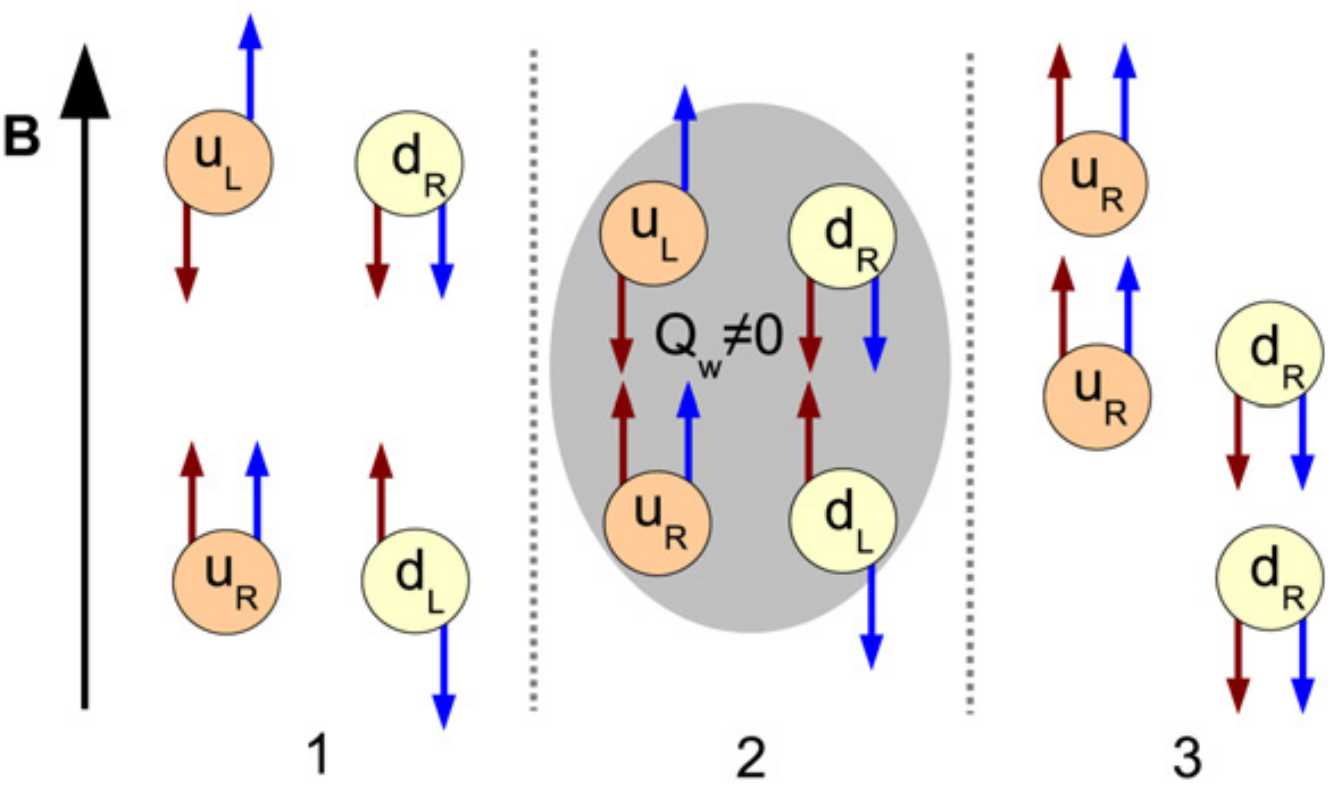}}
  \caption{(Color online) Illustration of the CME. The red arrows denote the direction of momentum, the blue arrows the spin of the quarks. (1) Initially there are as many left-handed as right-handed quarks. Due to the strong magnetic field the up and down quarks are all in the lowest Landau level and can only move along the magnetic field. (2) The quarks interact with a topological gluon field with nonzero $Q_w$, converting left-handed quarks into right-handed ones (in this case $Q_w<0$) by reversing the direction of momentum. (3) The right-handed up quarks will move upward, and the right-handed down quarks will move downward, resulting in a charge separation. Adapted from Ref.~\cite{Kharzeev:2007jp}.}
  \label{fig:cme}
\end{figure*}

The chirality imbalance can have experimental consequences if submerged in a strong enough magnetic field ($\Bvec$), with magnitude on the order of $eB\sim m_\pi^2$ where $m_\pi$ is the pion mass~\cite{Kharzeev:2007jp,Fukushima:2008xe}. The lowest Laudau level energy $eB/2m_q\sim1$~GeV is much larger than the typical transverse momentum of the quarks (determined by the system temperature) so the quarks and antiquarks are all locked in the lowest Laudau level. Here $m_q\sim$ a few MeV/$c^2$ are the light quark masses under the approximate chiral symmetry. The quark spins are locked either parallel or anti-parallel to the magnetic field direction, depending on the quark charge. This would lead to an experimentally observable charge separation in the final state, an electric current along the direction of the magnetic field~\cite{Kharzeev:2007jp,Fukushima:2008xe,Muller:2010jd,Liu:2011ys}. Such a charge separation phenomenon is called the chiral magnetic effect, CME~\cite{Kharzeev:2013ffa,Kharzeev:2015znc,Huang:2015oca}. The cartoon in Fig.~\ref{fig:cme} illustrates the physics of the CME. Quarks will eventually hadronize into (charged) hadrons, so the CME would lead to an experimentally observable charge separation in the final state. 

\subsection{Relativistic heavy-ion collisions}
\begin{figure*}[!htb]
  \centerline{\includegraphics[width=0.4\hsize]{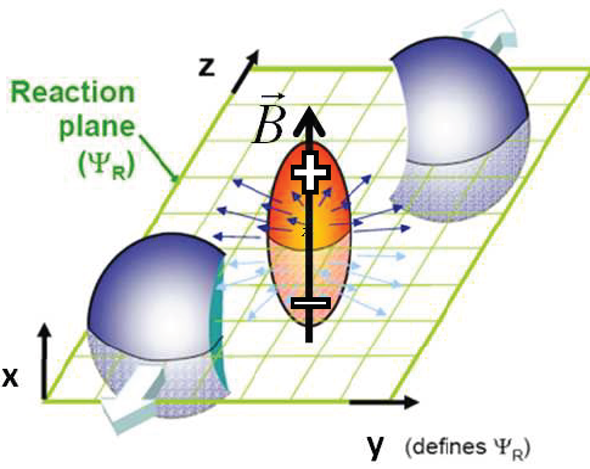}}
  \caption{Illustration of a non-central heavy-ion collision, where the overlap participant region is an ellipse (on the transverse plane) with anisotropic expansion (indicated by the radial arrows) and a strong magnetic field pointing upward generated by the spectator protons. The reaction plane is defined by the impact parameter direction and the beam direction.}
  \label{fig:hi}
\end{figure*}

Relativistic heavy-ion collisions have been conducted at BNL's Relativistic Heavy-Ion Collider (RHIC) and CERN's Large Hadron Collider (LHC).
The primary goal of relativistic heavy-ion collisions is to create a state of high temperature and energy density, where the matter exists in the form of the QGP~\cite{Shuryak:2008eq}. It recreates the condition similar to that in the early universe.
Relativistic heavy-ion collisions may provide a suitable environment for the realization of the CME.
The approximate chiral symmetry, which is spontaneously broken under normal conditions~\cite{Nambu:1960xd,Nambu:1960tm}, is likely restored in relativistic heavy-ion collisions and the relevant degrees of freedom are quarks and gluons~\cite{Adams:2005dq,Adcox:2004mh,Arsene:2004fa,Back:2004je,Muller:2012zq}.
In non-central heavy-ion collisions, an extremely strong magnetic field is produced mainly by the fast moving spectator protons in the early times of those collisions, as illustrated by the cartoon in Fig.~\ref{fig:hi}.
The magnitude of the initial magnetic field produced in \AuAu\ collisions at RHIC is estimated to be on the order of $B\sim10^{14}$~Tesla ($eB\sim m_\pi^2$)~\cite{Kharzeev:2007jp,Kharzeev:2004ey,Fukushima:2008xe}.

The quantitative prediction for the magnitude of the CME in heavy-ion collisions is theoretically challenging. Although QCD vacuum fluctuations are well founded theoretically, the magnitude of the fluctuation effects is quantitatively less known. Semi-quantitative estimates proceed as follows.
The variance of the net topological charge change is proportional to the total number of topological charge changing transitions. The probability of forming topologically charged domains is not suppressed in the deconfined phase. So, if sufficiently hot matter is created in heavy-ion collisions, one would expect on average a finite amount [$\mathcal{O}(1)$] of topological charge change in each event~\cite{Kharzeev:2007jp}. Since in heavy-ion collisions the typical number of pions in on the order of 100, one may expect the CME magnitude to be on the order of $10^{-2}$~\cite{Kharzeev:2004ey,Kharzeev:2007tn}. 
The authors of Refs.~\cite{Yin:2015fca,Jiang:2016wve,Shi:2017cpu} estimated the initial axial charge density in heavy ion collisions and applied it to their Anomalous Viscous Fluid Dynamics (AVFD) on top of a realistic hydrodynamic evolution. They found the CME signal to be also on the order of $10^{-2}$.
The authors of Ref.~\cite{Muller:2010jd} estimated, by assuming the winding number transition density of 8~fm$^{-3}$ and an temperature of 350~MeV, that the CME magnitude is only approximately $6\times10^{-4}$, an order of magnitude smaller than other estimates.

It seems that the various estimates of the CME span a wide range of magnitudes. Furthermore, most estimates are based on the quark level, and are expected to suffer from further uncertainties toward final-state observables~\cite{Kharzeev:2015znc,Selyuzhenkov:2005xa}. These uncertainties arise from parton-parton and hadron-hadron scatterings, and perhaps also from the hadronization process. It is well established that significant final-state interactions are present in heavy-ion collisions. For example, many experimental observations, such as particle $\pt$ spectra and yields, can be well described by the String-Melting version of A Multi-Phase Transport (AMPT) model~\cite{Zhang:1999bd,Lin:2001zk,Lin:2004en} which incorporates significant parton-parton and hadron-hadron interactions~\cite{Lin:2014tya}. An estimate by AMPT simulation indicates that the quark charge separation magnitude at the initial time is reduced by a factor of ten in the final-state freeze-out quarks before hadronization~\cite{Ma:2011uma}. Final-state hadronic interactions would likely reduce the charge separation effect further. However, the hadronic scattering effects cannot be readily studied in AMPT, because the hadron cascade currently implemented in AMPT does not conserve electric charge~\cite{Zhang:1999bd,Lin:2004en}, which is essential to the charge separation studies.

It should be noted that the CME can come about in many ways in a pure hadronic scenario~\cite{Muller:2010jd,Asakawa:2010bu}. It thus does not necessarily mean local $\mathcal{P}$ and \CP\ violations. The quantitative predictions of the CME will also depend on the hadronic physics mechanisms and can have a wide range of magnitudes. It is fair to say that a quantitative understanding of the CME, although extensively studied, is not yet in hand theoretically. 

\subsection{Magnetic field in heavy-ion collisions}\label{sec:B}
\begin{figure}[!htb]
  \centerline{
    \includegraphics[width=0.4\hsize]{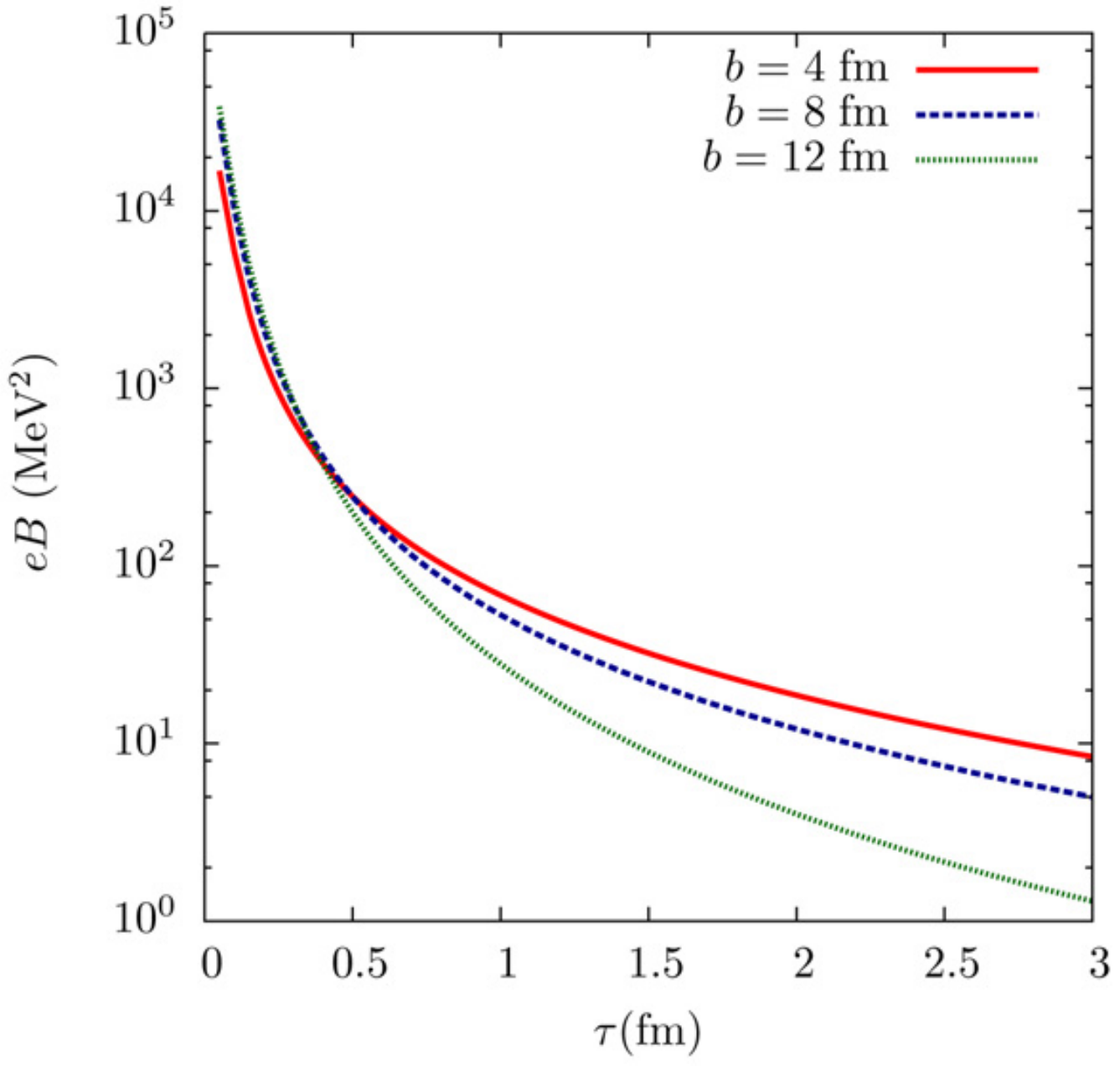}
    \hspace{0.05\hsize}
    \includegraphics[width=0.5\hsize]{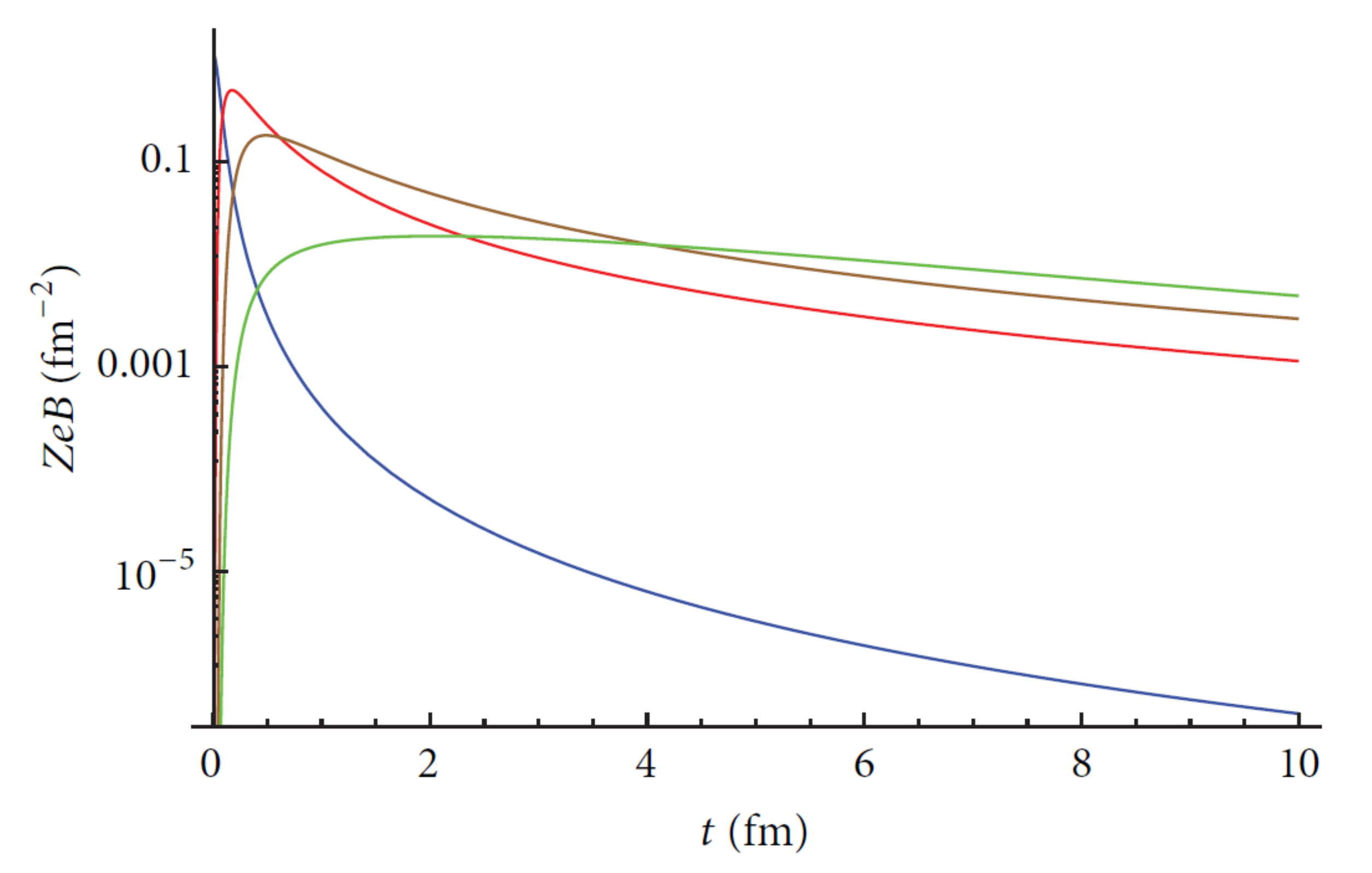}
  }
  \caption{(Color online) The magnetic field magnitude at the center of 200~GeV \AuAu\ collisions (left panel) for various impact parameters ($b$)~\cite{Kharzeev:2007jp} and (right panel) for $b=7$~fm calculated with electric conductivity of the plasma at $\sigma = 0$ in vacuum (blue), in static conducting medium at $\sigma = 5.8$~MeV (red) and at $\sigma = 16$~MeV (brown), and in the expanding medium (green)~\cite{Tuchin:2013ie}. The magnetic field magnitude is expressed in terms of $eB$ in both panels although the right panel coordinate is labeled as $ZeB$.}
  \label{fig:B}
\end{figure}

Another difficulty to the quantitative prediction of the CME is that the time dependence of the magnetic field created in heavy-ion collisions is poorly understood. There seems no doubt that the initial magnetic field in heavy-ion collisions is strong, the larger the collision energy the stronger the magnitude of the magnetic field~\cite{Kharzeev:2007jp,Skokov:2009qp,Tuchin:2015oka}. Under normal conditions the magnetic field dies quickly, as shown in the left panel of Fig.~\ref{fig:B}. This is because the spectator protons, which are primarily responsible for the magnetic field, quickly recede from each other. The magnetic field dies more quickly at higher collision energy~\cite{Kharzeev:2007jp,Sun:2019hao}. For the effect of the magnetic field, both the magnitude and the duration of the magnetic field are relevant. As a result, the effect of magnetic field may have a relative weak dependence on the collision energy~\cite{Sun:2019hao}. If the magnetic field dies quickly, then the CME could be too small to be experimentally observable~\cite{Muller:2018ibh,Sun:2018idn}. On the other hand, it is postulated that the magnetic field could sustain for a relatively long time in a conducting QGP produced in heavy-ion collisions~\cite{Kharzeev:2009pj,Tuchin:2010vs,Tuchin:2013ie,She:2017icp,Li:2018ufq}, as shown in the right panel of Fig.~\ref{fig:B}. It is therefore possible that the magnetic field may have a more significant effect at higher collision energies where the QGP may be produced with high conductivity~\cite{Kharzeev:2009pj,Li:2018ufq}. If the strong magnetic field and the parity-violating local domains are on similar time scales in relativistic heavy-ion collisions, then the magnetic field would be large enough to turn the CME into an experimentally observable of charge separation along the magnetic field.

On average the magnetic field is perpendicular to the reaction plane, RP (span by the beam and impact parameter directions of the colliding nuclei); see Fig.~\ref{fig:hi} for an illustration. Because of fluctuations of the proton distributions in the colliding nuclei, the magnetic field can have both perpendicular and parallel components with respect to the RP, varying from collision to collision~\cite{Bzdak:2011yy,Deng:2012pc,Bloczynski:2012en,Zakharov:2017yst}. Within the same collision, the magnetic field also varies from location to location in the collision fireball~\cite{Bzdak:2011yy,Deng:2012pc,Bloczynski:2012en}.
Besides the magnetic field, strong electric fields are also produced in heavy-ion collisions. Those electric fields affect the motions of charged particles, and therefore have influence on experimental observables in the search of the CME~\cite{Bzdak:2011yy,Deng:2012pc,Bloczynski:2012en}. It was found that in asymmetric Cu+Au collisions compared to symmetric \AuAu\ collisions, the electric field can even reverse the sign of the CME observable due to the magnetic field~\cite{Deng:2014uja}. Such effects may, in turn, be used to improve our understanding of the CME by comparing different collision systems.

The magnetic field cannot be readily measured in experiment. The RP may also be hard to assess in experiment. Often symmetry planes are constructed experimentally based on final-state particle azimuthal distributions~\cite{Poskanzer:1998yz}. Such symmetry planes are affected by fluctuations of nucleons participating in the collision, and thus fluctuate about the RP\cite{Alver:2006wh}. Those fluctuations can impact in a number of ways the experimental search for the CME.
On the other hand, there are several ways to probe the effect of the magnetic field in heavy-ion collisions. It is predicted that the magnetic field affects the directed flows of positively and negatively charged particles in opposite directions~\cite{Gursoy:2014aka,Das:2016cwd}. The effect is stronger for heavier particles, and may be observable for the charmed $D^\pm$ mesons~\cite{Das:2016cwd}. The directed flows of $D^\pm$ have been measured by STAR in 200~GeV \AuAu\ collisions at RHIC~\cite{Singha:2018cdj,He:2019gcx} and by ALICE in \PbPb\ collisions at 5.02~TeV at the LHC~\cite{Grosa:2018zix}. The statistical precisions are presently too poor to draw conclusions. The results, with large error bars, are consistent with no magnetic field effect but also with a strong magnetic field of $10^{14}$~Tesla.

Another way to probe the effect of the magnetic field is via global polarization measurements of $\Lambda$ and $\bar{\Lambda}$ hyperons~\cite{Becattini:2016gvu,Han:2017hdi}. Global polarizations of $\Lambda$ and $\bar{\Lambda}$ hyperons arise from the coupling of their spin to the vorticity of the collision system~\cite{Liang:2004ph,Becattini:2007sr,Pang:2016igs,Csernai:2018yok,Xia:2018tes}. Finite $\Lambda$ and $\bar{\Lambda}$ global polarizations have been measured in \AuAu\ collisions at RHIC~\cite{STAR:2017ckg}, suggesting a strong vorticity attained in those collisions.
It is predicted that the strong magnetic field would lift the degeneracy between $\Lambda$ and $\bar{\Lambda}$, making the global polarization of $\bar{\Lambda}$ larger than that of the $\Lambda$~\cite{Becattini:2016gvu}. Current measurements hint at a difference of the correct order, but the statistical precision prevents a firm conclusion~\cite{STAR:2017ckg}.

It was recently pointed out that low-$\pt$ dileptons could be affected by the strong magnetic field~\cite{Adam:2018tdm}. The back-to-back dilepton pair will experience the magnetic force and bend in the opposite directions, increasing the total $\pt$ of the pair. The magnetic field may also reduce the $\pt$ of the pair that is not strictly back-to-back, depending on the details of the pair configuration. The net effect would be a broadening of the dilepton pair $\pt$ distribution. A strong broadening, on the order of 30~\mevc, may have been observed by STAR~\cite{Adam:2018tdm} by comparing the dilepton $\pt$ spectrum in hadronic peripheral collisions to model calculations. The effect is consistent with an integral of $\int eBd\ell\sim 0.3m_\pi^2\cdot{\rm fm}$, or effectively a magnetic field of $0.3m_\pi^2$ acting over a distance of 1~fm~\cite{Adam:2018tdm}. The inferred broadening by STAR~\cite{Adam:2018tdm} is statistically significant, and hence could be a good indication of a long-lived strong magnetic field. Other physical reasons are nevertheless also possible. For example, ATLAS has measured an angular broadening of the back-to-back muon pairs and attributed their observation to Coulomb scatterings in the QGP medium~\cite{Aaboud:2018eph}. Calculations confirm that the acoplanarity of dilepton pairs can be attributed to QED radiation in the QGP~\cite{Klein:2018fmp}.
Moreover, the simple radial Coulomb field due to the net positive charge in the collision fireball may also be a possible explanation. This is because the positively (negatively) charged lepton will gain (lose) an overall radial $\pt$. Since the leptons are initially back-to-back, this Coulomb effect will give a net $\pt$ to the pair, on the order of 10~\mevc\ for a typical fireball size.

Because the magnetic field is on average in the $y$ direction, the $\pt$ broadening should happen only in the $x$ direction, i.e.~$p_x$ broadening. Because of fluctuations, the $x$ and $z$ components of the magnetic field do not vanish, so there may also be broadening in $p_y$. However, over the path of the dilepton trajectories through the QGP, the $x$ and $z$ components of the magnetic field should average to approximately zero, resulting in little net broadening in $p_y$. It is, therefore, expected that the $\pt$ broadening due to the magnetic field should primarily happen in the $x$ direction than the $y$ direction. On the other hand, elliptic flow, which can also result in a larger $p_x$ than $p_y$, should be negligibly small at small dilepton pair $\pt$. Thus, a larger $p_x$ than $p_y$ broadening would be a crucial test for the validity of the magnetic field broadening mechanism.

Theoretically many uncertainties prevent a full understanding of the magnetic field in the conductive QGP. The time evolution of the magnetic field in heavy-ion collisions is far from settled~\cite{McLerran:2013hla}. The difference between many theoretical approaches to the CME lies in the assumptions on the length of persistence of the magnetic field generated by the colliding nuclei~\cite{Muller:2018ibh}. A relatively large magnitude of the CME is thus not theoretically guaranteed. Whether the CME exists and how large it is will have to be answered experimentally. On the other hand, an observation of the CME-induced charge separation in heavy-ion collisions would confirm several fundamental properties of QCD~\cite{Lee:1973iz,Lee:1974ma,Morley:1983wr,Kharzeev:1998kz}, namely, the approximate chiral symmetry restoration, topological charge fluctuations, and local $\mathcal{P}$ and $\mathcal{CP}$ violations. It may also solve the long-standing strong \CP\ problem. It is therefore clearly of paramount importance.

\subsection{The CME in condensed matter physics}
The CME phenomenon is not unique to heavy-ion collisions and QCD. It is also an important topic in condensed matter physics where high-energy physics concepts of Dirac and Weyl fermions are adapted~\cite{Miransky:2015ava}. Dirac fermions are spin-1/2 massless particles described by the Dirac equation, and in the condensed matter context represented by the spin-degenerate valence and conduction bands crossing in single points. A Dirac semimetal possesses spin-orbit four-fold degenerate Dirac nodes at the Fermi surface level~\cite{Armitage:2017cjs}. These Dirac nodes correspond to chiral quasi-particles and have been realized in several (topological) materials~\cite{Li:2014bha,Lv:2015pya}. 
Weyl fermions are solutions to the Weyl equation~\cite{Weyl:1929fm}, derived from the Dirac equation. Weyl fermions as fundamental particles have not been discovered in elementary particle physics. Analogs of Weyl fermions may have been observed in so-called Weyl semimetals, where the chirality degeneracy of Dirac nodes are lifted~\cite{Lv:2015pya,Xu:2015cga,Huang:2015eia}. These nodes are called Weyl nodes and the corresponding quasiparticles behave like massless Weyl fermions with definite chirality. 
By applying parallel external electric and magnetic fields to those topological materials, the originally equal populations of left and right chirality Weyl fermions are now out-balanced. This resembles the CME.
Weyl fermions could be realized as an emergent phenomenon by breaking either inversion or time-reversal symmetry in Dirac semimetals, and therefore may intrinsically violate the $\mathcal{P}$ and \CP\ symmetries~\cite{Lv:2015pya}. 

The physics of the CME in heavy-ion collisions and QCD and the physics of the CME in condense matter are different. They may share the same aspects in mathematics and may be fundamentally connected in physics in terms of topology and symmetry breaking~\cite{Mizher:2018dtf}. However, the observation of the CME in condense matter materials does not bear implications on the existence, or not, of the CME in heavy-ion collisions and QCD. The efforts on the CME search in heavy-ion collisions are unique and indispensable.

\subsection{Other chiral effects}
Besides the CME, several other chiral effects have been predicted, most notably, the Chiral Magnetic Wave (CMW)~\cite{Burnier:2011bf,Kharzeev:2010gd} and the Chiral Vortical Effect (CVE)~\cite{Kharzeev:2009fn,Son:2009tf}.
The CMW is a collective excitation formed by the CME and the chiral separation effect (CSE). The CSE is a separation of the chiral charge along the magnetic field in the presence of a finite vector charge density, e.g.~at finite baryon number density and electric charge density~\cite{Son:2004tq,Metlitski:2005pr,Mueller:2016ven}. It is a propagation of chiral charge density in a long wave-length hydrodynamic mode~\cite{Burnier:2011bf,Kharzeev:2010gd,Newman:2005hd,Gorbar:2011ya}, and would result in a finite electric quadrupole moment in heavy-ion collisions. 

The CVE refers to charge separation along the direction of the vorticity, which is large in non-central heavy-ion collisions with a large total angular momentum~\cite{Kharzeev:2009fn,Son:2009tf}. This is due to an effective interaction between the quark spin and the vorticity causing the spin to align up with the vorticity. This interaction is similar to the interaction of a magnetic moment in magnetic field but is charge blind. The CVE results in vector charge density (particularly baryon density) separation. Therefore, one would have, just like charge separation in CME, a quark-antiquark (baryon-antibaryon) separation along the direction of the total angular momentum~\cite{Kharzeev:2015znc}.

In this review, we will concentrate on the CME, touching only briefly on the CMW and CVE. This review focuses mainly on the {\em experimental} searches for the CME. The reader is referred to the literature (for example, Refs.~\cite{Kharzeev:2015znc,Kharzeev:2013ffa,Huang:2015oca} and references therein) for a thorough review of the theoretical aspects of the CME and other chiral anomaly effects. The rest of the review is organized as follow. Sect.~\ref{sec:early} reviews the early measurements of charge correlations in search for the CME. Sect.~\ref{sec:bkgd} discusses the physics backgrounds in those early measurements. Sect.~\ref{sec:attempts} discusses some of the early efforts to remove those physics backgrounds. Sect.~\ref{sec:efforts} describes recent innovative efforts that we believe have the best capability to date to quantify the CME. Sect.~\ref{sec:future} gives future perspectives on the search for the CME. Sect.~\ref{sec:summary} summarizes our review.

\section{Early measurements}\label{sec:early}
The unique signature of the CME is the charge separation along the strong magnetic field in heavy-ion collisions, perpendicular on average to the reaction plane. Experiments at RHIC and the LHC are well equipped to measure charge separation effect with respect to the RP. Intensive efforts have been invested to search for the CME in heavy-ion collisions at RHIC and the LHC~\cite{Kharzeev:2015znc,Zhao:2018ixy,Zhao:2018skm}.

\subsection{The three-point correlator}\label{sec:gamma}
Among various observables~\cite{Adamczyk:2013kcb,Ajitanand:2010rc,Magdy:2017yje,Bozek:2017plp,Feng:2018chm}, a commonly used observable to measure the CME-induced charge separation in heavy-ion collisions is the three-point correlator~\cite{Voloshin:2004vk}.
In non-central heavy-ion collisions, the overlap interaction region is of an almond shape. High energy or matter densities are build up during the collision due to compression and conversion of kinetic energy of the colliding nuclei into thermal energy in the central rapidity region. This high energy density region expands anisotropically because of the anisotropic geometry of the overlap region. This is commonly attributed to hydrodynamic expansion~\cite{Ollitrault:1992bk,Heinz:2013th}, but other contributions may not be negligible, perhaps even dominant in some cases~\cite{He:2015hfa,Lin:2015ucn,Romatschke:2015dha,Romatschke:2018wgi,Kurkela:2018ygx,Kurkela:2018qeb}. The anisotropic expansion results in an anisotropic distribution of particles in momentum space. The particle azimuthal distribution (in momentum space) is often described by a Fourier decomposition~\cite{Voloshin:1994mz},
\be
  \frac{dN}{d\phi} \propto 1 + 2v_1\cos(\phi-\psiRP) + 2v_2\cos2(\phi-\psiRP) + ... 
  \label{eq:fourier}
\ee
where $\phi$ is the particle azimuthal angle and $\psiRP$ is that of the RP direction.
The Fourier coefficients $v_1$ and $v_2$ are referred to as the directed and elliptic flow parameters~\cite{Reisdorf:1997fx,Herrmann:1999wu}. The directed flow accounts for the sidewards deflection of particles in the impact parameter direction after the nuclei pass each other. The elliptic flow stems from the transverse expansion of the almond-shape overlap region.

In order to model the particle emission along the magnetic field arising from the CME, perpendicular on average to the RP direction, a sine term is introduced in the Fourier expansion of the particle azimuthal distribution,
\be
    \frac{dN}{d\phi} \propto 1 + 2v_1\cos(\phi-\psiRP) + 2a_1\sin(\phi-\psiRP) + 2v_2\cos2(\phi-\psiRP) + ...
  \label{eq:sine}
\ee
The parameter $a_1$ can be used to describe the charge separation effect.
Positively and negatively charged particles have opposite $a_1$ values, $a_1^{+}=-a_1^{-}$.
However, they average to zero  because of the random topological charge fluctuations from event to event~\cite{Kharzeev:2004ey}, 
making a direct observation of this parity violating effect impossible.
It is possible only via two-particle correlations, e.g.~by measuring $\mean{a_{\alpha}a_{\beta}}$ with the average taken over particle pairs ($\alpha$ and $\beta$) over all events in a given event sample.
The three-point $\gamma$ correlator is designed for this purpose~\cite{Voloshin:2004vk}, 
\be
  \gamma = \mean{\cos(\phia+\phib-2\psiRP)}\,,
  \label{eq:gammaRP}
\ee
where $\phia$ and $\phib$ are the azimuthal angles of two particles. Charge separation along the magnetic field, which is perpendicular to $\psi$ on average, would yield different values of $\gamma$ for particle pairs of same-sign (SS) and opposite-sign (OS) charges: $\gSS=-1$ and $\gOS=+1$, respectively, opposite in sign and same in magnitude. Here OS ($+-$, $-+$) and SS ($++$, $--$) stand for the charge sign combinations of the $\alpha$ and $\beta$ particles.
Since the $\gamma$ correlators are two-particle correlation measurements averaged over all particle pairs, the final CME signals would be $\gOS=a_1^2$ and $\gSS=-a_1^2$.
It is worthwhile to note that $a_1$ describes an overall effect of the CME from possibly multiple metastable domains of nonzero topological charges. The sign of the topological charge is random from domain to domain. So the overall $a_1$ is not proportional to the number of domains, $\Ndomain$, but only to $\sqrt{\Ndomain}$. 

To assess the RP, experimentally one constructs an event plane (EP) from the azimuthal distribution of final-state particles (see Sect.~\ref{sec:EP}). The $\gamma$ correlator can then be obtained by
\be
  \gamma=\mean{\cos(\phia+\phib-2\psi)}\approx\mean{\cos(\phia+\phib-2\psiEP)}/\mathcal{R}_{\rm EP}\,.
  \label{eq:gammaEP}
\ee
In Eq.~(\ref{eq:gammaEP}), in place of $\psiRP$, we have written generally $\psi$ to stand for harmonic plane. While the correlator should ideally be measured relative to $\psiRP$ (or more precisely the direction perpendicular to $\psiB$), experimentally many different ways have been used to determine the harmonic plane $\psi$ to measure the $\gamma$. 
The $\mathcal{R}_{\rm EP}$ in Eq.~(\ref{eq:gammaEP}) is the resolution factor to correct for the inaccuracy in determining the $\psi$ by $\psiEP$ (see Sect.~\ref{sec:EP}).
Often the EP is constructed from mid-rapidity particles produced in heavy-ion collisions. Then the $\psi$ in Eq.~(\ref{eq:gammaEP}) is the azimuthal angle of the so-called participant plane (PP), $\psiPP$~\cite{Alver:2006wh}. Sometimes the EP is constructed from spectator neutrons, then the $\psi$ is the spectator plane (SP) angle, which is essentially $\psiRP$~\cite{Xu:2017qfs}.

The $\gamma$ correlator can also be calculated by the three-particle correlation method without an explicit determination of $\psiEP$~\cite{Voloshin:2004vk},
\be
  \gamma=\mean{\cos(\phia+\phib-2\psi)}\approx\mean{\cos(\phia+\phib-2\phi_c)}/\vc\,.
  \label{eq:gamma}
\ee
The role of the reaction plane (or harmonic plane in general) is instead fulfilled by the third particle, $c$, and $\vc$ is the elliptic flow parameter of the particle $c$, serving as the resolution of using a single particle to measure the RP. See Sect.~\ref{sec:ebye} for details of $v_2$ calculation.
Often the particle $c$ is a produced particle in heavy-ion collisions, either in the same phase space of the $\alpha$ and $\beta$ particles or from a different phase space. In these cases, the three-point correlator measures charge correlations with respect to the PP. 
The two sides in Eq.~(\ref{eq:gamma}) would be equal if the particle $c$ is correlated with particles $\alpha$ and $\beta$ only through the common correlation to the $\psi$, without contamination of nonflow (few-particle) correlations between $c$ and $\alpha$ and/or $\beta$.
The same can be said about Eq.~(\ref{eq:gammaEP}) because the $\psiEP$ is constructed essentially from particles.

\subsection{Harmonic planes and azimuthal anisotropies}\label{sec:EP}
As aforementioned, experiments have measured the $\gamma$ correlator, as in Eq.~(\ref{eq:gamma}), with respect to different definitions of $\psi$. In fact, the particle azimuthal distribution is expressed, instead of Eq.~(\ref{eq:fourier}), more generally by
\be
\frac{dN}{d\phi} \propto 1 + \sum_{n=1}^{\infty} 2v_n\cos n(\phi-\psi_n)\,.
\label{eq:Fourier}
\ee
Experimentally, to assess $\psi_n$, an EP angle is constructed from the azimuthal distribution of the final-state particle density using the fact that the particle density is the largest along the short axis of the collision overlap geometry [see Fig.~\ref{fig:hi} and Eq.~(\ref{eq:fourier})]~\cite{Poskanzer:1998yz}. The \nth-order ($n=1,2,3,...$) harmonic EP is often constructed by the so-called $Q$ flow vector method, where a Q-vector is defined as
\be
  Q_n=\frac{1}{M}\sum_{j=1}^{N} w_j e^{in\phi_j}\,.
  \label{eq:Q}
\ee
$Q_n$ sums over the particles within a particular phase space in the event ($M$ is the particle multiplicity); $\phi_{j}$ is the azimuthal angle of the $j$-th particle, and $w_j$ is the weight.
Depending on experiments and detectors, the weight is applied in order to account for finite detector granularity or efficiency. 
The \nth-order harmonic EP azimuthal angle ($\psi_{\nEP}$) is then calculated by
\be
e^{in\psi_{\nEP}}= Q_n/|Q_n|\equiv\hat{Q}_{\nEP}\,.\label{eq:qEP}
\ee
The 2nd harmonic EP is the main component in heavy ion collisions and is often simply written as $\psi_{\rm EP}\equiv\psi_{2,{\rm EP}}$, as in Eq.~(\ref{eq:gammaEP}).
Due to finite multiplicity of particles used in the EP calculation, the reconstructed $\psi_{\nEP}$ is not identical to the true harmonic plane $\psi_n$, but has a resolution reflecting its reconstruction accuracy. In order to improve the resolution, a $\pt$ weight is sometimes included in the weight $w_j$. This is because higher $\pt$ particles are more anisotropically distributed and thus more powerful in determining the harmonic plane. The EP resolution is often obtained by the sub-event method, using an iterative procedure~\cite{Poskanzer:1998yz}.

The $\psi_n$ is the azimuthal direction of maximum particle emission probability of the \nth\ harmonic component in the limit of infinite multiplicity. The first-order $\psi_1$ is called directed flow harmonic plane, and the second-order $\psi_2$ is called elliptical harmonic plane. They correspond to the short axis of each harmonic component of the overlap geometry, i.e.~the PP. For example, $\psi_2$ is the elliptical PP, and since it is the main component, it is often simply called PP, $\psiPP\equiv\psi_2$.
Because of fluctuations of the nucleon positions in the colliding nuclei, the reconstructed PP unnecessarily coincides with the RP, but fluctuating about it event-by-event~\cite{Alver:2006wh}. The $\psi_2$ is often reconstructed from produced particles. The RP, on the other hand, is more accurately represented by the spectator plane, which can be determined by the spectator neutrons measured by zero-degree calorimeters (ZDC)~\cite{Adler:2003sp} (usually labeled as $\psi_1$ as in Fig.~\ref{fig:star_y7}) because of a slight side kick they receive from the collision~\cite{Reisdorf:1997fx,Herrmann:1999wu}. For simplicity and without loss of generality, we use $\psiRP$ for the present discussion. Section~\ref{sec:plane} will discuss how to utilize these different planes to better extract the possible CME signal.

Various methods are available to measure the azimuthal anisotropies $v_n$~\cite{Poskanzer:1998yz}. The most obvious one is to calculate the Fourier coefficient,
\be
v_n=\mean{\cos n(\phi-\psi_n)}=\mean{\cos n(\phi-\psi_{\nEP})}/\mathcal{R}_{\nEP}\,,
\label{eq:vn}
\ee
where $\psi_{\nEP}$ is given by Eq.~(\ref{eq:qEP}), and $\mathcal{R}_{\nEP}$ is the \nth\ harmonic EP resolution.
In Eq.~(\ref{eq:vn}), $\phi$ is the particle azimuthal angle and the average is taken over all particles of interest (POI). 
To avoid self-correlation~\cite{Poskanzer:1998yz}, the POI is excluded from the $\psi_{\nEP}$ calculation of Eq.~(\ref{eq:qEP}) via Eq.~(\ref{eq:Q}). This is often achieved by taking the particles for the $\psi_{\nEP}$ calculation and the POIs from different phase spaces. If not separated in phase space, then the $\psi_{\nEP}$ has to be recalculated for each of the POIs before taking the $\cos n(\phi-\psi_{\nEP})$ in Eq.~(\ref{eq:vn}), so the $\psi_{\nEP}$ angles are slightly different for different POIs. 
  
Another method to obtain $v_n$ is via two-particle cumulant~\cite{Poskanzer:1998yz,Bilandzic:2010jr,Bilandzic:2013kga},
\be
v_n=\sqrt{\frac{M|Q_n|^2-1}{M-1}}\,,
\label{eq:c2}
\ee
where $Q$ is given by Eq.~(\ref{eq:Q}) for the POIs.
One can also form the two-particle cumulant between POI and reference particles from different phase spaces,
\be
c_{2,n}=Re(Q_n^{*}Q_{n,{\rm ref}})\,,
\ee
and $v_n$ is obtained from
\be
v_n=c_{2,n}/v_{n,{\rm ref}}
\ee
where $v_{n,{\rm ref}}$ is obtained by Eq.~(\ref{eq:c2}).
This is sometimes called the differential $v_n$ method~\cite{Poskanzer:1998yz} in that it is useful to obtain $v_n$ in a small phase space (such as a narrow $\pt$ bin) with reference particles from a wide phase space (so that the statistics are good).
The two-particle cumulant method and the EP method are almost the same, both relying on two-particle correlations. The two-particle method treats all pairs the same way and obtain the $v_n$ from the average of them. The EP method correlates a particle with the EP reconstructed from all other particles, so it equivalently takes into account all pairs in obtaining the $v_n$. The two methods therefore yield approximately equal $v_n$, with the EP-method $v_n$ relatively smaller than the two-particle cumulant $v_n$ by a few percent.

The two-particle correlations are contaminated by effects other than the global correlation of all particles to the harmonic plane. Those correlations are often dubbed as ``nonflow'' correlations, and include resonance decays and jet correlations. Nonflow correlations are usually short ranged, so an $\eta$ gap between the two particles can effectively reduce those nonflow contributions~\cite{Borghini:2000cm,Borghini:2001vi,Kikola:2011tu,Xu:2012ue,Jia:2013tja,Abdelwahab:2014sge}. Most of the EP and two-particle cumulant analyses apply an $\eta$ gap between the POI and the reference particles (or EP). The $\eta$-gap method is sometimes called sub-event method~\cite{Poskanzer:1998yz}. For the reference particle $v_n$, often two regions symmetric about midrapidity are used; this is, however, only useful to obtain the $v_{n,{\rm ref}}$ in symmetric collision systems. In this case,
\be
v_{n,{\rm ref}}=\sqrt{Q^{*}_{n,{\rm ref}1}Q_{n,{\rm ref}2}}\,,
\ee
where $Q_{n,{\rm ref}1}$ and $Q_{n,{\rm ref}2}$ are the \nth-harmonic Q-vectors of the reference particles from two symmetric phase spaces.
There are nonflow correlations that are long-ranged, e.g.~the back-to-back away-side jet correlations. To reduce this nonflow, three-subevent method has been proposed~\cite{Jia:2017hbm}.

The four-particle cumulant method is also used to analyze $v_n$~\cite{Poskanzer:1998yz}. The four-particle cumulant $v_n$ is smaller than the two-particle one because the flow fluctuation effect in $\vnf$ is negative while it is positive in $\vnt$~\cite{Poskanzer:1998yz}. For the $\vc$ in CME analyses, the EP method and the two-particle cumulant method are used, not the four-particle cumulant method, because the $v_2$ in the $\gamma$ variable is of two-particle correlation nature.

\subsection{First measurements at RHIC}\label{sec:STAR}
\begin{figure}[!htb]
  \centerline{\includegraphics[width=0.45\hsize]{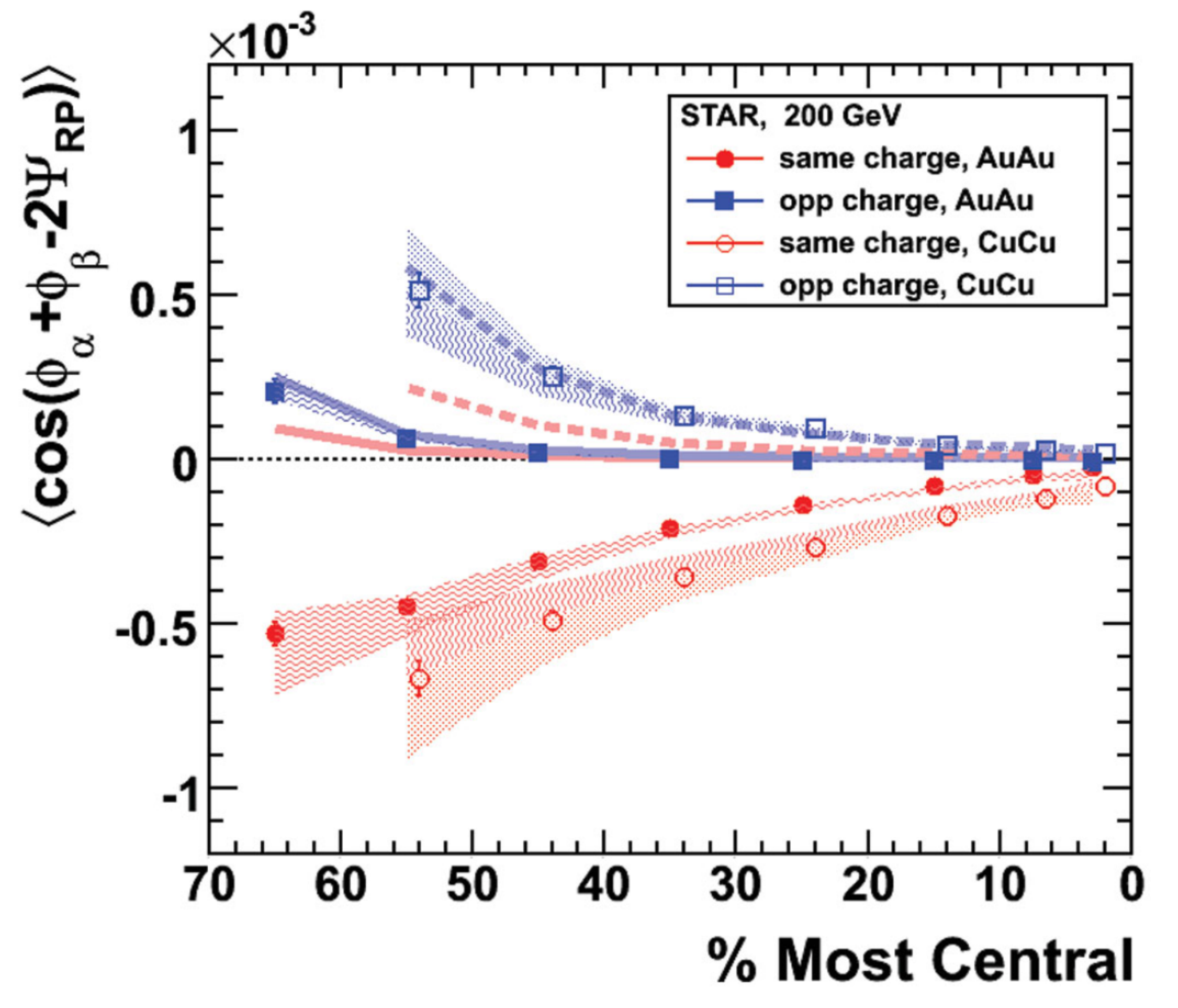}}
  \caption{The azimuthal $\gamma$ correlators as functions of centrality in \AuAu\ and Cu+Cu collisions at $\snn=200$~GeV from STAR. Shaded bands represent uncertainty from the $v_2$ measurement. The thick solid (\AuAu) and dashed (Cu+Cu) lines represent HIJING calculations of the contributions from three-particle correlations. Adapted from Ref.~\cite{Abelev:2009ac,Abelev:2009ad}.}
  \label{fig:star}
\end{figure}

The STAR experiment at RHIC made the first measurement of charge correlations in \AuAu\ collisions at the nucleon-nucleon center-of-mass energy of $\snn=200$~GeV taken in 2004~\cite{Abelev:2009ac,Abelev:2009ad}.
Figure~\ref{fig:star} shows the $\gamma$ correlators as functions of the collision centrality in \AuAu\ and Cu+Cu collisions at $\snn=200$~GeV from STAR~\cite{Abelev:2009ad}. The $\gOS$ and $\gSS$ correlators decrease with increasing centrality, mainly because of the combinatorial dilution effect by the multiplicity. This is also responsible for the larger correlator values in Cu+Au than \AuAu\ collisions. 
Although the OS and SS results are not the same in magnitude and opposite in sign as would be expected from the CME, the OS result is larger than the SS result. This OS-SS difference is qualitatively consistent with the CME expectations~\cite{Abelev:2009ac,Abelev:2009ad}. A CME signal is also expected to decrease with centrality because the magnetic field strength decreases with increasing centrality~\cite{Kharzeev:2007jp,Fukushima:2008xe}.
So the decrease of the correlators with increasing centrality may also contain influence from the CME.
Given the particle azimuthal distribution of Eq.~(\ref{eq:sine}), the $\gamma$ observables would be $\pm a_1^2$ (see Sect.~\ref{sec:gamma}). The measured $\gamma$ magnitudes of the order of $10^{-4}$ therefore agree with the predictions of the CME signal of the order of $a_1=10^{-2}$ in Refs.~\cite{Kharzeev:2004ey,Kharzeev:2007tn,Yin:2015fca,Jiang:2016wve,Shi:2017cpu}. Meanwhile, other predictions~\cite{Muller:2010jd,Asakawa:2010bu} are significantly smaller.

\begin{figure}[!htb]
  \centerline{\includegraphics[width=0.45\hsize]{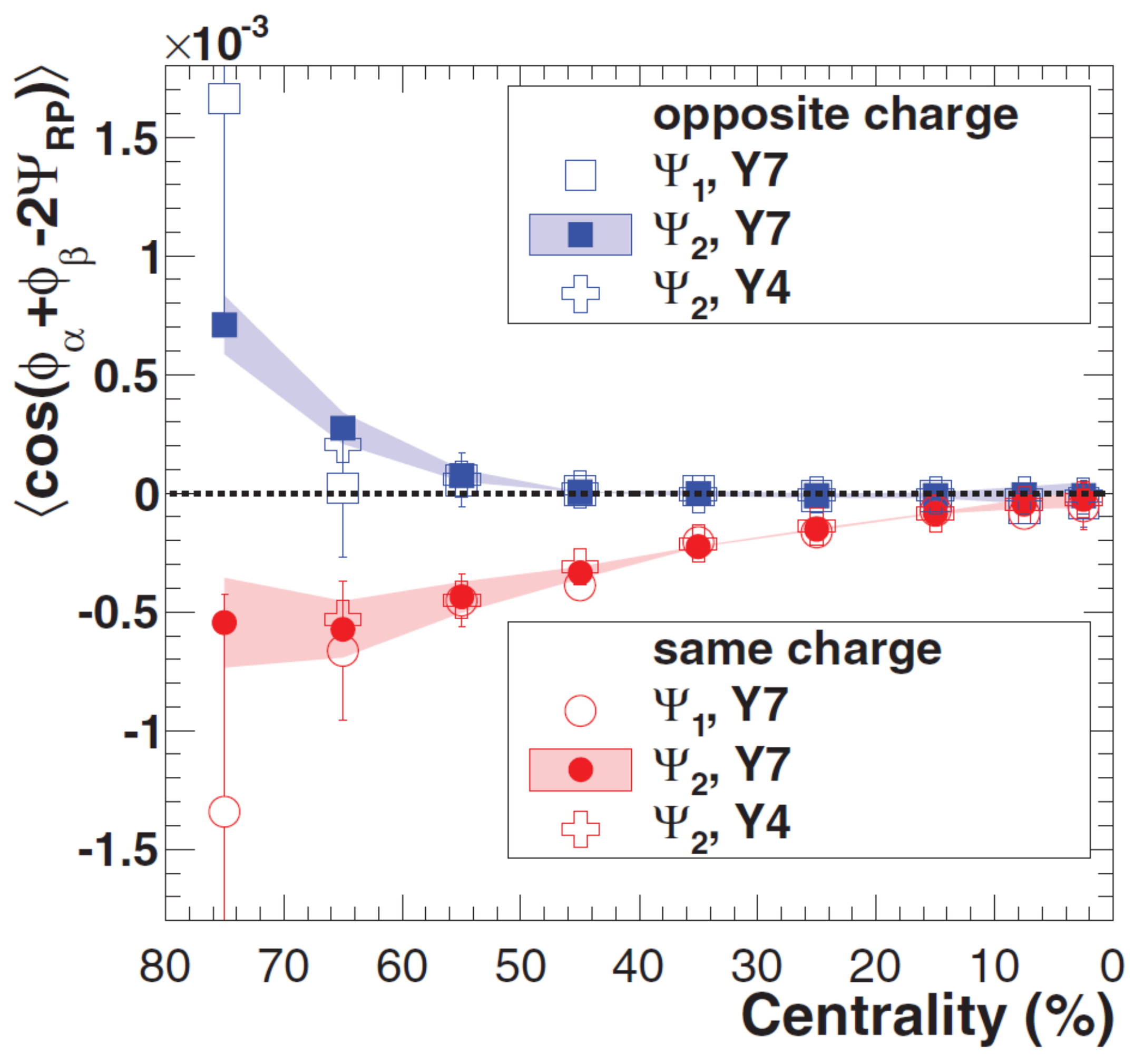}}
  \caption{(Color online) The azimuthal correlator $\gamma$ measured with the first-order event plane $\psi_1$ from the ZDC and the second-order event plane from the time projection chamber (TPC) as functions of centrality in \AuAu\ collisions at $\snn=200$~GeV from STAR. The Y4 and Y7 represent the results from the 2004 and 2007 RHIC run. Shaded areas for the results measured with $\psi_2$ represent the systematic uncertainty of the event plane determination. Systematic uncertainties for the results with respect to $\psi_1$ are negligible compared to the statistical ones shown. Adapted from Ref.~\cite{Adamczyk:2013hsi}.}
  \label{fig:star_y7}
\end{figure}

STAR further analyzed the \AuAu\ data from RHIC Run-7 (taken in year 2007), with both the first-order harmonic plane $\psi_1$ measured by the ZDCs and the second-order harmonic plane $\psi_2$ measured by mid-rapidity hadrons in the TPC~\cite{Adamczyk:2013hsi}. The data are shown in Fig.~\ref{fig:star_y7} and suggest that the CME is a possible explanation for the data. The results for $\psi_1$ and $\psi_2$ are equal within statistical uncertainties. It was thought that the two results should be equal and their numerical difference was used as an assessment of the systematic uncertainty~\cite{Adamczyk:2013hsi}. However, as will be shown in Sect.~\ref{sec:plane}, this understanding was incorrect and the two results should physically differ because the magnetic field (and flow) projections onto $\psi_1$ and $\psi_2$ are different~\cite{Xu:2017qfs}.

\subsection{Measurements at the LHC}
\begin{figure}[!htb]
  \centerline{\includegraphics[width=0.45\hsize]{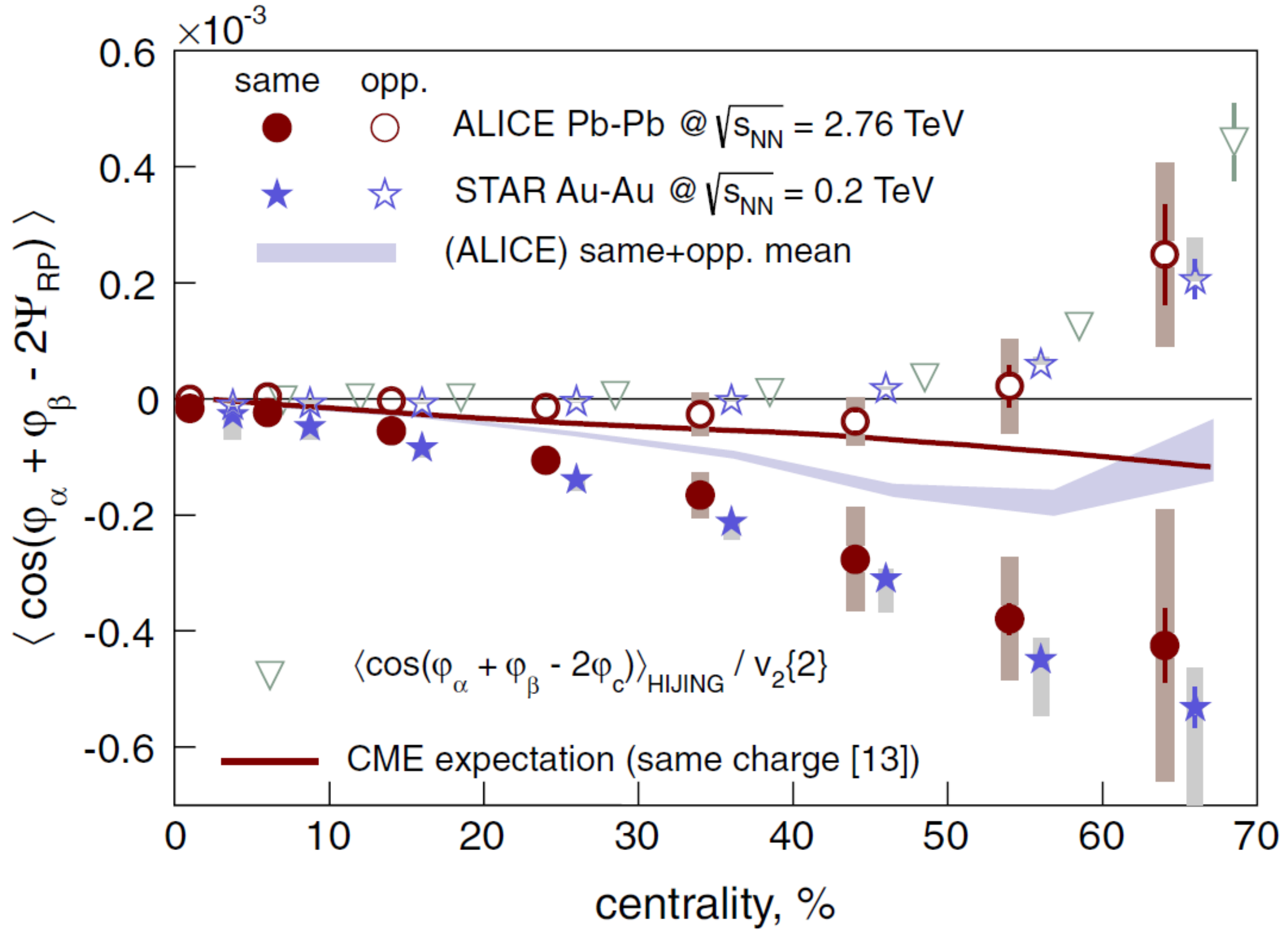}}
  \caption{The centrality dependence of the three-point correlator. The circles indicate the ALICE results obtained from the cumulant analysis. The stars show the STAR data from~\cite{Abelev:2009ac,Abelev:2009ad}. The shaded boxes represent the systematic uncertainties. The triangles and the curve represent the three-point correlations from model calculations. The shaded band represents the centrality dependence of the charge-independent correlations. Adapted from Ref.~\cite{Abelev:2012pa}.}
  \label{fig:alice}
\end{figure}

The $\gamma$ correlators were also measured in \PbPb\ collisions at 2.76 TeV at the LHC by the ALICE experiment~\cite{Abelev:2012pa}. The results are shown in Fig.~\ref{fig:alice}. The results are found to be similar to those measured at RHIC~\cite{Abelev:2009ac,Abelev:2009ad,Adamczyk:2013hsi}. As discussed in Sect.~\ref{sec:B}, the initial magnetic field strength is larger at the LHC than at RHIC, but the field duration may be shorter. The net effect may thus weakly depend on the collision energy. The experimental data at RHIC and the LHC are consistent with this expectation from the CME.

\subsection{Beam-energy dependent measurements}
\begin{figure}[!htb]
  \centerline{\includegraphics[width=0.5\hsize]{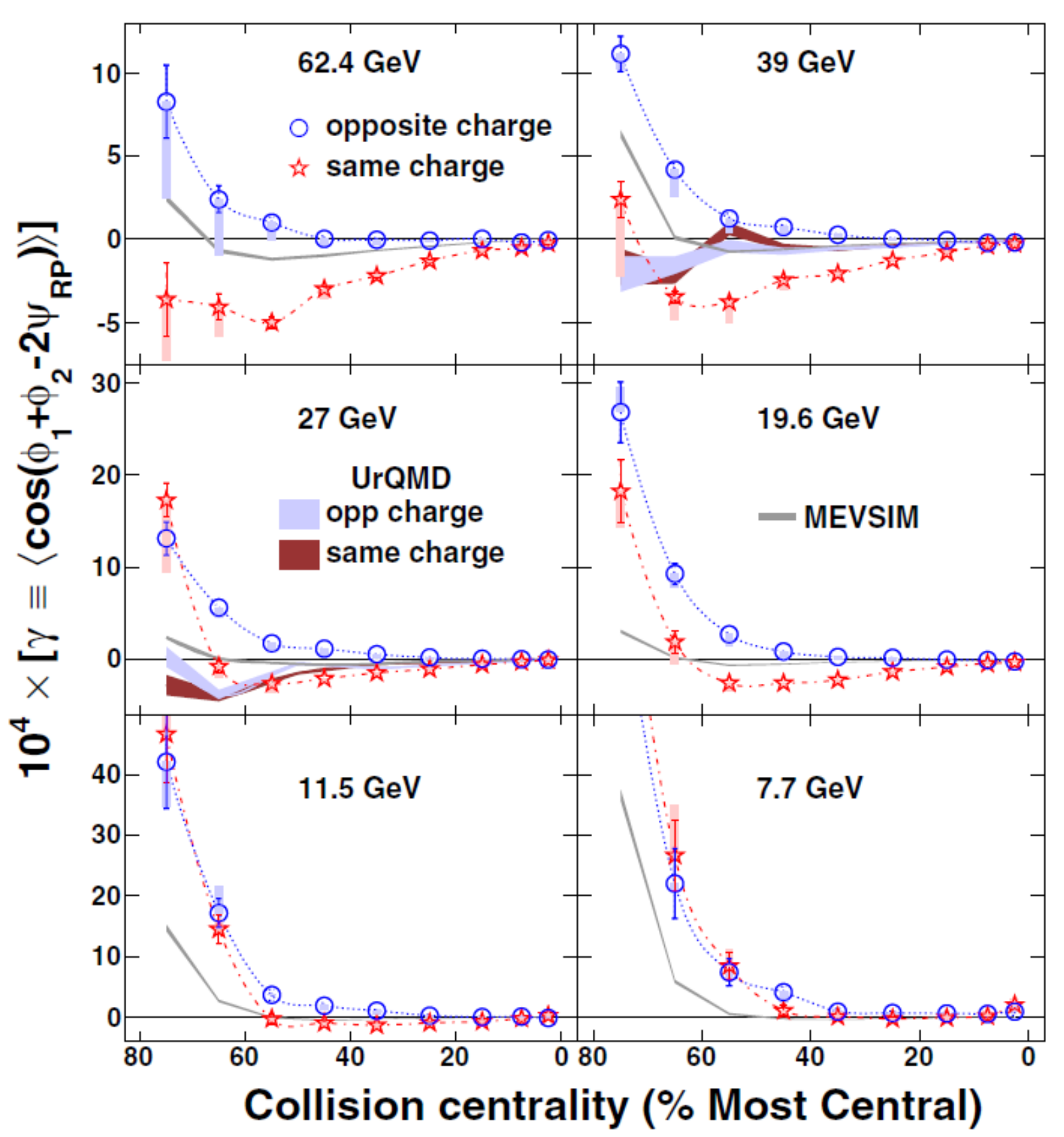}}
  \caption{The three-point $\gamma$ correlators as functions of centrality in \AuAu\ collisions at $\snn=7.7-62.4$~GeV. The filled boxes (starting from the central values) represent the range of results suppressing effects from HBT and the final-state Coulomb interaction. The curves and shaded bands are model calculations. Adapted from Ref.~\cite{Adamczyk:2014mzf}.}
  \label{fig:star_bes}
\end{figure}

STAR has measured the $\gOS$ and $\gSS$ correlators in lower energy \AuAu\ collisions from the Beam Energy Scan (BES) data at $\snn=\sim7.7$-62.4~GeV~\cite{Adamczyk:2014mzf}. The results are shown in Fig.~\ref{fig:star_bes}. The results are generally similar to the 200~GeV data, except at the low collision energy of $\snn =7.7$~GeV. There, the difference between $\gOS$ and $\gSS$ disappears. This is suggestive of the disappearance of the CME at this energy, which is expected because hadronic interactions should dominate at this low energy~\cite{Adamczyk:2014mzf}.

\subsection{Measurements related to other chiral effects}
The CMW, closely related to the CME, would result in a finite electric quadrupole moment of the collision system at finite charge density~\cite{Burnier:2011bf,Kharzeev:2010gd}. It would thus alter the elliptic flow anisotropies of hadrons charge-dependently, yielding a split of the $v_2$'s of $\pip$ and $\pim$ dependent on the charge multiplicity asymmetry ($\Ach$)~\cite{Burnier:2011bf}. 
STAR has analyzed the $v_2$ of charged pions as a function of the $\Ach$ measured in the same phase space of the pions~\cite{Adamczyk:2015eqo}. The $v_2$ difference between $\pim$ and $\pip$ was found to be linear in $\Ach$, and a slope parameter of the order of 3\% was extracted for mid-central Au+Au collisions at 200~GeV. The data are consistent with the CMW expectation. 

ALICE~\cite{Adam:2015vje} and CMS~\cite{Sirunyan:2017tax} have also measured the $\Ach$-dependent $v_2$ splitting between $\pip$ and $\pim$. The results are similar to measurements at RHIC and consistent with the CMW expectation.

In the experimental measurements, the same set of particles are used for both $v_2$ and $\Ach$~\cite{Adamczyk:2015eqo,Adam:2015vje,Sirunyan:2017tax}, possible self-correlations are present. It is found that when the $v_2$ and $\Ach$ were measured using exclusive sets of particles so that self-correlation effects are excluded, the effect of $v_2$ splitting is reduced by approximately a factor of three but remains finite~\cite{Adamczyk:2015eqo,Sirunyan:2017tax}. 
It should also be noted that the measured $\Ach$ is affected (perhaps dominated) by statistical fluctuations of finite multiplicities. Because of those statistical fluctuations, the face value of the selected $\Ach$ bin does not directly correspond to the true charge density asymmetry. This affects numerically the extracted slope parameter which, therefore, may not be directly comparable to theoretical calculations of the CMW~\cite{Burnier:2011bf,Burnier:2012ae}.

The CVE would result in a baryon-antibaryon separation along the direction of the total angular momentum, analogous to the CME-induced charge separation. This would yield a distinct hierarchy in the magnitudes of the correlation differences $\dg$: the $p$-$p$ and $p$-$\bar{p}$ correlation difference (containing both CVE and CME) is stronger than the $p$-$\Lambda$ one (containing only CVE) and $\pi$-$\pi$ one (containing only CME), which in turn are stronger than the $\Lambda$-$\pi$ or $K_S^0$-$\pi$ correlation difference (containing neither CVE nor CME). Preliminary data are available from STAR but not finalized~\cite{Zhao:2014aja,Wen:2017ibm,Zhao:2017ckp}. 

\section{Physics backgrounds}\label{sec:bkgd}
\begin{figure*}[!htb]
  \centerline{\includegraphics[width=0.8\hsize]{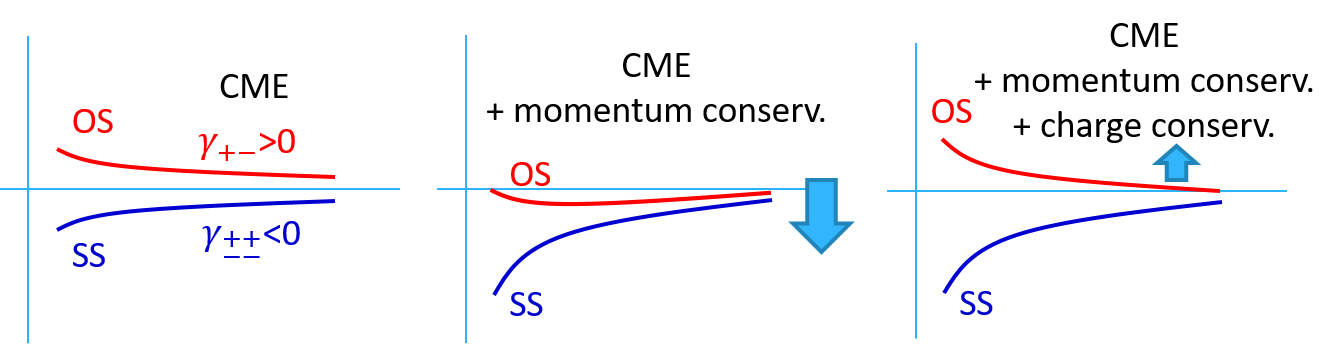}}
  \caption{Left: expected CME signals [Eq.~(\ref{eq:gammaRP})] for opposite-sign (OS) and same-sign (SS) particle pairs, opposite in sign and equal in magnitude. Center: effect of momentum conservation is negative and equal for OS and SS. Right: effect of local charge conservation (e.g.~neutral resonance decays) is positive and only applies to OS.}
  \label{fig:bkgd}
\end{figure*}

All $\gamma$ correlator measurements are qualitatively consistent with the CME expectations.
There, however, exist background correlations unrelated to the CME~\cite{Wang:2009kd,Bzdak:2009fc,Liao:2010nv,Bzdak:2010fd,Schlichting:2010qia,Pratt:2010zn,Petersen:2010di,Toneev:2012zx}. For example, the global transverse momentum conservation induces correlations among particles that enhance back-to-back pairs~\cite{Bzdak:2009fc,Liao:2010nv,Bzdak:2010fd,Schlichting:2010qia,Pratt:2010zn}. Since more pairs are emitted in the RP direction ($\phia+\phib-2\psiRP\approx\pi$), the net effect of this background is negative. This would drag the CME-induced $\gSS$ and $\gOS$, originally symmetric about zero (as illustrated by the left sketch of Fig.~\ref{fig:bkgd}), both down in the negative direction (as illustrated by the central sketch of Fig.~\ref{fig:bkgd}).
This background is, fortunately, independent of particle charges, affecting SS and OS pairs equally and cancels in the difference,
\be
  \dg\equiv\gOS-\gSS\,.
  \label{eq:dg}
\ee
Experimental searches have thus focused on the $\dg$ observable~\cite{Kharzeev:2015znc,Zhao:2018ixy,Zhao:2018skm}; the CME would yield $\dg>0$.

\subsection{Nature of charge-dependent backgrounds}
\begin{figure*}[!htb]
  \centerline{\includegraphics[width=0.30\hsize]{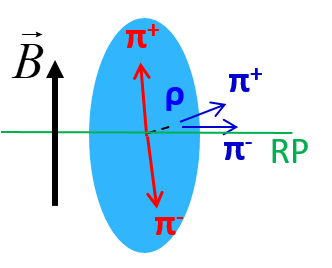}}
  \caption{Illustration that the decay $\pip\pim$ pair from a $\rho$ resonance moving in the RP direction has the same effect on the $\dg$ observable [Eq.~(\ref{eq:gammaRP})] as a CME $\pip\pim$ pair perpendicular to the RP.}
  \label{fig:rho}
\end{figure*}

There are, unfortunately, also mundane physics that differ between OS and SS pairs. One such physics is resonance/cluster decays~\cite{Voloshin:2004vk,Wang:2009kd,Bzdak:2009fc,Liao:2010nv,Bzdak:2010fd,Schlichting:2010qia,Pratt:2010zn}, more significantly affecting OS pairs than SS pairs (as illustrated by the right sketch of Fig.~\ref{fig:bkgd}). This background is positive and arises from the coupling of elliptical anisotropy $v_2$ of resonances/clusters and the angular correlations between their decay daughters (nonflow)~\cite{Voloshin:2004vk,Wang:2009kd,Bzdak:2009fc,Schlichting:2010qia,Wang:2016iov}. 
Take $\rho\rightarrow\pip\pim$ decay as an example (Fig.~\ref{fig:rho}). The effect on $\gOS$ from the decay of a $\rho$ in the RP direction is identical to a back-to-back pair from the CME in the magnetic field direction perpendicular to the RP~\cite{Wang:2016iov}. In other words, the $\dg$ variable is ambiguous between a back-to-back OS pair from the CME perpendicular to the RP ($\phia+\phib-2\psiRP\approx2\pi$) and an OS pair from a resonance decay along the RP ($\phia+\phib-2\psiRP\approx0$ or $2\pi$).
Since there are more $\rho$ resonances in the RP direction than the perpendicular direction because of the finite $v_2$ of the $\rho$ resonances, the overall effect on $\gOS$ is positive. They would produce the same effect as the CME in the $\dg$ variable~\cite{Wang:2009kd,Bzdak:2009fc,Schlichting:2010qia,Wang:2016iov}. 

There are of course more sources of particle correlations except that from $\rho$ decays, such as other resonances and jet correlations. We can generally refer to those as cluster correlations~\cite{Wang:2009kd}.
In general, those backgrounds are generated by two-particle correlations coupled with elliptic flow of the parent sources (clusters)~\cite{Voloshin:2004vk,Wang:2016iov}:
\be
\dg_{\bkg}\approx\frac{N_{\alpha\beta,\clust}}{N_{\pi}^2}\mean{\cos(\phia+\phib-2\phiclust)}\vclust\,.
\label{eq:bkgd}
\ee
Here $N_{\alpha\beta,\clust}$ is the number of pairs from cluster decays and $N_{\pi}$ is the number of single-charge pions ($N_{\pi}\approx N_{\pip}\approx N_{\pim}$), respectively. The $\vclust\equiv\mean{\cos2(\phiclust-\psiRP)}$ is the $v_2$ of the clusters, and $\mean{\cos(\phia+\phib-2\phiclust)}$ is the two-particle angular correlation from the cluster decay.
The factorization of $\mean{\cos(\phia+\phib-2\phiclust)}$ with $\vclust$ is only approximate, because both depend on the $\pt$ of the clusters~\cite{Wang:2016iov}.
A simple estimate~\cite{Wang:2016iov}, again using the $\rho$ resonance as an example, indicates that the background magnitude is $\dg_{\bkg}\approx\frac{20}{100^2}\times0.65\times0.1\approx10^{-4}$ for mid-central \AuAu\ collisions, comparable to the experimental data in Fig.~\ref{fig:star}.
In fact, the magnitude of the possible resonance decay backgrounds was estimated before, but unfortunately with wrong values, and was thus incorrectly thought to be negligible~\cite{Voloshin:2004vk,Abelev:2009ac,Abelev:2009ad}. 
This has led to the premature claim that the CME must be invoked to explain the experimental data~\cite{Kharzeev:2015znc,Wang:2016mkm}. 

\begin{figure}[!htb]
  \centerline{\includegraphics[width=0.5\hsize]{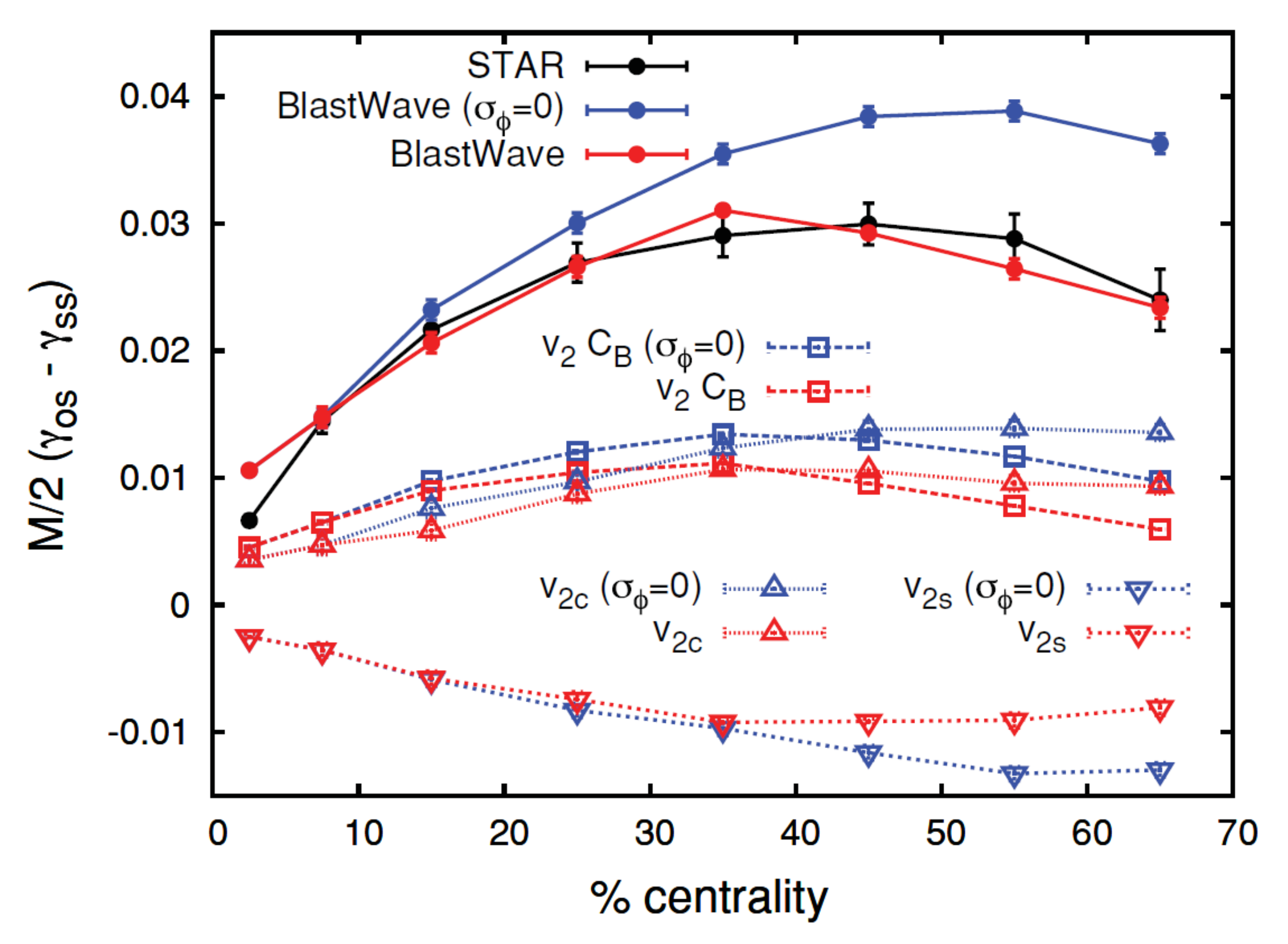}}
  \caption{(Color online) Blast-wave calculations of the $\dg$ observable (multiplied by half of the multiplicity) for realistic local charge separation at freeze-out (red dots) and perfectly local charge conservation (blue dots), compared to the STAR measurement (black dots). The dashed lines of the corresponding color represent the decomposition of three contributions. Adapted from Ref.~\cite{Schlichting:2010qia}.}   
  \label{fig:pratt}
\end{figure}

When the first measurements became available from STAR, one of us (Wang~\cite{Wang:2009kd}) showed, by using the available dihadron angular correlation measurements~\cite{Adler:2002tq,Adams:2004pa,Adams:2005ph,Abelev:2008ac,Daugherity:2008su,Abelev:2009jv}, that the measured $\dg$ magnitudes could be explained by those existing dihadron correlation data.
Bzdak \etal~\cite{Bzdak:2009fc,Liao:2010nv,Bzdak:2010fd} showed that the measured correlation signal is in-plane rather than out-of-plane as would be expected from the CME. This was also concluded by the STAR experiment using the charge multiplicity asymmetry observable~\cite{Adamczyk:2013kcb}. The authors~\cite{Bzdak:2009fc,Liao:2010nv,Bzdak:2010fd} also showed that the global momentum conservation contributes significantly to the measured $\gamma$ signal, and pointed out the importance to also measure the $\delta$ correlator (see Sect.~\ref{sec:kappa}) besides the $\gamma$ correlator.
Pratt \etal~\cite{Pratt:2010zn} also found significant contributions of global momentum conservation to the $\gamma$ observables.
Schlichting and Pratt~\cite{Schlichting:2010qia} showed that the $\dg$ signal by STAR can be fully described by local charge conservation.
Figure~\ref{fig:pratt} shows Blast-wave calculations of the $\dg$ observable incorporating local charge conservation and momentum conservation effects~\cite{Schlichting:2010qia}. An ideal case and a more realistic case of local charge conservation are shown. For each case, the three contributions are shown individually representing more balancing pairs in-plane than out-of-plane, more tightly correlated in-plane pairs in $\dphi$ than out-of-plane, and more balancing charge in-plane than out-of-plane. As seen from Fig.~\ref{fig:pratt}, the realistic case of local charge conservation can almost fully account for the STAR data.
Toneev \etal~\cite{Toneev:2012zx} came to the same conclusion using the parton hadron string dynamics (PHSD) model.
Petersen \etal~\cite{Petersen:2010di} investigated the effect of jet correlations on the CME-sensitive multiplicity asymmetry observable~\cite{Adamczyk:2013kcb} and found it less significant than the effects due to momentum and local charge conservations.

AMPT model simulations can also largely account for the measured $\dg$ signal~\cite{Ma:2011uma,Shou:2014zsa,Zhao:2017nfq}. In the AMPT studies the hadron rescattering is not included because it is known that the hadron cascade in AMPT does not conserve charge~\cite{Zhang:1999bd,Lin:2004en}, which is essential to the charge correlations. However, the hadronic rescattering, while responsible for the majority of the mass splitting of the azimuthal anisotropies~\cite{Li:2016flp,Li:2016ubw}, is not important for the main development of $v_2$, and thus may not be important for the CME backgrounds. Quantitatively, AMPT does not fully account for the measured $\dg$ magnitude. The reason may be that the model does not fully account for the resonance production in real data~\cite{Zhao:2017nfq}. 

\subsection{A background-only three-point correlator}\label{sec:gamma123}
As aforementioned, the CME signal and the cluster-induced correlation background are ambiguous in the $\dg$ observable. They in principle cannot be distinguished by the $\dg$ measurement alone. One needs extra information.

The CME signal is pertinent to the RP. Due to event-by-event geometry fluctuations, there is a triangular component in the azimuthal distributions of the participant nucleons that is mostly uncorrelated with respect to the RP. The triangular component generates a triangular (third-order harmonic) flow ($v_3$) in the final-state momentum space. Since the third-order harmonic plane ($\psi_3$) is random with respect to the RP~\cite{Lacey:2010av}, the CME signal is averaged to zero when analyzed with respect to $\psi_3$ by measuring $\mean{\cos(\phia+\phib-2\psi_3)}$. The background, on the other hand, is due to intrinsic particle correlations and is coupled to the harmonic plane by anisotropic flow. With respect to $\psi_3$, the background would persist in the measurement of the following three-point correlator, 
\be
\gamma_{123}\equiv\mean{\cos(\phia+2\phib-3\psi_3)}\,,
\label{eq:gamma123}
\ee
first suggested by the CMS experiment~\cite{Sirunyan:2017quh}. 
This is different from $\mean{\cos(\phia+\phib-2\psi_3)}$, in which the CME would already average to zero. From $\mean{\cos(\phia+\phib-2\psi_3)}$ to the $\gamma_{123}$ of Eq.~(\ref{eq:gamma123}), there is an additional ``randomization'' by $\phib-\psi_3$, so there would be surely no CME signal surviving in $\gamma_{123}$.
The background, on the other hand, couples now to $\psi_3$ through $v_3$ and therefore persists in $\gamma_{123}$. The background, in the OS and SS difference of $\gamma_{123}$, is similar to Eq.~(\ref{eq:bkgd}) and is proportional to $v_3$, :
\be
  \dg_{123}=\dg_{123}^{\bkg}=\mean{\cos(\phia+2\phib-3\phiclust)}v_{3,\clust}\,.
  \label{eq:dg123}
\ee
So the $\dg_{123}$ three-point correlator is sensitive only to the background, not to the CME.
The study of $\gamma_{123}$ would therefore provide further insights into the background issue in the $\dg$ observable. To distinguish from $\gamma_{123}$, we sometimes denote $\gamma_{112}$ for the $\gamma$ defined in Eq.~(\ref{eq:gamma}), but will use $\gamma$ and $\gamma_{112}$ interchangeably.

\subsection{Small-system collisions}\label{sec:small}
In non-central heavy-ion collisions, the PP, although fluctuating~\cite{Alver:2006wh}, is generally aligned with the RP, thus generally perpendicular to the magnetic field. The $\dg$ measurement with respect to the PP (i.e.~via the experimentally constructed EP) is thus $\emph{entangled}$ by the possible CME signal and the $v_2$-induced background. In small-system proton-nucleus (\pA) and deuteron-nucleus (\dA) collisions, however, the PP arises from geometry fluctuations, uncorrelated to the impact parameter direction~\cite{Khachatryan:2016got,Belmont:2016oqp,Tu:2017kfa}. As a result, any CME signal would average to zero in the $\dg$ measurements with respect to the PP. On the other hand, background sources contribute to small-system collisions similarly as to heavy-ion collisions: resonance/cluster decay correlations are similar, and the collective azimuthal anisotropies seem also similar~\cite{Dusling:2015gta,Nagle:2018nvi}. Small-system \pA\ collisions thus provide a control experiment, where the CME signal can be ``turned off,'' whereas the $v_2$-related backgrounds remain.
It was recently suggested~\cite{Kharzeev:2017uym} that, because of proton size fluctuations, the PP in \pA\ collisions may still have some correlations with the impact parameter direction. In such a case, some CME signal would survive, but the magnitude would be significantly reduced from its original one because of the relatively weak correlation between the PP and the magnetic field direction in \pA\ collisions.

Even the CME could be perfectly measured, there would still be difference between heavy-ion collisions and small-system \pA\ collisions. This is because the magnetic field in small-system collisions is smaller than that in heavy-ion collisions, the approximate chiral symmetry is less likely restored, and the QGP is less likely created. The CME would thus be of smaller magnitude in \pA\ collisions than in heavy-ion collisions. 
It can further our understanding of the CME signal and the background issue in the $\dg$ measurements by comparing the small-system \pA\ collisions to heavy-ion collisions.

\begin{figure}[!htb]
  \centerline{
    \includegraphics[width=0.40\hsize]{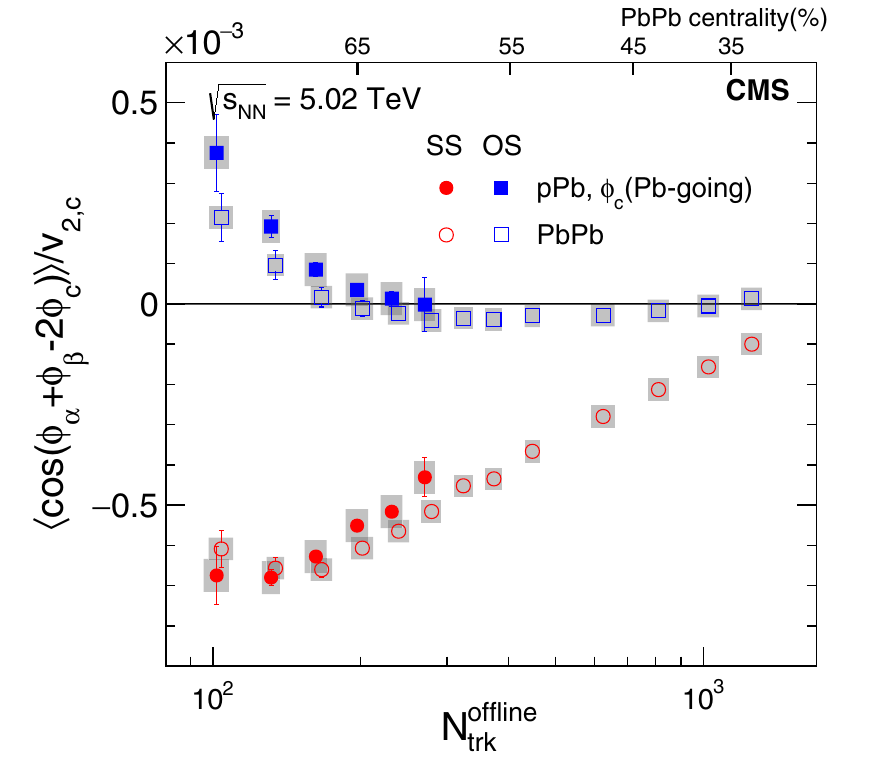} 
    \hspace{0.05\hsize}
    \includegraphics[width=0.45\hsize]{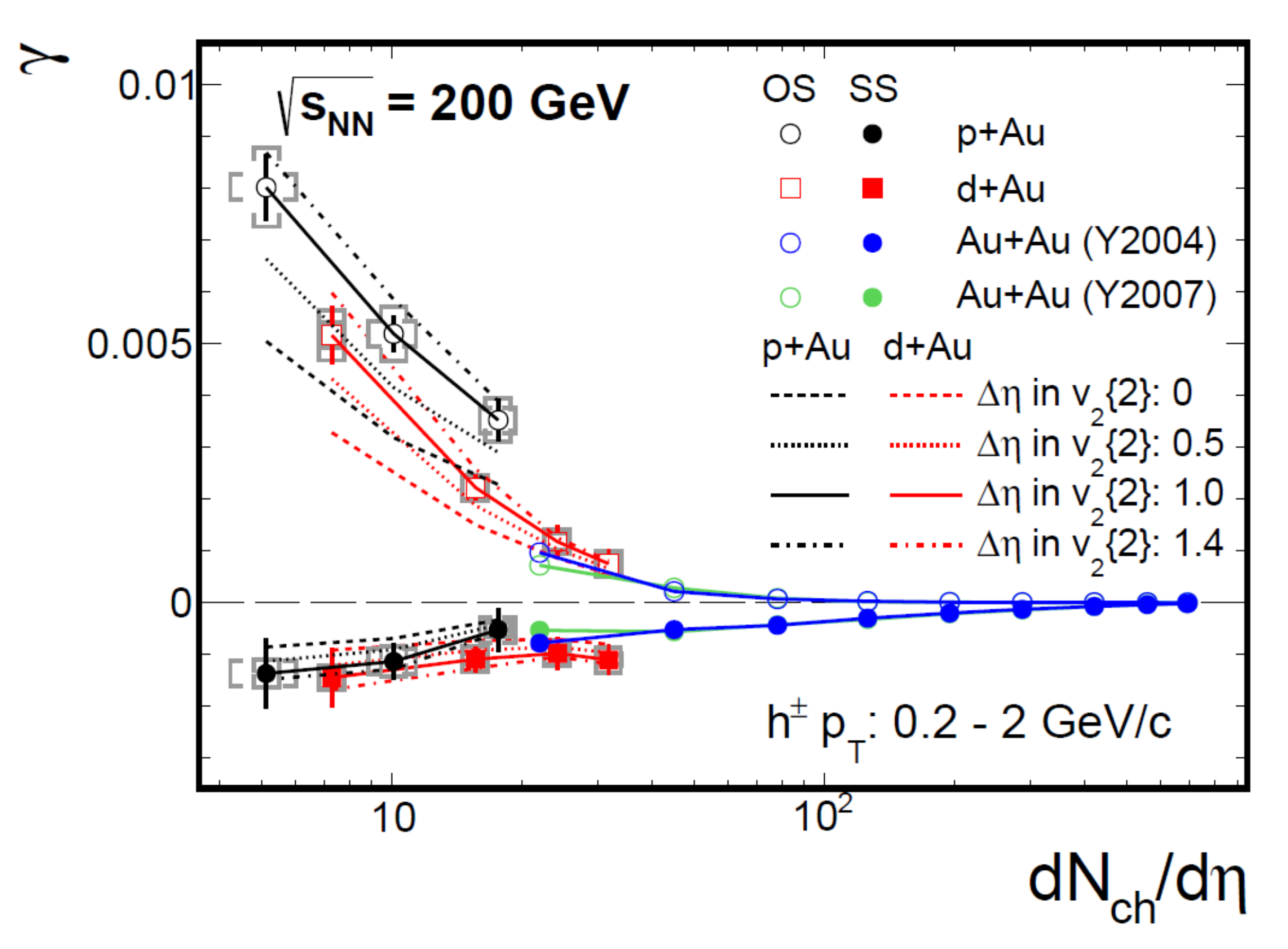}
  }
  \caption{(Color online) The opposite-sign (OS) and same-sign (SS) three-point correlators in \pPb\ and \PbPb\ collisions at $\snn=5.02$~TeV as a function of the offline track multiplicity ($\Noff$) from CMS~\cite{Khachatryan:2016got} (left panel) and in \pAu\ and \dAu\ collisions at $\snn=200$~GeV as a function of the mid-rapidity charged hadron density ($\dNdeta$) from STAR~\cite{ZhaoQM17,Zhao:2017wck,Zhao:2018pnk,STAR:2019xzd} (right panel). The CMS data are averaged over $|\eta_{\alpha}-\eta_{\beta}| < 1.6$; particles $\alpha$ and $\beta$ are from the midrapidity tracker and particle $c$ from the forward/backward hadronic calorimeters. All three particles of the STAR data are from the TPC pseudorapidity coverage of $|\eta|<1$ with no $\eta$ gap applied. The $\vct$ is obtained by two-particle cumulant with $\eta$ gap of $\Delta\eta > 1.0$ (results with several other $\eta$ gaps are also shown as dashed lines). Statistical uncertainties are indicated by the error bars and systematic ones by the shaded regions (CMS) and caps (STAR), respectively.}   
  \label{fig:small}
\end{figure}

Figure~\ref{fig:small} left panel shows the $\dg$ measurements in small-system \pPb\ collisions at 5.02 TeV by CMS~\cite{Khachatryan:2016got}, compared to \PbPb\ collisions at the same energy. 
Within uncertainties, the SS and OS correlators in \pPb\ and \PbPb\ collisions exhibit the same magnitude and trend as a function of the offline track multiplicity ($\Noff$). The CMS data further show that the $|\Delta\eta|=|\eta_{\alpha}-\eta_{\beta}|$ and multiplicity dependences of the $\dg$ correlators are highly similar between \pPb\ and \PbPb\ collisions~\cite{Khachatryan:2016got}.
The $|\Delta\eta|$ dependence shows a short-range correlation structure, similar to that observed in the early STAR data~\cite{Abelev:2009ad}. This suggests that the correlations may come from the late hadronic stage of the collision, while the CME is expected to be a long-range correlation arising from the early stage.
The similarity seen between high-multiplicity \pPb\ and peripheral \PbPb\ collisions strongly suggests a common physical origin, challenging the attribution of the observed charge-dependent correlations to the CME~\cite{Khachatryan:2016got}. 

Similar analysis has also been carried out at RHIC, using \pAu\ and d+Au collisions~\cite{ZhaoQM17,Zhao:2017wck,Zhao:2018pnk,STAR:2019xzd}. Figure~\ref{fig:small} right panel shows the $\gSS$ and $\gOS$ correlators as functions of mid-rapidity charged hadron multiplicity density ($\dNdeta$) in \pA\ and \dA\ collisions at $\snn=200$~GeV, compared to \AuAu\ collisions at the same energy~\cite{Abelev:2009ac,Abelev:2009ad,Adamczyk:2013hsi}. 
The trends of the correlators are similar, decreasing with increasing multiplicity. Similar to LHC, the small-system data at RHIC are found to be comparable to \AuAu\ results at similar multiplicities, although quantitative details may differ. 
Given the large differences in the collision energies and the multiplicity coverages, the similarities between the RHIC and LHC data in terms of the systematic trends from small-system to heavy-ion collisions are astonishing.

\begin{figure}[!htb]
  \centerline{\includegraphics[width=0.50\hsize]{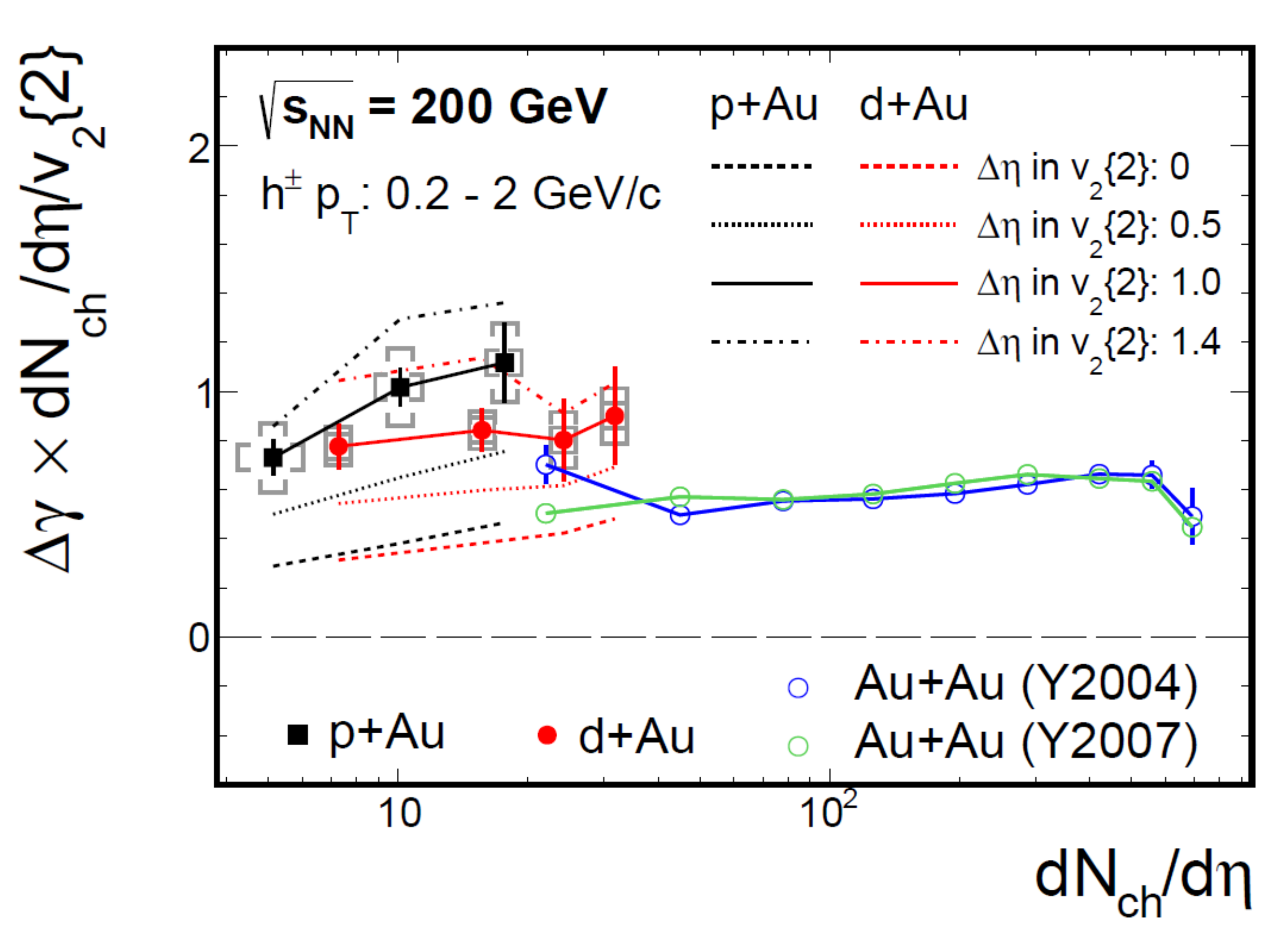}}
  \caption{(Color online) The scaled $\dgscale$ correlator in \pAu, d+Au, and \AuAu\ collisions as functions of $\dNdeta$ at RHIC by STAR. Dash lines represent the results using $\vc$ with different $\deta$ gaps. Error bars are statistical and caps are systematic uncertainties. Only statistical errors are plotted for the \AuAu\ results. Adapted from Refs.~\cite{Zhao:2017wck,Zhao:2018pnk,STAR:2019xzd}.}
  \label{fig:dgscale}
\end{figure}

Since the small-system data are dominated by background contributions, the $\dg$ observable should follow Eq.~(\ref{eq:bkgd}), proportional to the averaged $v_2$ of the background sources, and, in turn, likely also the $v_2$ of final-state particles. It should also be proportional to the number of background sources, and, because $\dg$ is a pair-wise average, inversely proportional to the total number of pairs. As the number of background sources likely scales with the final-state hadron $\dNdeta$, Eq.~(\ref{eq:bkgd}) reduces to $\dg\propto v_2/N$. 
It is thus instructive to investigate the scaled $\dg$ correlator,
\be \dgscale=\dg/v_2\times\dNdeta\,, \ee
which is shown in Fig.~\ref{fig:dgscale} as function of $\dNdeta$ in \pAu, \dAu, and \AuAu\ collisions by STAR~\cite{Zhao:2017wck,Zhao:2018pnk,STAR:2019xzd}. Indeed, the $\dgscale$ is rather constant. Similar conclusion can be drawn for \pPb\ and \PbPb\ data from CMS~\cite{Sirunyan:2017quh,Khachatryan:2016got}.
It is interesting that the scaled $\dg$ is rather insensitive to the event multiplicity for both the RHIC and LHC data. This can be understood if $\dg$ is dominated by backgrounds because, according to Eq.~(\ref{eq:bkgd}), the $\dgscale$ should essentially be the decay correlation, $\mean{\cos(\phia+\phib-\phiclust)}$. The decay correlation should depend only on the parent kinematics, insensitive to the event centrality or collision energy. The $\dgscale$ in \pA\ and \dA\ collisions are compatible to that in heavy-ion collisions. Since in \pA\ and \dA\ collisions essentially only backgrounds are present, the data strongly suggest that the heavy-ion measurements may be largely, if not all, backgrounds.

\subsection{Backgrounds to other chiral effects}
Local charge conservation can produce not only a background $\dg$ signal, but also an $\Ach$-dependent $v_2$ splitting between $\pip$ and $\pim$~\cite{Bzdak:2013yla}. Decay particles from a lower $\pt$ resonance tend to have a larger rapidity separation, resulting in one of the decay daughters to more likely fall outside the detector acceptance, leading to a nonzero $\Ach$. This process would generate a correlation between $\Ach$ and the average $\pt$ of charged particles, and therefore also between $\Ach$ and the $v_2$ coefficient, since $v_2$ depends on $\pt$. The resonance $v_2$ decreasing with increasing rapidity would also add to the effect. This local charge conservation mechanism would produce the same effect for $v_3$~\cite{Bzdak:2013yla}, whereas the CMW would produce no $v_3$ splitting. This would be a crucial test of this background mechanism.

The authors of Ref.~\cite{Hatta:2015hca} have shown that the standard viscous hydrodynamic could also produce $\Ach$-dependent $v_2$ splitting between $\pip$ and $\pim$. This came directly out of the analytical result that the anisotropic Gubser flow~\cite{Gubser:2010ui} coupled with conserved currents led to a $v_2$ splitting proportional to the isospin chemical potential~\cite{Hatta:2015era}. The finite isospin chemical potential also result in a finite $\Ach$, causing an indirect correlation between the $v_2$ splitting magnitude and $\Ach$. This mechanism would also yield an $\Ach$-dependent $v_3$ splitting, as well as an effect that is opposite in sign in the kaon and proton-antiproton sectors~\cite{Hatta:2015hca}. These would be good tests of this background mechanism

An anomalous transport model calculation suggests that including the Lorentz Force on quarks and antiquarks could even flip the sign of the elliptic flow difference between positively and negatively charged pions~\cite{Sun:2016nig,Sun:2016mvh}. It was pointed out~\cite{Shovkovy:2018tks} that the propagation of the long-wavelength CMW could be badly interrupted by high electrical conductivity in dynamically induced electromagnetic fields and become a diffusive one. Even at small electrical conductivity, the CMW is still strongly over-damped due to the effects of electrical conductivity and charge diffusion.
It was shown~\cite{Gursoy:2018yai} that the overall positive charge in the collision fireball produces a radial Coulomb field that is generally stronger in the out-of-plane than in-plane direction. This would reduce (increase) the $v_2$ of positively (negatively) charged particles without invoking the CMW, and the magnitude of the effect seems to be on the same order of the STAR measurement~\cite{Adamczyk:2015eqo}. 

The CVE is assessed by the difference between baryon-antibaryon and baryon-baryon correlations. Except charge conservation, an additional constraint comes into play, namely, net-baryon conservation. Furthermore, unlike charge-charge (dominated by pion-pion) correlations, baryon-antibaryon annihilation can have a large effect on baryon-antibaryon correlations. These effects will make the identification of the CVE harder than the CME. Not many efforts have been investigated into background studies of the CVE, partially because experimental measurements are not extensive. The only measurement~\cite{Zhao:2014aja,Wen:2017ibm,Zhao:2017ckp} is so far preliminary.

\section{Early efforts to remove backgrounds}\label{sec:attempts}
There is no doubt that the early $\dg$ measurements~\cite{Abelev:2009ac,Abelev:2009ad,Abelev:2012pa,Adamczyk:2013hsi,Adamczyk:2014mzf} are dominated by backgrounds. Experimentally, there have been many proposals and attempts to remove the backgrounds~\cite{Adamczyk:2013kcb,Ajitanand:2010rc,Wang:2016iov,Bzdak:2011np,Wen:2016zic}. Since the main background sources of the $\dg$ measurements are from the $v_2$-induced effects, most of those early efforts focused on $v_2$. In this section, we describe those early efforts. As we will show, none of those efforts can completely eliminate, but only reduce the background contributions to $\dg$, some better than other.

\subsection{Event-by-event $v_2$ method}\label{sec:ebye}
The main background sources to the $\dg$ observable are from the $v_2$-induced effects. Those backgrounds are expected to be proportional to $v_2$; see Eq.~(\ref{eq:bkgd}). One possible way to eliminate or suppress these $v_2$-induced backgrounds is to select ``spherical'' events, exploiting the statistical and dynamical fluctuations of the event-by-event $\vexe$ such that $\vexe=0$. Due to finite multiplicity fluctuations, one can easily vary the shape of the measured particle azimuthal distribution in final-state momentum space. This measured shape is directly related to the $v_2$ backgrounds in the measured $\dg$ correlator~\cite{Adamczyk:2013kcb,Wang:2016iov}. 

\begin{figure}[!htb]
  \centerline{ 
    \includegraphics[width=0.48\hsize]{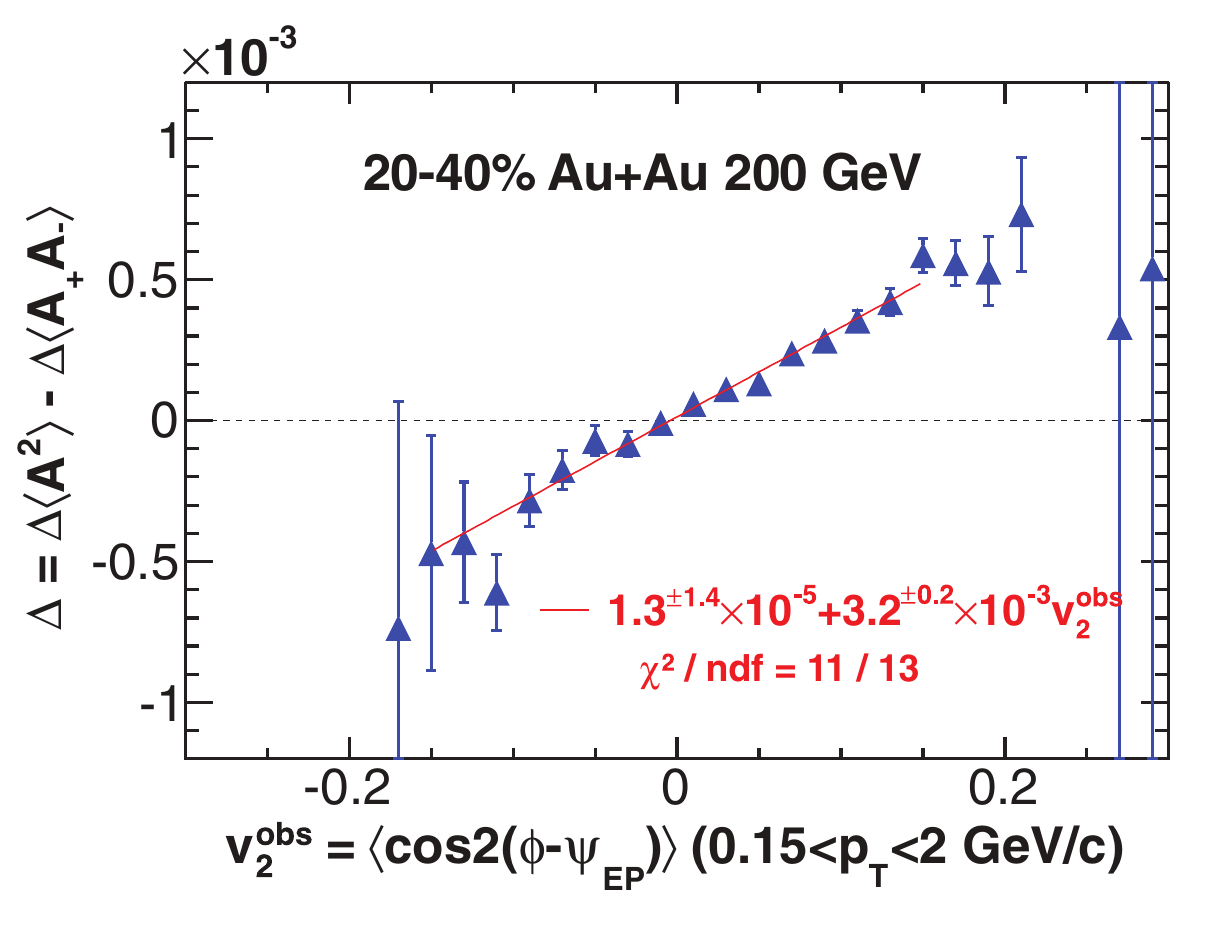}
    \hspace{0.05\hsize}
    \includegraphics[width=0.40\hsize]{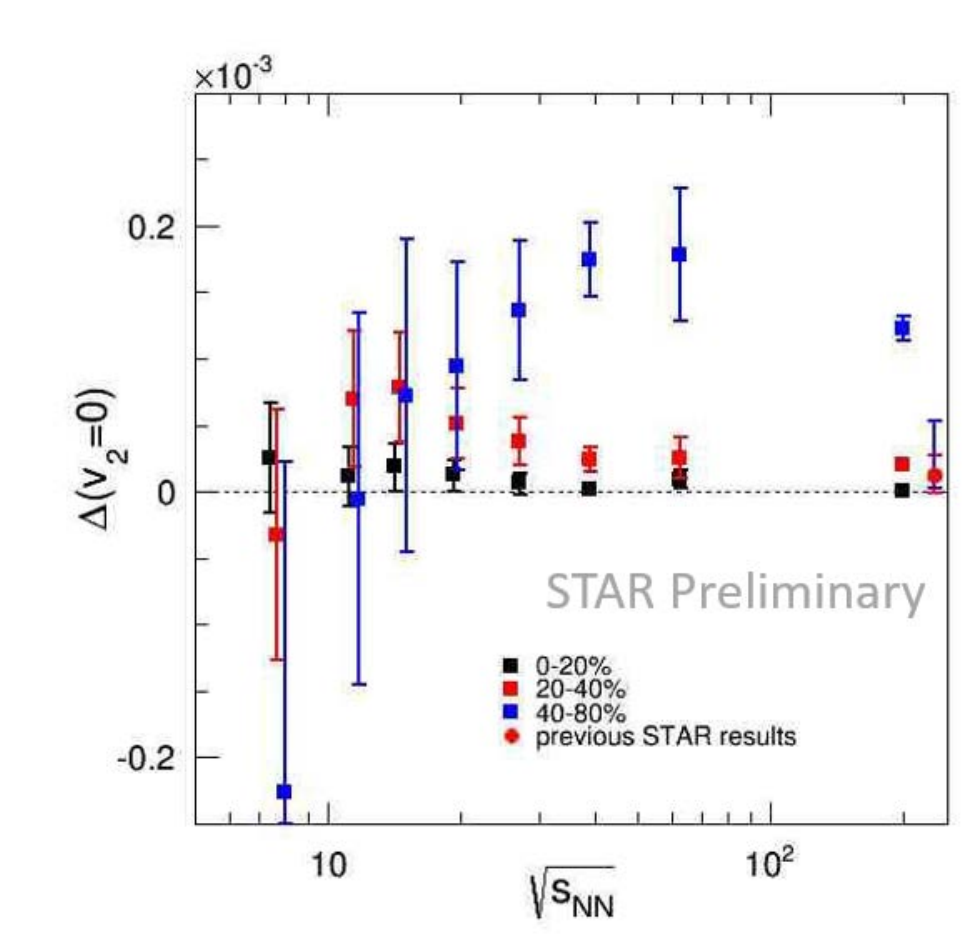}
  }
  \caption{(Color online) Left panel: charge multiplicity asymmetry correlation ($\Delta$) as a function of $\vexe$ in 20-40\% \AuAu\ collisions at $\snn = 200$~GeV from Run-4 data~\cite{Adamczyk:2013kcb}. Right panel: the $\Delta$ intercept at $\vexe=0$ in various centralities of \AuAu\ collisions from the Beam Energy Scan data as well as from the higher statistics 200~GeV data~\cite{TuBiaoQM15}. The 200~GeV Run-4 data~\cite{Adamczyk:2013kcb} are labeled as ``previous STAR results'' and are plotted at a slightly shifted $\snn$ value for clarity. Errors shown are statistical in both panels.}
  \label{fig:STARese}
\end{figure}

By using the event-by-event $\vexe$ shape selection, STAR~\cite{Adamczyk:2013kcb} has carried out the first attempt to remove the backgrounds in their measurement of the charge multiplicity asymmetry correlations, called the $\Delta$ observable (which is similar to the $\gamma$ correlator).
The event-by-event $\vexe$ can be measured by the $Q$ vector method, where $Q_2$ is given by Eq.~(\ref{eq:Q}) by summing over all POIs (used for the $\Delta$ measurement) in each event. In the STAR analysis, half of the TPC is used for the POI. The $\vexe$ is given by
\be \vexe = Q_2^{*}\hat{Q}_{2,{\rm EP}}\,, \label{eq:vexe} \ee
where $\hat{Q}_{2,{\rm EP}}$ is given by Eq.~(\ref{eq:qEP}), using particles from the other half of the TPC. 
Figure~\ref{fig:STARese} left panel shows the $\Delta$ as a function of $\vexe$ in 20-40\% \AuAu\ collisions at $\snn=200$~GeV~\cite{Adamczyk:2013kcb}.
A distinctive linear dependence is observed, as would be expected from backgrounds. By selecting the events with $\vexe=0$, the backgrounds in the $\Delta$ observable should be largely reduced~\cite{Adamczyk:2013kcb,TuBiaoQM15}. 
The background-suppressed $\dg$ signal can be extracted from the intercept at $\vexe=0$.
With the limited statistics from Run-4 data (taken in year 2004 by STAR), the extracted intercept is consistent with zero in 200~GeV \AuAu\ collisions~\cite{Adamczyk:2013kcb}. Analysis of higher statistics data from later runs indicates that the intercept is finite, greater than zero~\cite{TuBiaoQM15}.
This is shown in the right panel of Fig.~\ref{fig:STARese} where the extracted intercept is plotted as a function of centrality for \AuAu\ collisions of different beam energies~\cite{TuBiaoQM15}. Positive intercepts are observed, also at $\snn=200$~GeV, with the high statistics data.

\begin{figure}[!htb]
  \centerline{
    \includegraphics[width=0.45\hsize]{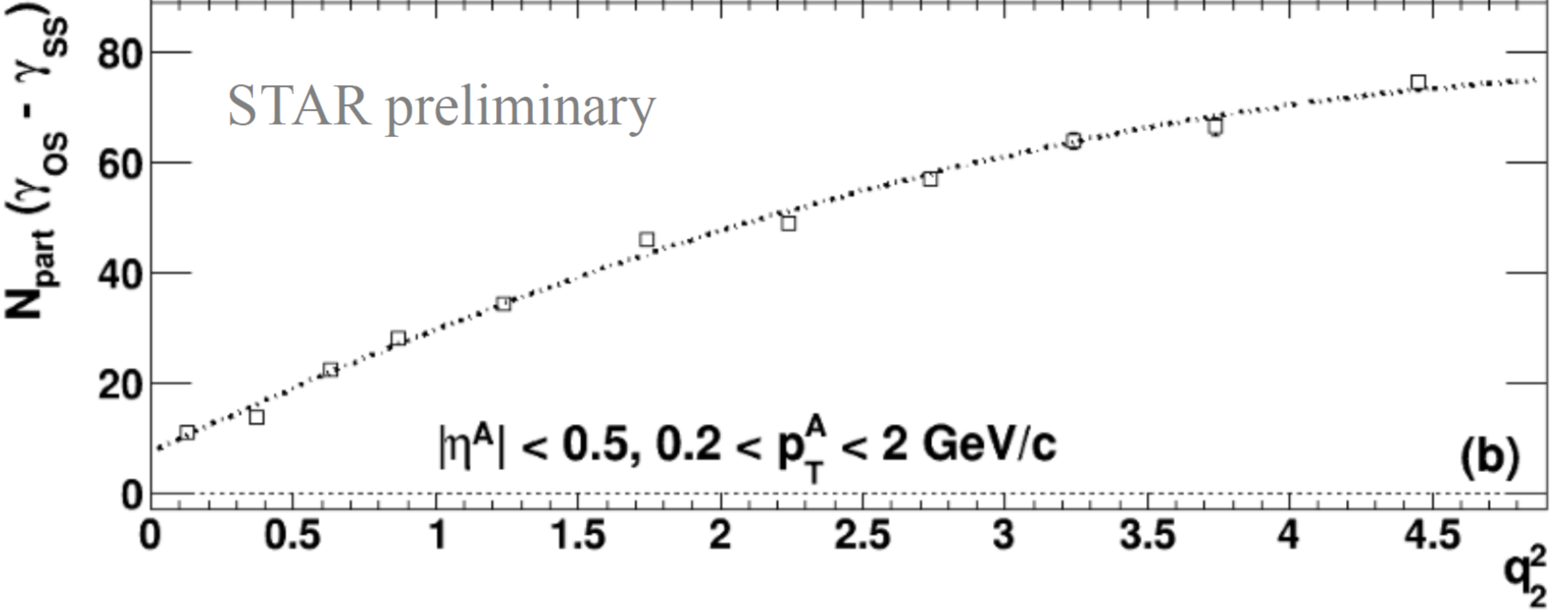}
    \hspace{0.05\hsize}
    \includegraphics[width=0.45\hsize]{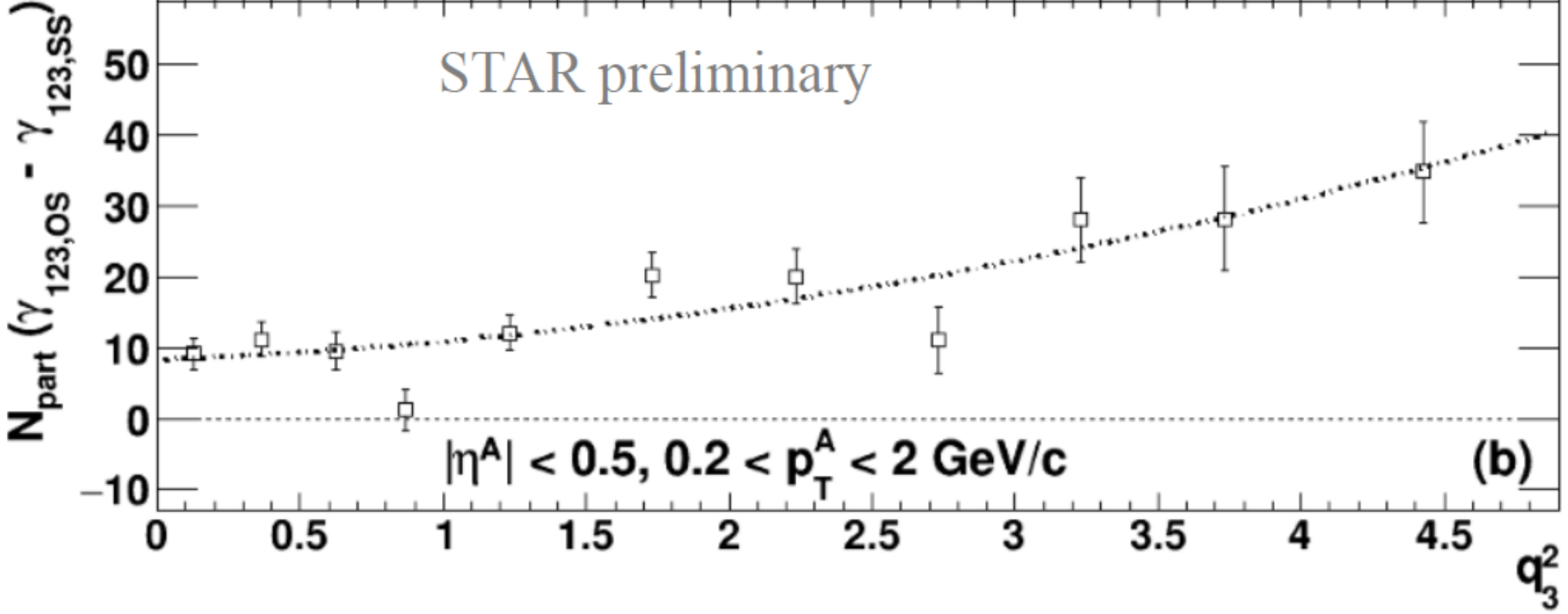}
  }
  \caption{The $\dg_{112}$ correlator multiplied by the number of participants (\Npart) as a function of the event-by-event $q_2^2$ (left panel), and that with respect to the third harmonic plane ($\dg_{123}$) as a function of $q_3^2$ in 20-60\% \AuAu\ collisions at $\snn = 200$~GeV. Errors shown are statistical uncertainties. Adapted from Ref.~\cite{WangGang_q2q3}.}
  \label{fig:Gang}
\end{figure}

A similar method selecting events with the event-by-event $q_n$ variable has been recently proposed~\cite{Wen:2016zic}. Here $q_n$ is the magnitude of the reduced flow vector~\cite{Adler:2002pu}, defined as
\be
    q_n = \sqrt{M}|Q_n|\;\;\;{\rm where}\;n=1,2,3,...\,,
  \label{eq:q}
\ee
and is related to $v_n$.
To suppress the $v_2$-induced background, a tight cut, $q_2 = 0$, is proposed. The cut is tight because $q_2 = 0$ corresponds to a zero 2nd-order harmonic to any plane, while $\vexe = 0$ corresponds to the zero 2nd-order harmonic with respect only to the reconstructed EP in another phase space of the event.
This $q_2$ method is therefore more difficult than the event-by-event $\vexe$ method because the extrapolation to zero $q_2$ is statistically limited and because it is unclear whether the background is linear in $q_2$ or not.
Figure~\ref{fig:Gang} shows the preliminary STAR data analyzed by the event-by-event $q_2$ method~\cite{WangGang_q2q3}. An extrapolation to zero $q_2$ indicates a positive intercept (see Fig.~\ref{fig:Gang} left panel). A similar study using the third harmonic (via the variable $\dg_{123}$ as discussed in Sect.~\ref{sec:kappa}) indicates a positive intercept as well (see Fig.~\ref{fig:Gang} right panel), comparable in magnitude to that from the $q_2$ method, while only background is expected in $\dg_{123}$.

\begin{figure}[!htb]
  \centerline{
    \includegraphics[width=0.40\hsize]{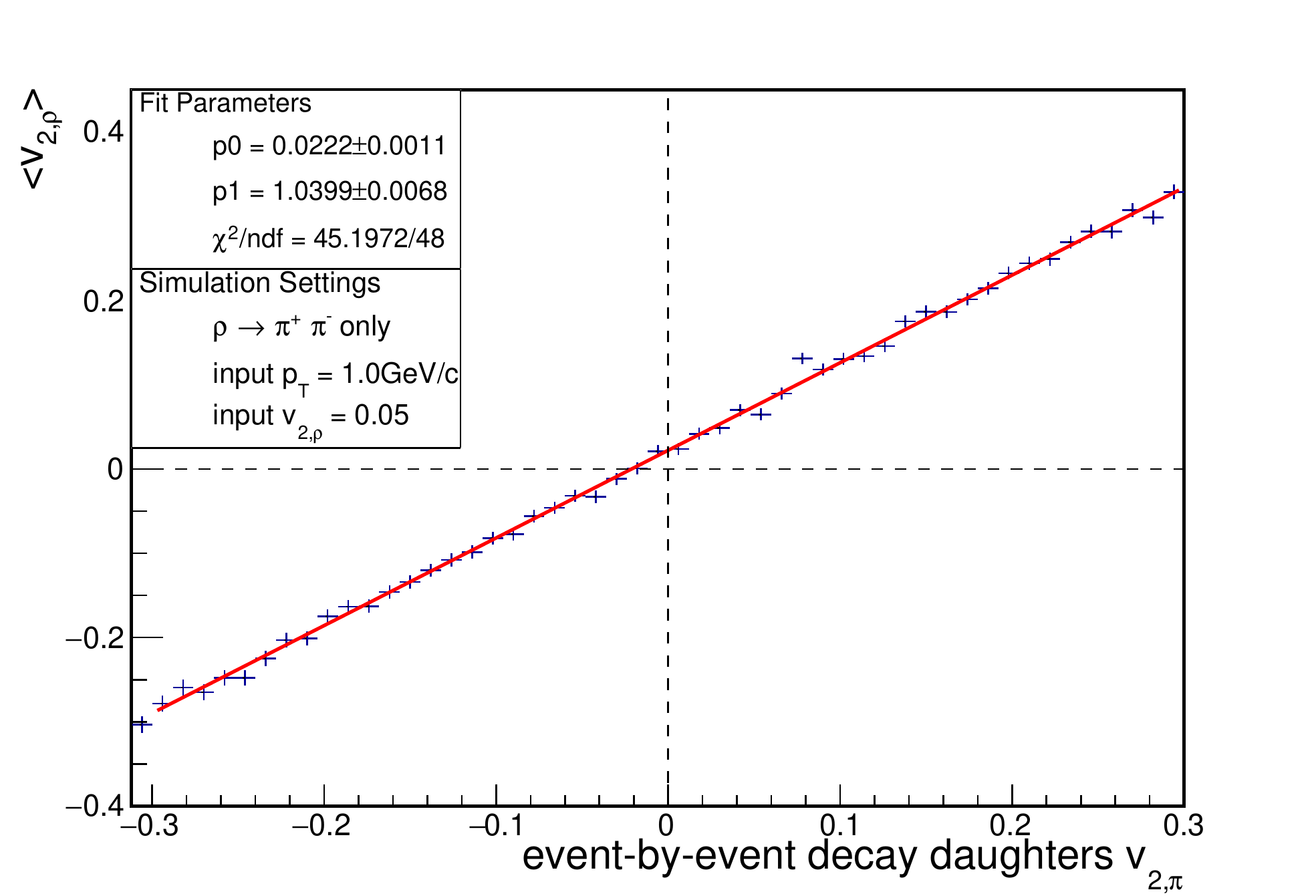}
    \hspace{0.05\hsize}
    \includegraphics[width=0.40\hsize]{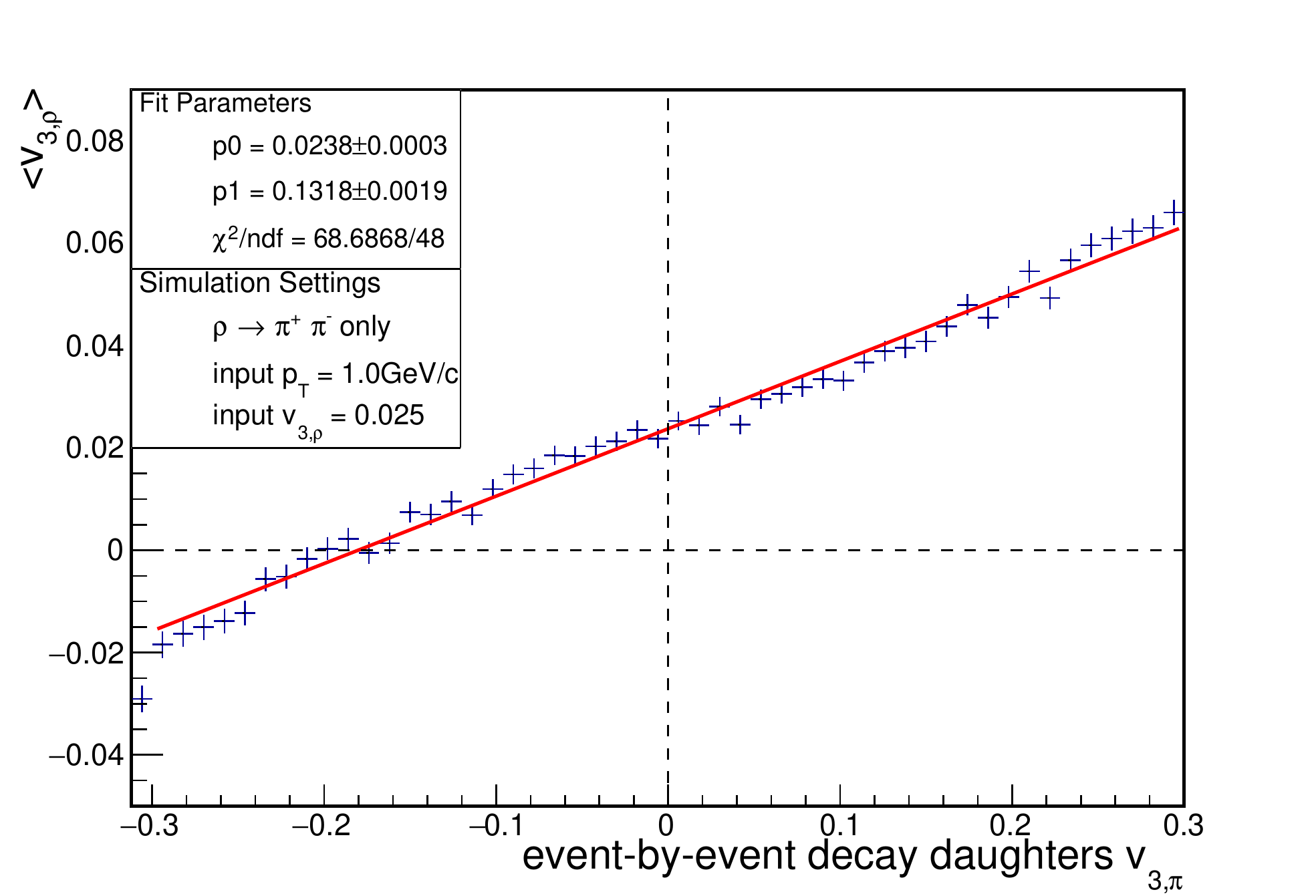}
  }
  \caption{The $\mean{v_{2,\rho}}$ versus $v_{2,\pi,{\rm ebye}}$ (left panel) and $\mean{v_{3,\rho}}$ versus $v_{3,\pi,{\rm ebye}}$ (right panel) from toy-model simulations of $\rho$ resonances with fixed $p_{T,\rho}=1.0$~\gevc, $v_{2,\rho}=5$\% and $v_{3,\rho}=2.5$\%. The finite $\mean{v_{2,\rho}}$ and $\mean{v_{3,\rho}}$ values are the reasons why flow backgrounds cannot be completely removed by $v_{2,\pi,{\rm ebye}}=0$ or $v_{3,\pi,{\rm ebye}}=0$. Adapted from~\cite{Wang:2016iov}.}
  \label{fig:rhov2piv2}
\end{figure}

These methods extract the $\dg$ signal at zero $\vexe$ or $q_2$ of the {\em final-state} particles. However, the backgrounds arise from resonance/cluster decay correlations coupled with the $v_2$ of the {\em parent} sources of the resonances/clusters, not that of all final-state particles. 
Since the $\vexe$ and $q_2$ quantities in these methods are the event-by-event quantities, the $v_2$ of the correlation sources (resonances/clusters), i.e.~the $\vclust$ in Eq.~(\ref{eq:bkgd}), are not necessarily zero when the final-state particle $\vexe$ or $q_2$ is selected to be zero. This is shown in Fig.~\ref{fig:rhov2piv2} in a resonance toy model simulation~\cite{Wang:2016iov}; the average $\mean{v_{n,\rho}}$ of the $\rho$ resonances in events with $\vnexe=0$ are found to be nonzero. It is interesting to note that the intercepts are similar for $v_2$ and $v_3$, and the slope for $v_3$ is significantly smaller than that for $v_2$. This would explain the features seen in Fig.~\ref{fig:Gang} for the preliminary STAR data where the inclusive $\dg_{123}$ is much smaller than the inclusive $\dg_{112}$ but the $q_2=0$ and $q_3=0$ projection intercepts are similar. We thus conclude that the positive intercept results from the event-by-event $v_2$ and $q_2$ methods are likely still contaminated by flow backgrounds~\cite{Wang:2016iov}.

It is difficult, if not at all impossible, to ensure the $\vexe$ of one resonance species to be zero on event-by-event basis. It would be nearly impossible to ensure the event-by-event $v_2$'s of all the background sources to be zero. 
Therefore, it is practically impossible to completely remove the flow backgrounds by using the event-by-event $v_2$ or $q_2$ method~\cite{Wang:2016iov}.

\subsection{Comments on the $\kappa$ parameter}\label{sec:kappa}
It was pointed out~\cite{Liao:2010nv,Bzdak:2012ia} that, besides the $\gamma$ correlator, the CME is also contained in another azimuthal correlator, 
\be
\delta\equiv\langle\cos(\phia-\phib)\rangle\,,\;\;\;
\dd\equiv\dOS-\dSS\,.
\ee
This can be easily seen by a two-component decomposition of the event made of CME particles and the majority rest of background particles. The back-to-back OS pairs from the CME contribute positively to $\gamma$ and negatively to $\delta$, while the same-direction SS pairs from the CME contribute negatively to $\gamma$ and positively to $\delta$. In other words, the CME contribution to $\gamma$ and $\delta$ are opposite in sign and same in magnitude: $\dg_{\cme}=-\Delta H$ and $\dd_{\cme}=\Delta H$.
The background particle pair correlations contribute to $\delta$ and there is a large difference between OS and SS, $\dd_{\bkg}=\Delta F$. 
In terms of the flow contribution to $\gamma$, one may naively write:
\be
\gamma\equiv\mean{\cos(\phia+\phib-2\psiRP)} = \mean{\cos(\phia-\phib)}\cdot\mean{\cos2(\phib-\psiRP)} = v_2\delta \,. \label{eq:kappa_g}
\ee
So the flow contribution to $\dg$ is $\dg_{\bkg}=v_2\Delta F$.
Hence, we have:
\bea
\dg &=& \kappa v_2 \Delta F - \Delta H \,, \label{eq:kappa_dg}\\
\dd &=& \Delta F + \Delta H \,. \label{eq:kappa_dd}
\eea
The parameter $\kappa$ in Eq.~(\ref{eq:kappa_dg}) is supposed to be unity if Eq.~(\ref{eq:kappa_g}) holds, but is included to absorb correlation (non-factorization) effects that may have been neglected in Eq.~(\ref{eq:kappa_g}).

Unfortunately, Eq.~(\ref{eq:kappa_g}) does not hold because the terms $\cos(\phia-\phib)$ and $\cos2(\phib-\psiRP)$ both contain $\phib$ and cannot be factorized as done in the equation. The correct algebra is in Eq.~(\ref{eq:bkgd}); there, although the terms $\cos(\phia+\phib-2\phiclust)$ and $\cos2(\phiclust-\psiRP)$ both contain $\phiclust$, they are two separate physics processes and therefore decoupled: the former is decay kinematics that does not depend on the parent azimuthal angle relative to $\psiRP$, and the latter is the cluster azimuthal anisotropy that does not affect the decay topology. (The two may be slightly correlated because both depend on the parent cluster $\pt$~\cite{Wang:2016iov}, but this must be secondary.) Hence the $\kappa$ parameter is actually equal to
\be
\kappa_2\equiv\kappa=\frac{\mean{\cos(\phia+\phib-2\phiclust)}}{\mean{\cos(\phia-\phib)}_{\clust}}\cdot\frac{\vclust}{v_2}\,, \label{eq:kappa2}
\ee
where we have taken $\mean{\cos(\phia-\phib)}_{\clust}$ to be the average quantity for only those pairs from cluster decays, i.e.~$\mean{\cos(\phia-\phib)}=\frac{N_{\alpha\beta,\clust}}{N_{\pi}^2}\mean{\cos(\phia-\phib)}_{\clust}$.
One can easily see why $\kappa$ can be very different from unity. Take again the resonance decay as an example for the background. The quantities $\mean{\cos(\phia+\phib-2\phiclust)}$ and $\mean{\cos(\phia-\phib)}_{\clust}$ are the resonance decay angular correlation properties. Numerically, the former may be significantly larger than the latter. The $v_2$ in Eq.~(\ref{eq:kappa_g}) is that of the resonance decay daughters, and in practice is taken as that of all final-state particles; the two can be different. In Eq.~(\ref{eq:bkgd}), the $v_2$ is that of the resonances (or correlation sources), which can be easily a factor of two of that of the inclusive particles if one assumes the number-of-constituent-quark (NCQ) scaling for hadron $v_2$, because the $\mean{\pt}$ of resonances can be a factor of two larger than that of charged hadrons (mainly charged pions). So the range for the value of the parameter $\kappa$ is wide open, and can depend on the collision centrality and beam energy. It is clear from the above discussion that the $\kappa$ parameter is ill-defined and has several issues that are mixed up. Its value is unknown a priori; even the range of its value is uncertain.

Ref.~\cite{Adamczyk:2014mzf} took Eqs.~(\ref{eq:kappa_dd}) and (\ref{eq:kappa_dg}) literally, and assigned a range of $\kappa=1$-$2$ to obtain the $\Delta H$ ``signal.'' This would be a useful exercise if the value of $\kappa$ is theoretically constrained or experimentally measured. However, as discussed above, the value of $\kappa$ is not at all theoretically constrained. It is neither experimentally measured as that would constitute an experimental measurement of the backgrounds. The postulated value of $\kappa=1$-$2$ in Ref.~\cite{Adamczyk:2014mzf} is a misconception. Without the knowledge of the $\kappa$ values, the presented results in Ref.~\cite{Adamczyk:2014mzf} with the various values of $\kappa$ do not give additional information other than those already in the $\gamma$ measurements.

A variation of the $\kappa$ analysis is to take the ratio of the measured $\dg$ to the ``expected'' elliptic flow background~\cite{Adamczyk:2014mzf,Bzdak:2012ia,Wen:2017mol}, and study its behavior as functions of centrality and particle species. This is dubbed $\kappa_{\rm kill}$, indicating that the CME would be zero if the $\kappa$ turns out to be as large as $\kappa_{\rm kill}$. However, as discussed above, the $\kappa$ is rather ill-defined, so such a study has yielded limited insights.

\begin{figure}[!htb]
  \centerline{\includegraphics[width=0.45\hsize]{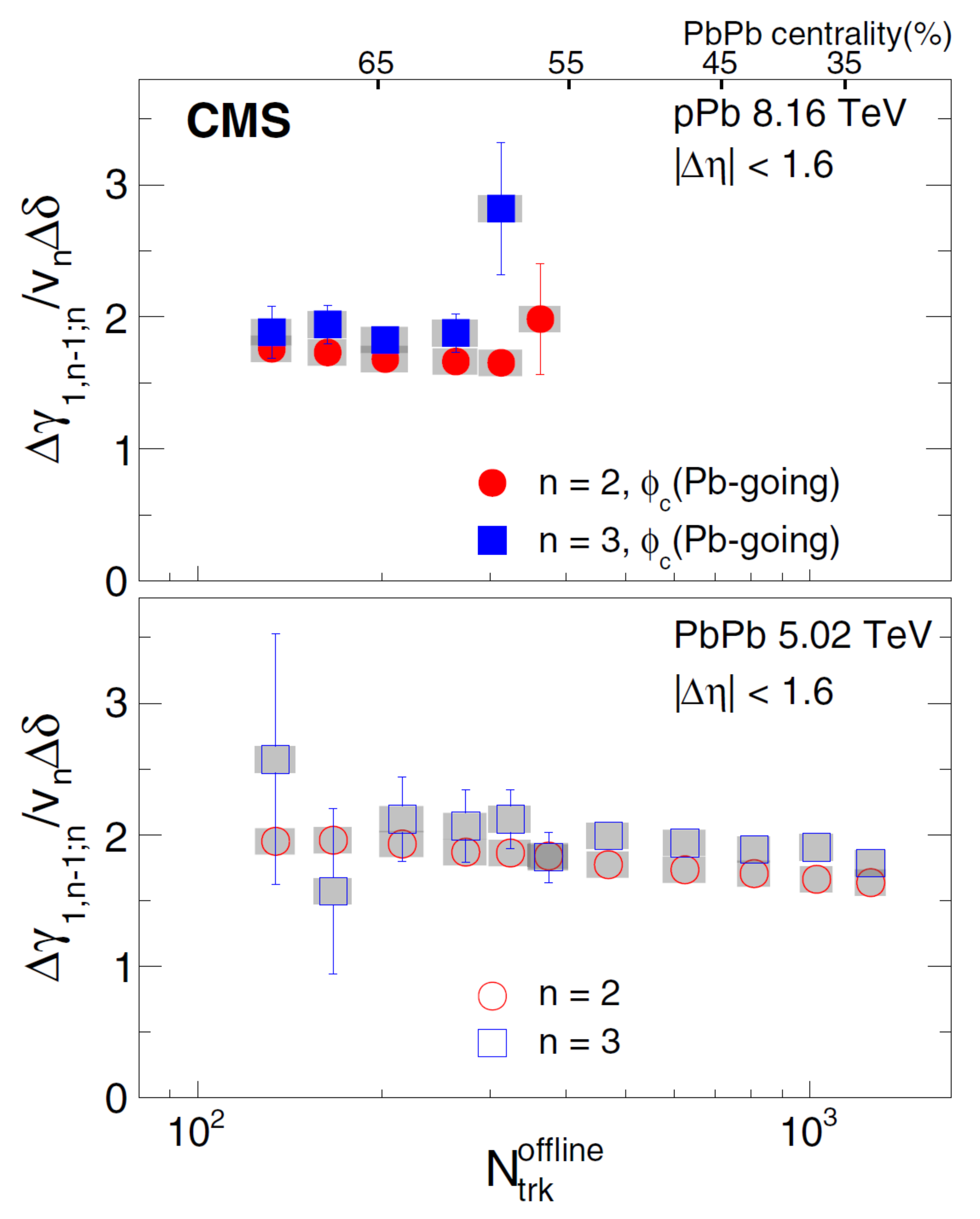}}
  \caption{The $\kappa_{112}\equiv\dg_{112}/v_2\delta$ and $\kappa_{123}\equiv\dg_{123}/v_3\delta$ variables measured by CMS as functions of $\Noff$. The measurements are averaged over $|\eta|<1.6$ in \pPb\ collisions at $\snn=8.16$~TeV (upper panel) and \PbPb\ collisions at 5.02 TeV (lower panel). Statistical and systematic uncertainties are indicated by the error bars and shaded regions, respectively. Adapted from~\cite{Sirunyan:2017quh}.}
  \label{fig:kappa}
\end{figure}

With the $\gamma_{123}$ variable with respect to $\psi_3$ (see Sect.~\ref{sec:gamma123})~\cite{Sirunyan:2017quh}, we have a set of equations analogous to Eqs.~(\ref{eq:kappa_g}) and (\ref{eq:kappa_dg}), except that there is no CME contribution of $\Delta H$ to $\dg_{123}$:
\be \gamma_{123}\equiv\mean{\cos(\phia+2\phib-3\psi_3)} = \mean{\cos(\phia-\phib)}\cdot\mean{\cos3(\phib-\psi_3)} = v_3\delta \,, \label{eq:kappa_g123} \ee
\be \dg_{123} = \kappa_3 v_3 \Delta F \,. \label{eq:kappa_dg123} \ee
CMS studied the ratio of $\dg_{112}/v_2\dd$ and $\dg_{123}/v_3\dd$ in \pPb\ and \PbPb\ collisions at the LHC~\cite{Sirunyan:2017quh}. The results are shown in Fig.~\ref{fig:kappa}. Absent of CME, these ratios would equal to $\kappa_2$ and $\kappa_3$. In \pPb\ collisions, there should be negligible CME contributions to both the $\dg_{112}$ and $\dg_{123}$ measurements. The ratios appear to be approximately equal in \pPb\ collisions indicating that $\kappa_2\approx\kappa_3$. The ratios in \PbPb\ collisions are also approximately equal, strongly suggesting that the CME contributions in \PbPb\ collisions are indeed small.

Similar to Eq.~(\ref{eq:kappa2}), the $\kappa_3$ parameter is 
\be
\kappa_3=\frac{\mean{\cos(\phia+2\phib-3\phiclust)}}{\mean{\cos(\phia-\phib)}_{\clust}}\cdot\frac{v_{3,\clust}}{v_3}\,. \label{eq:kappa3}
\ee
Comparing Eqs.~(\ref{eq:kappa2}) and (\ref{eq:kappa3}), it is clear that $\kappa_2$ and $\kappa_3$ do not have to be equal. It is probably a good approximation that $v_{2,\clust}/v_2\approx v_{3,\clust}/v_3$. The $\kappa_2\approx\kappa_3$ result from CMS thus suggests that $\mean{\cos(\phia+2\phib-3\phiclust)}\approx\mean{\cos(\phia+\phib-2\phiclust)}$. This may not be unexpected because most of the resonances decay into more or less collimated daughter particles so these averages are similar.

Inspired by the CMS work of $\gamma_{123}$, many correlators can be devised~\cite{Schenke:2019ruo}, for example, $\gamma_{132}\equiv\mean{\cos(\phia-3\phib+2\psi_2)}$. One may express it as $\gamma_{132}=\kappa_{132}\mean{\cos(\phia-\phib)}v_2$, but again because factorization does not hold, the value of $\gamma_{132}$ is not known a priori. The value is determined by $\kappa_{132}=\frac{\mean{\cos(\phia-3\phib+2\phiclust)}}{\mean{\cos(\phia-\phib)}_{\clust}}\cdot\frac{v_{2,\clust}}{v_2}$ and could be anything. Combining $\gamma_{112}$ and $\gamma_{132}$, one can easily obtain $\mean{\cos(\phia-\phib)\cos2(\phib-\psi_2)}=\frac{1}{2}(\kappa_{132}+\kappa_{112})v_2\delta$ and $\mean{\sin(\phia-\phib)\sin2(\phib-\psi_2)}=\frac{1}{2}(\kappa_{132}-\kappa_{112})v_2\delta$ (here $\kappa_{112}\equiv\kappa_2$). However, because the left sides cannot factorize, such mathematical decompositions do not seem to offer much insights.

\subsection{Deformed U+U collisions}
It has been suggested~\cite{Voloshin:2010ut} that, because the Uranium (U) nucleus is strongly deformed, U+U collisions could give insights into the background issue. In very central U+U collisions, the magnetic field is negligible but the elliptic flow is still appreciable because of the deformed nuclei in the initial state. This would yield appreciable $\dg$ measurement, dominated by $v_2$-induced background, in those very central collisions. Preliminary data from STAR indicates that the $\dg$ value vanishes in very central $\sim1$\% collisions~\cite{Wang:2012qs}. This is contrary to the expectation. If finite and positive background must exist in those central collisions because of the finite $v_2$, and the possible CME signal cannot be negative, then the data measurement of zero $\dg$ does not make sense. Since the data show finite $v_2$ but zero $\dg$, it has been argued that our current understanding of the $v_2$ backgrounds may be incorrect, but such an argument has not gained much support. 
In short, the U+U data from STAR~\cite{Wang:2012qs} are not fully understood. Nevertheless, because the data are still preliminary, one should exercise caution in their interpretation.

Various ways have been suggested to utilize the U+U deformed geometry to gauge the CME signal and flow background~\cite{Bloczynski:2013mca,Chatterjee:2014sea}.
However, as the initial geometry from random orientations of the colliding nuclei is difficult to disentangle experimentally~\cite{Wang:2012qs,Tribedy:2017hwn}, the U+U data have so far not yielded enough insights as anticipated.

\subsection{The sine-correlator observable}
A sine-correlator observable~\cite{Ajitanand:2010rc,Magdy:2017yje} has been proposed to identify the CME by examining the broadness of the event probability distribution in $\Delta S=\mean{\sin\phi_{+}}-\mean{\sin\phi_{-}}$, where $\phi_{\pm}$ are the azimuthal angles of positively and negatively charged particles relative to the RP and the averages are taken event-wise. For events with CME signals, charge separation along the magnetic field gives $\sin\phi_{\pm}\approx1$ and a maximal difference $\sin\phi_{+}-\sin\phi_{-}\approx\pm2$. The $\Delta S$ distribution would therefore become wider than its reference distribution, which can be constructed by randomizing the particle charges and by rotating the events by $\pi/2$ in azimuth~\cite{Ajitanand:2010rc,Magdy:2017yje}. 
The ratio of real event distribution to the reference distribution, $R(\Delta S)$, would thus be concave~\cite{Ajitanand:2010rc,Magdy:2017yje,Magdy:2018lwk}. For flow-induced background, the initial expectation was that the $R(\Delta S)$ curve would be convex~\cite{Ajitanand:2010rc}. However, more recent studies~\cite{Bozek:2017plp,Feng:2018chm} indicate that the $R(\Delta S)$ curve can also be concave for flow-induced backgrounds. Preliminary STAR data, on the other hand, show concave $R(\Delta S)$ curves in \AuAu\ collisions. 
However, in light of the model studies~\cite{Bozek:2017plp,Feng:2018chm}, it is unclear what the data try to reveal and whether the $R(\Delta S)$ variable would lead to unique conclusion regarding the CME. 

\section{Innovative background removal methods}\label{sec:efforts}
As discussed in the last section, none of the efforts described so far can eliminate the physics backgrounds entirely. Some of the methods can almost remove the backgrounds, but how much residual background still remains is hard to quantify. Given that the CME signal is likely very small, none of the methods discussed in the previous section seems probable to yield concrete conclusions on the CME.

Nevertheless, many insights have been learned from those early efforts. More thorough developments have recently emerged leading to analysis methods that, to our best judgment, can remove the backgrounds entirely. We believe those methods will likely lead to quantitative conclusions on the CME. In this section we discuss those new developments.

Examining Eq.~(\ref{eq:bkgd}), it is not difficult to identify innovative ways to remove backgrounds:
\begin{enumerate}
\item[(1)] One is to measure the $\dg$ observable where the elliptical anisotropy is zero, not by the event-by-event $\vexe$ or $q_2$ method exploiting statistical (and dynamical) fluctuations~\cite{Adamczyk:2013kcb,Wen:2016zic} as discussed in Sect.~\ref{sec:ebye}, but by the event-shape engineering (ESE) method exploiting only dynamical fluctuations in $v_2$~\cite{Schukraft:2012ah}. This has been applied in real data analyses~\cite{Sirunyan:2017quh,Acharya:2017fau}. We discuss this method in Sect.~\ref{sec:ese}.
\item[(2)] The second innovative method is to make measurements where resonance contributions are small or can be identified and removed~\cite{Zhao:2017nfq,Li:2018oot}. This can be achieved by differential measurements of the $\dg$ as a function of the particle pair invariant mass ($\minv$) to identify and remove the resonance decay backgrounds~\cite{Zhao:2017nfq,Li:2018oot}. This method has not been explored until recently~\cite{Zhao:2017wck,Zhao:2018pnk,Zhao:2018blc}. We discuss this method in Sect.~\ref{sec:mass}.
\item[(3)] The third innovative method~\cite{Xu:2017qfs,Xu:2018wvm} is not as obvious, but may present the best, most robust way to search for the CME~\cite{Zhao:2018blc}. It exploits comparative measurements of $\dg$ with respect to the RP and the PP~\cite{Xu:2017qfs,Xu:2018wvm} taking advantage of the geometry fluctuation effects of the PP and the magnetic field directions. We discuss this method in Sect.~\ref{sec:plane}.
\end{enumerate}

\subsection{Event-shape engineering}\label{sec:ese}
Since the background is proportional to the elliptic anisotropy, one way to remove the background is to select events with zero $v_2$. This was attempted by the event-by-event $\vexe$ and $q_2$ methods as discussed in Sect.~\ref{sec:ebye}, exploiting mainly the large statistical fluctuations due to finite multiplicities of individual events. However, these event-by-event shape methods do not completely remove the backgrounds which come from resonances/clusters. This is because the $\vexe$ or $q_2$ uses the same particles as those used for the $\gamma$ measurements, i.e.~the POIs. A zero anisotropy of those POIs does not guarantee a zero resonance anisotropy contribution to those same POIs on event-by-event basis~\cite{Wang:2016iov}. This shortcoming can be lifted by analyzing the $\dg$ observable of POIs as a function of the $q_2$~\cite{Schukraft:2012ah} calculated not using the POIs but particles from a different phase space, e.g.~displaced in pseudorapidity from the POIs~\cite{Acharya:2017fau,Sirunyan:2017quh}. This method is called ``event-shape engineering''~\cite{Schukraft:2012ah}.

Just like in the event-by-event method~\cite{Adamczyk:2013kcb,Wen:2016zic}, the $q_2$ variable [Eqs.~(\ref{eq:Q}),~(\ref{eq:q})] selects, within a given narrow centrality bin, different event shapes~\cite{Schukraft:2012ah}. 
A given $q_2$ cut-range samples a $v_2$ distribution of the POIs. In ESE, unlike the event-by-event method, the $q_2$ and the POI come from different phase spaces, so their statistical fluctuations are independent. The different average $v_2$ of the POIs resulted from different $q_2$ cut-ranges, therefore, assess only the dynamical fluctuations from the initial-state participant geometry within the given narrow centrality bin. The extrapolated zero average $v_2$ of the POIs will likely correspond to also zero average $v_2$ of all particle species, including the CME background sources of resonances/clusters. This is clearly advantageous over the event-by-event method in Sect.~\ref{sec:ebye}. The disadvantage is that an extrapolation to $v_2=0$ is required since the ESE $q_2$ sampling in its own phase space would not yield $v_2=0$ of the POI phase space. A dependence of the backgrounds on $v_2$ that is not strictly linear would introduce inaccuracy in the extracted CME signal.

\begin{figure}[!htb]
  \centerline{
    \includegraphics[width=0.4\hsize]{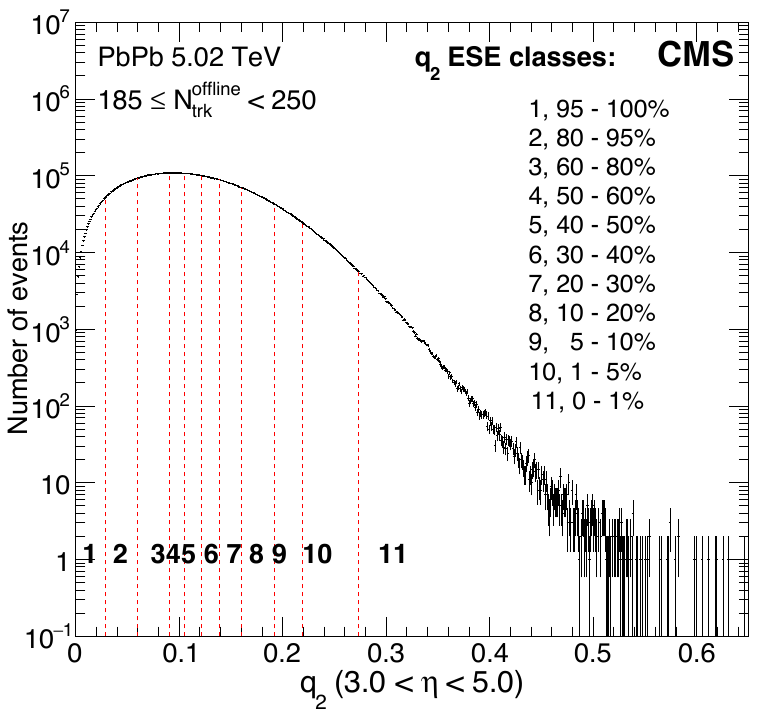} 
    \hspace{0.05\hsize}
    \includegraphics[width=0.4\hsize]{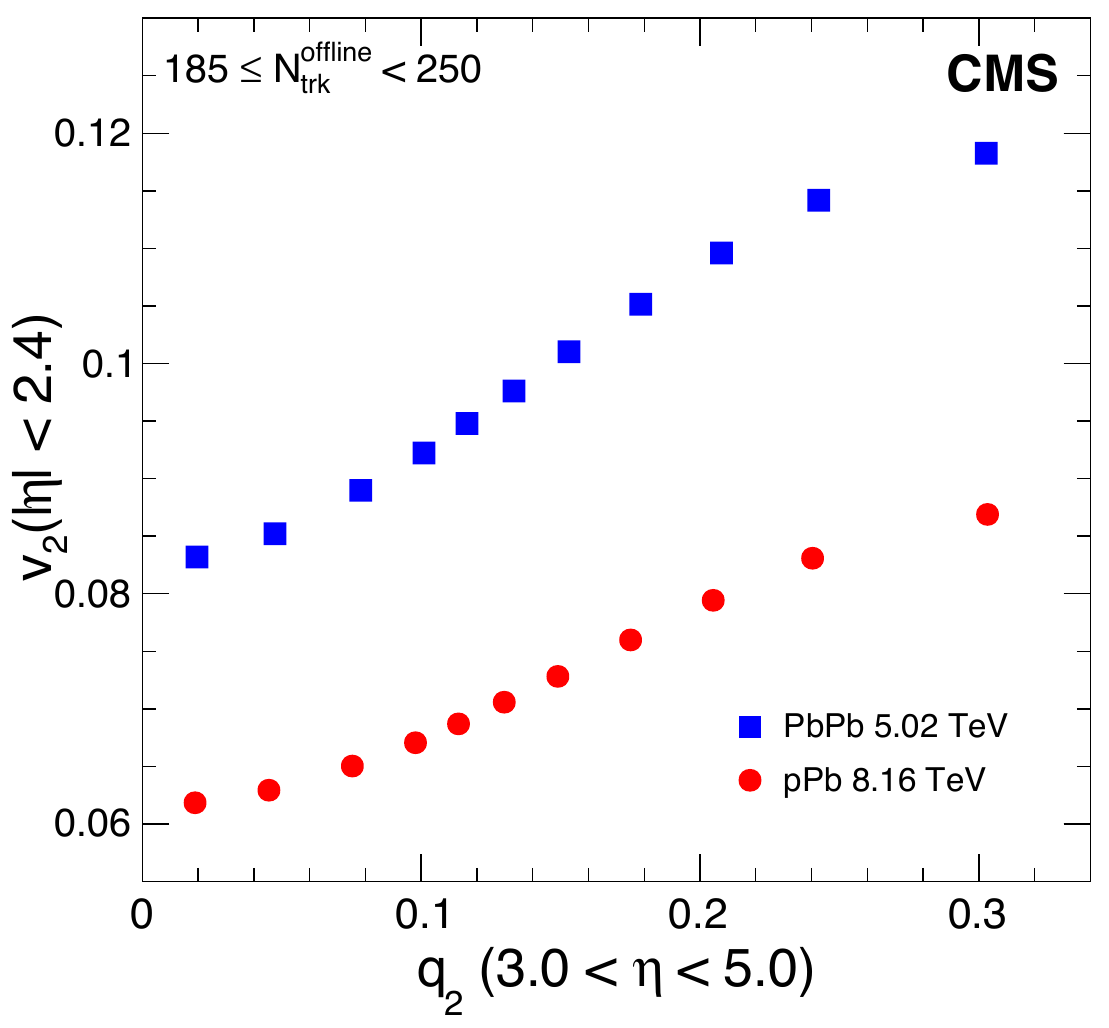}
  }
  \caption{Left panel: the $q_2$ distribution in multiplicity range $185 \leq \Noff < 250$ in \PbPb\ collisions by CMS. Dashed lines represent the selections used to divide the events into multiple $q_2$ classes. Right panel: the correlations between $v_2$ and $q_2$ in \pPb\ and \PbPb\ collisions based on the $q_2$ selections of the events. Adapted from Ref.~\cite{Sirunyan:2017quh}.}
  \label{fig:CMS_q2}
\end{figure}

Owing to the large acceptances of the LHC detectors, the large elliptic anisotropies, and the large event multiplicities of heavy-ion collisions at the LHC energies, the ESE method can be easily applied to LHC data and is proved to be powerful. It is, however, not easy to apply the ESE method to RHIC data because the acceptances of the RHIC experiments are limited and the event multiplicities are still not large enough even at the top RHIC energy.
Figure~\ref{fig:CMS_q2} (left panel) shows the $q_2$ distribution in \PbPb\ collisions from CMS for the multiplicity range of $185\leq\Noff<250$ as an example~\cite{Sirunyan:2017quh}. Events within a given multiplicity range are divided into several classes with each corresponding to a fraction of the full distribution, where the 0-1\% represents the class with the largest $q_2$ value. In Fig.~\ref{fig:CMS_q2} (right panel), the average $v_2$ values at mid-rapidity are presented in each selected $q_2$ class in both PbPb and pPb collisions of the same $\Noff$ range. The strong correlation between these two quantities indicates their underlying correlations to the initial-state geometry. The $\dg$ correlator within each multiplicity bin can now be studied as a function of $v_2$ explicitly using the $q_2$ selections. 

\begin{figure}[!htb]
  \centerline{\includegraphics[width=0.5\hsize]{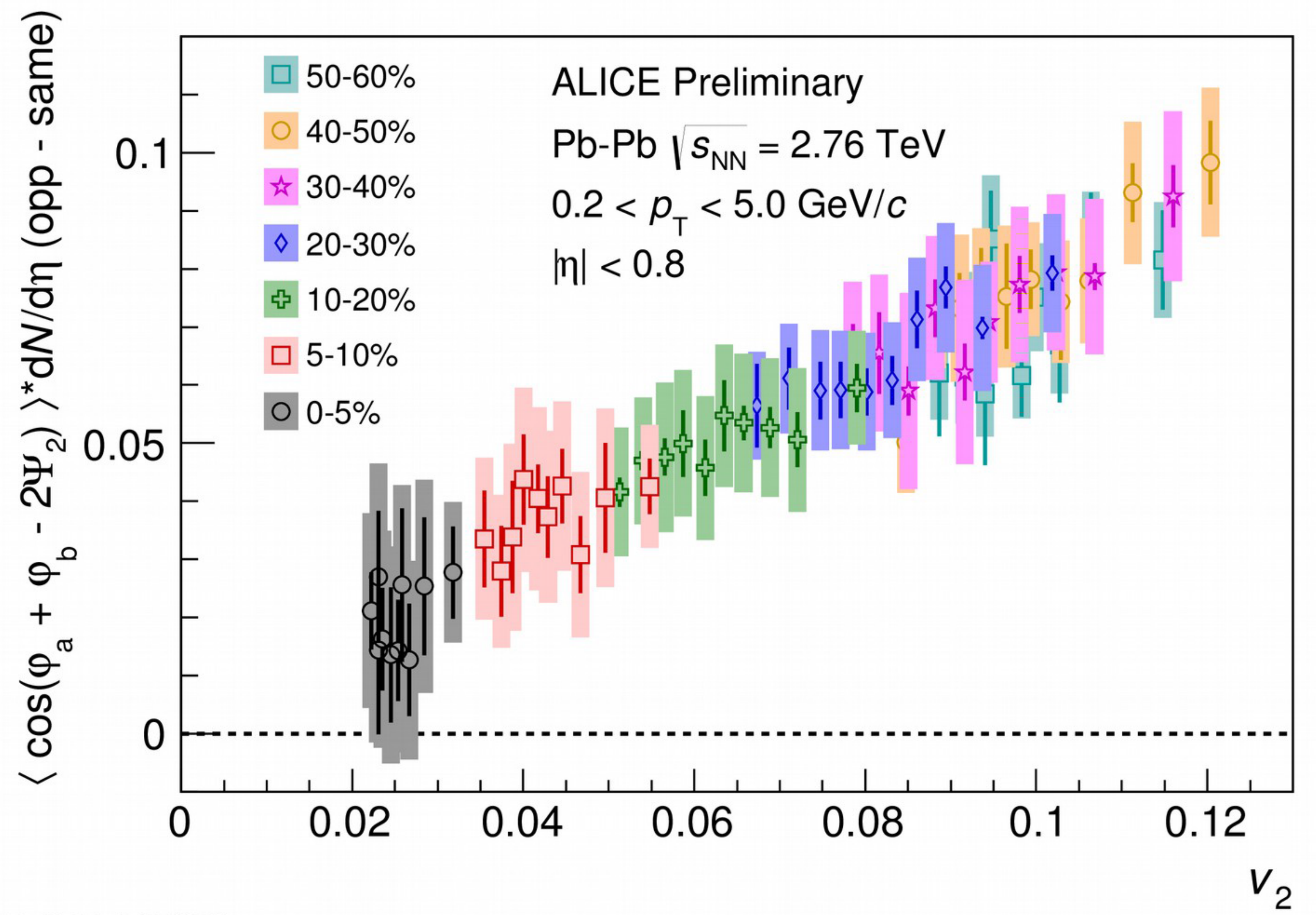}}
  \caption{(Color online) The charged-particle density scaled azimuthal correlator, $\dg\cdot\dNdeta$, as a function of $v_2$ for $q_2$ shape-selected events at various centralities in \PbPb\ collisions by ALICE. Error bars (shaded boxes) represent the statistical (systematic) uncertainties. Adapted from Ref.~\cite{Acharya:2017fau,ALICEQM17}.}
  \label{fig:ESE}
\end{figure}

Similarly, ALICE~\cite{Acharya:2017fau} divided their data in each collision centrality according to $q_2$. In order to remove the trivial multiplicity dilution effect, the correlator $\dg$ is scaled by the charged-particle density $\dNdeta$ in a given centrality.
The data are shown in Fig.~\ref{fig:ESE}.
The data indicate a strong linear dependence of the $\dg$ on the measured $v_2$ of the POIs, where different centralities fall onto the same linear trend after the multiplicity scaling. This observation is qualitatively consistent with the $v_2$-induced background scenario of Eq.~(\ref{eq:bkgd}).

\begin{figure}[!htb]
  \centerline{\includegraphics[width=0.5\hsize]{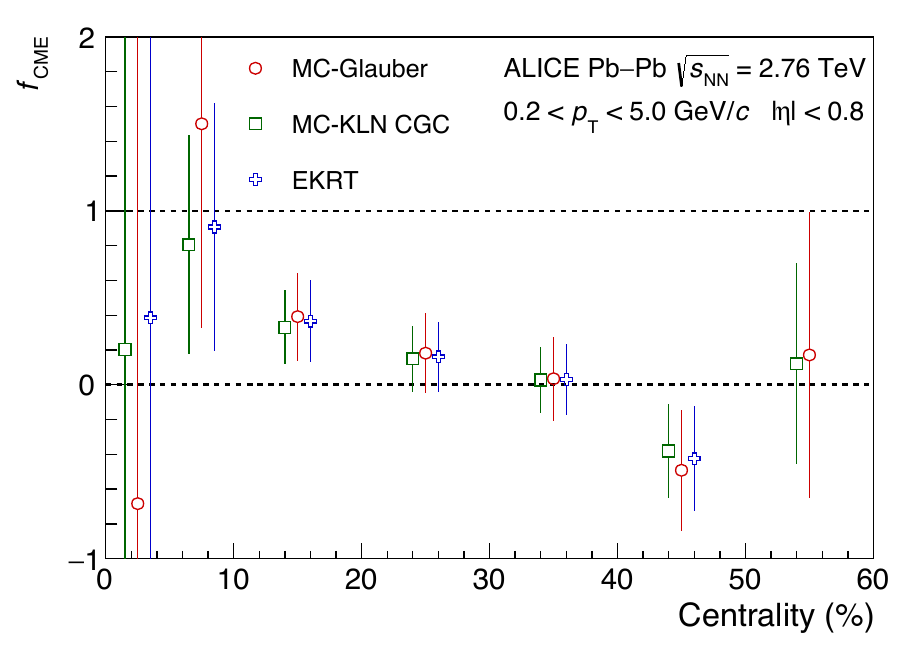}}
  \caption{Centrality dependence of the CME fraction extracted from the fits to data with different MC models for the magnetic field by ALICE. Points with the three models are slightly shifted along the horizontal axis for better visibility. Only statistical uncertainties are shown. Adapted from Ref.~\cite{Acharya:2017fau}.}
  \label{fig:ALICE_fcme}
\end{figure}

The advantage of using the ESE is to independently evaluate the $v_2$-dependent background in the $\dg$ correlator without significantly changing the CME signal due to the magnetic field.
A significant CME contribution would result in a non-zero intercept at $v_2=0$. 
One could fit the data with the linear function in $v_2$ and extract the possible CME signal by the fit intercept. 
However, within each centrality bin with different $q_2$ bins, the magnetic field could vary because the collision geometry may vary slightly by the $q_2$ selection within the centrality bin. Such a variation would be encoded in the variation of $q_2$ bins, hence the POI $v_2$. Thus, ALICE modeled the magnetic field as function of $v_2$, $B(v_2)$, using different {\em Monte Carlo} (MC) Glauber calculations: MC-Glauber, MC-KLN CGC and EKRT models~\cite{Acharya:2017fau}.
Specifically, the CME signal is considered to be proportional to $\mean{|\textbf B|^2 \cos2(\psiB - \psi_2)}$, where $|\textbf B|$ and $\psiB$ are the magnitude and azimuthal direction of the magnetic field (see Sect.~\ref{sec:plane}). 
With the $\dg$ signal dependence on $v_2$ from data, the residual CME signal can be extracted based on the different dependences of signal and background correlations on the measured $v_2$. 
Figure~\ref{fig:ALICE_fcme} presents the estimate of the fraction of the CME signal in the inclusive $\dginc$ measurement, $\fcme$. 
Averaging the 10-50\% centrality range gives a value of $\fcme=0.10\pm0.13$, $0.08\pm 0.10$, and $0.08\pm0.11$ using the three models for the magnetic field, where the quoted uncertainties are statistical. 
These results are consistent with zero CME fraction, and correspond to upper limits on $\fcme$ of 33\%, 26\% and 29\%, respectively, at 95\% confidence level (CL) for the 10-50\% centrality range~\cite{Acharya:2017fau}. 

\begin{figure}[!htb]
  \centerline{
    \includegraphics[width=0.42\hsize]{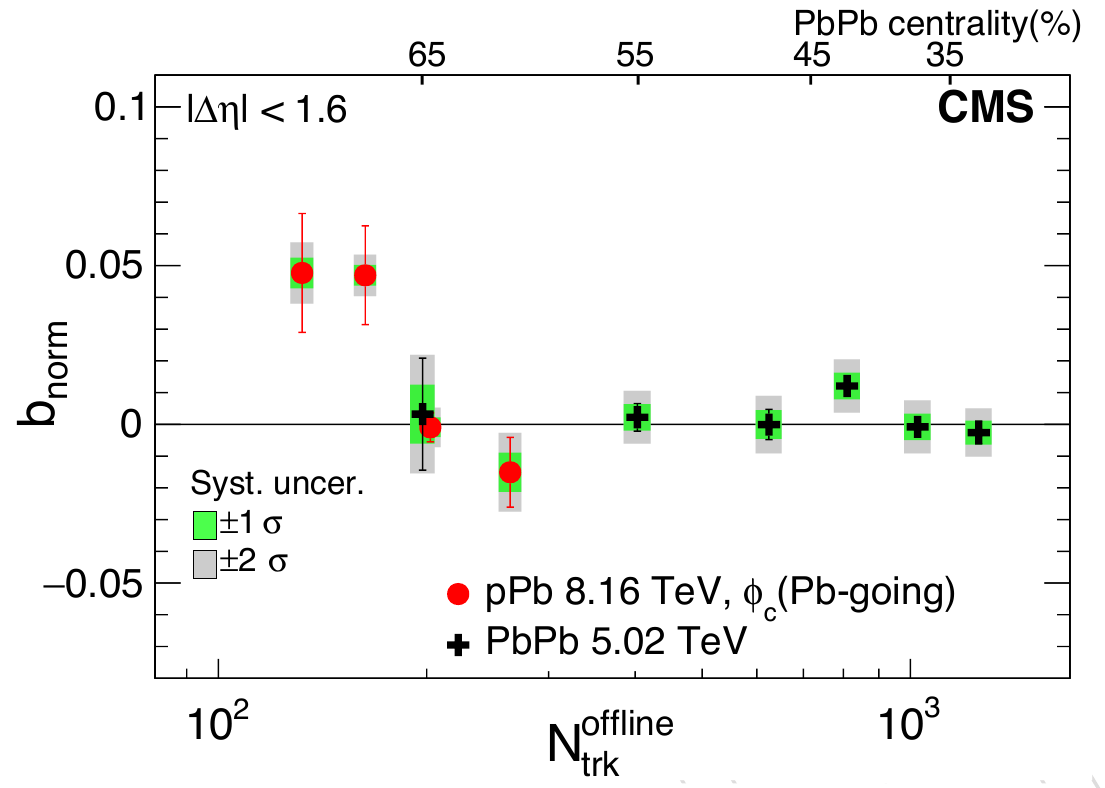}
    \hspace{0.05\hsize}
    \includegraphics[width=0.40\hsize]{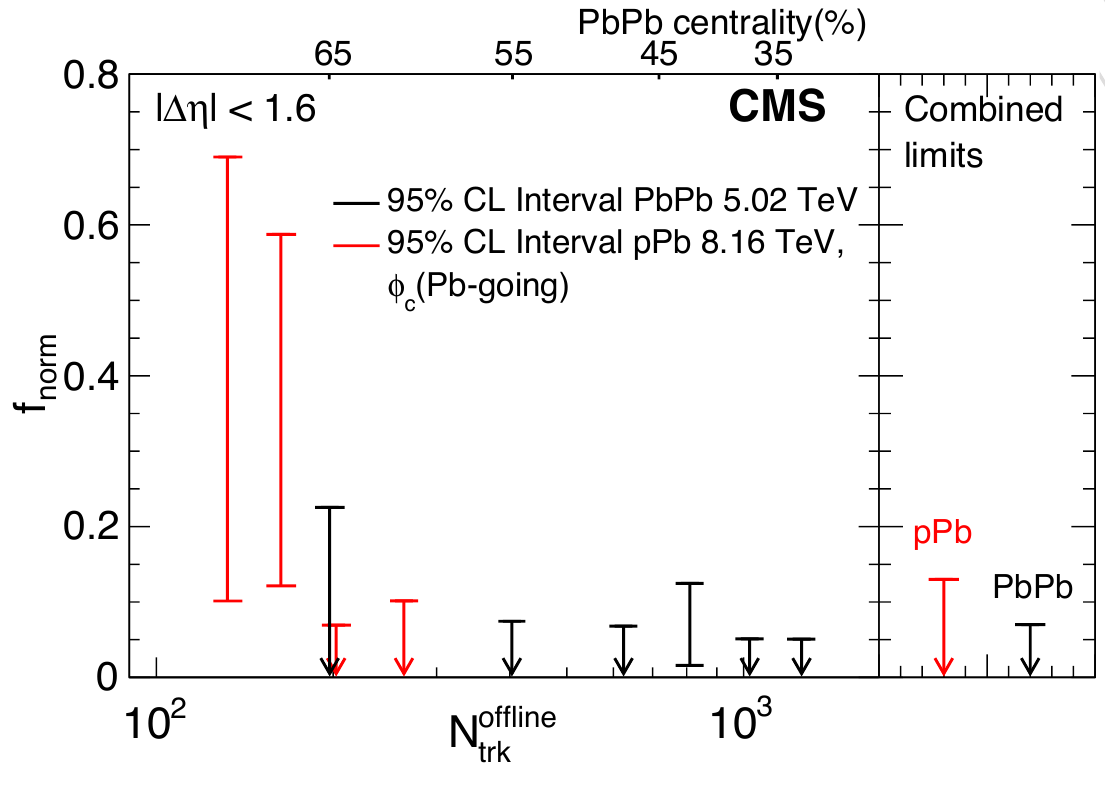}
  }
  \caption{(Color online) Extracted intercept parameter $\bnorm$ (left panel) and their corresponding upper limits of the fraction of the $v_2$-independent $\dg$ correlator component (right panel), averaged over $|\Delta\eta|< 1.6$, as a function of $\Noff$ in \pPb\ and \PbPb\ collisions by CMS. Adapted from Ref.~\cite{Sirunyan:2017quh}.}
  \label{fig:CMSeseB}
\end{figure}

CMS has also used the ESE to extract the possible CME signal by dividing their data into narrow multiplicity (centrality) bins. In the CMS approach, the signal and background contribution to the $\dg$ correlator are separated as~\cite{Bzdak:2012ia} (see Sect.~\ref{sec:kappa}): 
\be \dg=\kappa_2\dd v_2+\dg_{\rm{CME}}\,. \label{eq:dg_k} \ee
assuming that the magnetic field within each bin, and thus the possible CME signal $\dg_{\rm{CME}}$, does not change.
Using the ESE to select events with different $v_2$, the linear $v_2$ dependence in Eq.~(\ref{eq:dg_k}) can be explicitly tested and the $\dg_{\rm{CME}}$ be extracted. However, it is found that the $\dd$ is somewhat dependent on $v_2$ in peripheral events, mainly due to the multiplicity bias from the $q_2$ selection~\cite{Sirunyan:2017quh}. In order to remove this $v_2$ dependence, both sides of Eq.~(\ref{eq:dg_k}) are divided by $\dd$ and the equation becomes
\be \dg/\dd=\anorm v_2+\bnorm\,. \label{eq:dgd} \ee
Here $\bnorm$ represents the possible CME signal divided by $\dd$ which now has in principle a slight $v_2$ dependence. Since $\dg_{\cme}$ is small compared to the background contribution, $\bnorm$ can be simply treated as a constant in each multiplicity (centrality) bin.
The ratios of $\dg/\dd$ in \pPb\ and \PbPb\ collisions are indeed found to be linear in $v_2$ for different multiplicity (centrality) ranges~\cite{Sirunyan:2017quh}. The intercept parameter $\bnorm$ extracted from linear fits are shown as a function of $\Noff$ in Fig.~\ref{fig:CMSeseB} (left panel). Within statistical and systematic uncertainties, no significant positive value of $\bnorm$ is observed. 
Figure~\ref{fig:CMSeseB} (right panel) shows, at 95\% CL, the upper limit of the fraction $f_{\rm norm}\equiv\bnorm/(\mean{\dg}/\mean{\dd})$ (equivalently $\fcme$), as a function of $\Noff$. 
Combining all presented multiplicities and centralities, an upper limit on the possible CME signal fraction is estimated to be 13\% in \pPb\ and 7\% in \PbPb\ collisions, at 95\% CL. The results are consistent with a $v_2$-dependent background-only scenario, posing a significant challenge to the search for the CME in heavy-ion collisions using the $\gamma$ correlators~\cite{Sirunyan:2017quh}.

The attractive aspect of the ESE method is to be able to ``hold'' the magnetic field fixed and vary the event-by-event $v_2$~\cite{Adamczyk:2013kcb,Voloshin:2010ut,Chatterjee:2014sea}.
In reality, the magnetic field cannot really be held fixed and it is always possible that there is a variation of the magnetic field in an event sample as a function of the $v_2$, as ALICE has modeled. 
The ALICE analysis~\cite{Acharya:2017fau} is thus somewhat model-dependent which relies on the precise modeling of the correlations between the magnetic field and the $v_2$ in given centrality bins.
In the CMS approach~\cite{Sirunyan:2017quh}, narrow centrality bins are used and the CME signal is assumed to be constant within each of the narrow centrality bins. Thus the extraction of the CME signal does not depend on model assumptions about the magnetic field. However, the extracted CME signal is more vulnerable to systematics due to varying magnetic field.

\subsection{Invariant mass method}\label{sec:mass}
It has been known all along that the $\dg$ was contaminated by background from resonance decays coupled with the elliptic flow ($v_2$)~\cite{Wang:2009kd,Bzdak:2009fc,Liao:2010nv,Bzdak:2010fd,Schlichting:2010qia,Pratt:2010zn}.
The particle pair invariant mass ($\minv$) is a common tool to study resonances, however, the $\minv$ dependence of the $\dg$ observable has been examined only recently~\cite{Zhao:2017nfq,Li:2018oot}. Removing resonance decay backgrounds by $\minv$ cuts could enhance the sensitivity of the $\dg$ measurements to potential CME signals.

\begin{figure}[!htb]
  \centerline{
    \includegraphics[width=0.42\hsize]{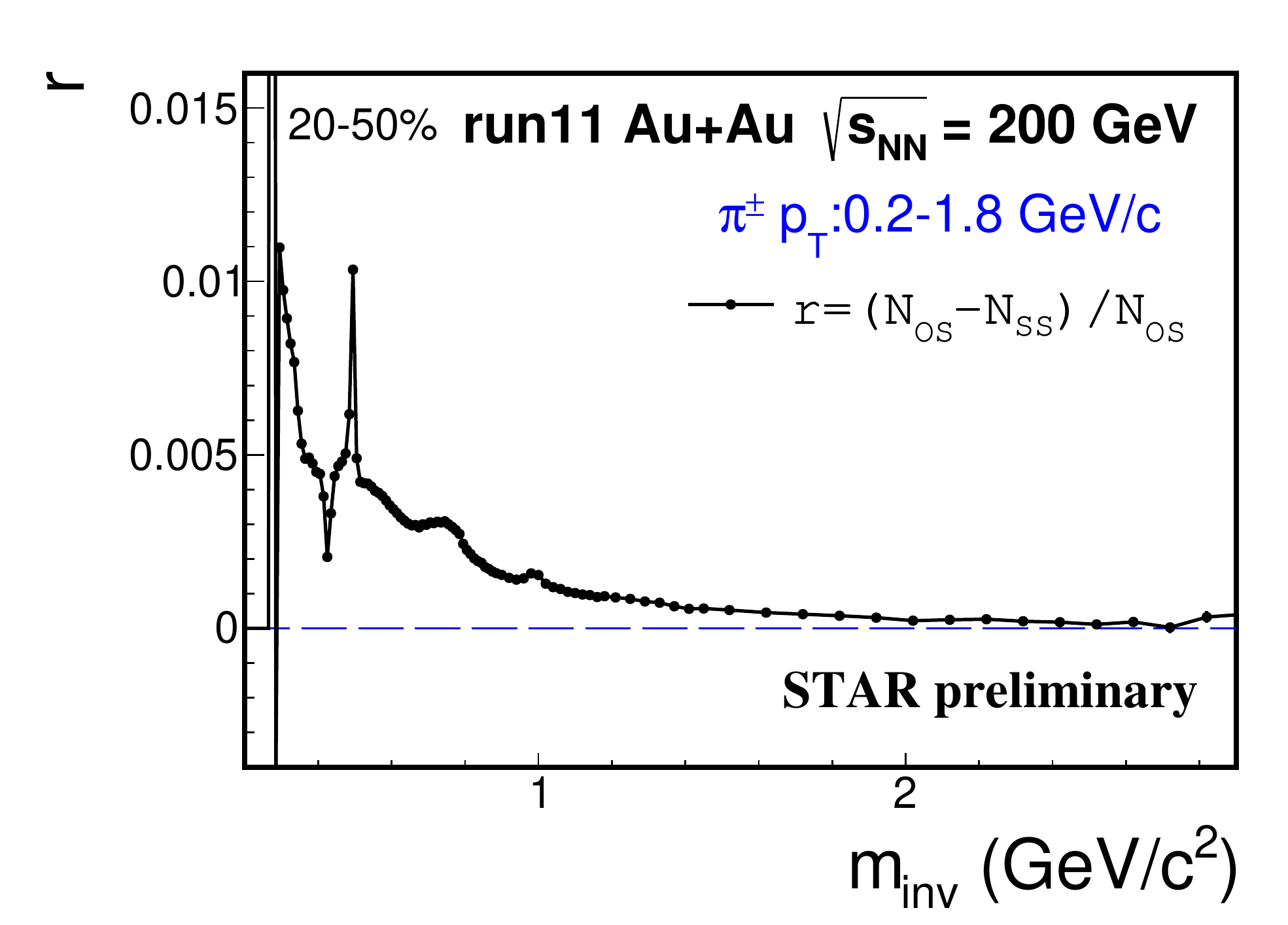}
    \hspace{0.05\hsize}
    \includegraphics[width=0.42\hsize]{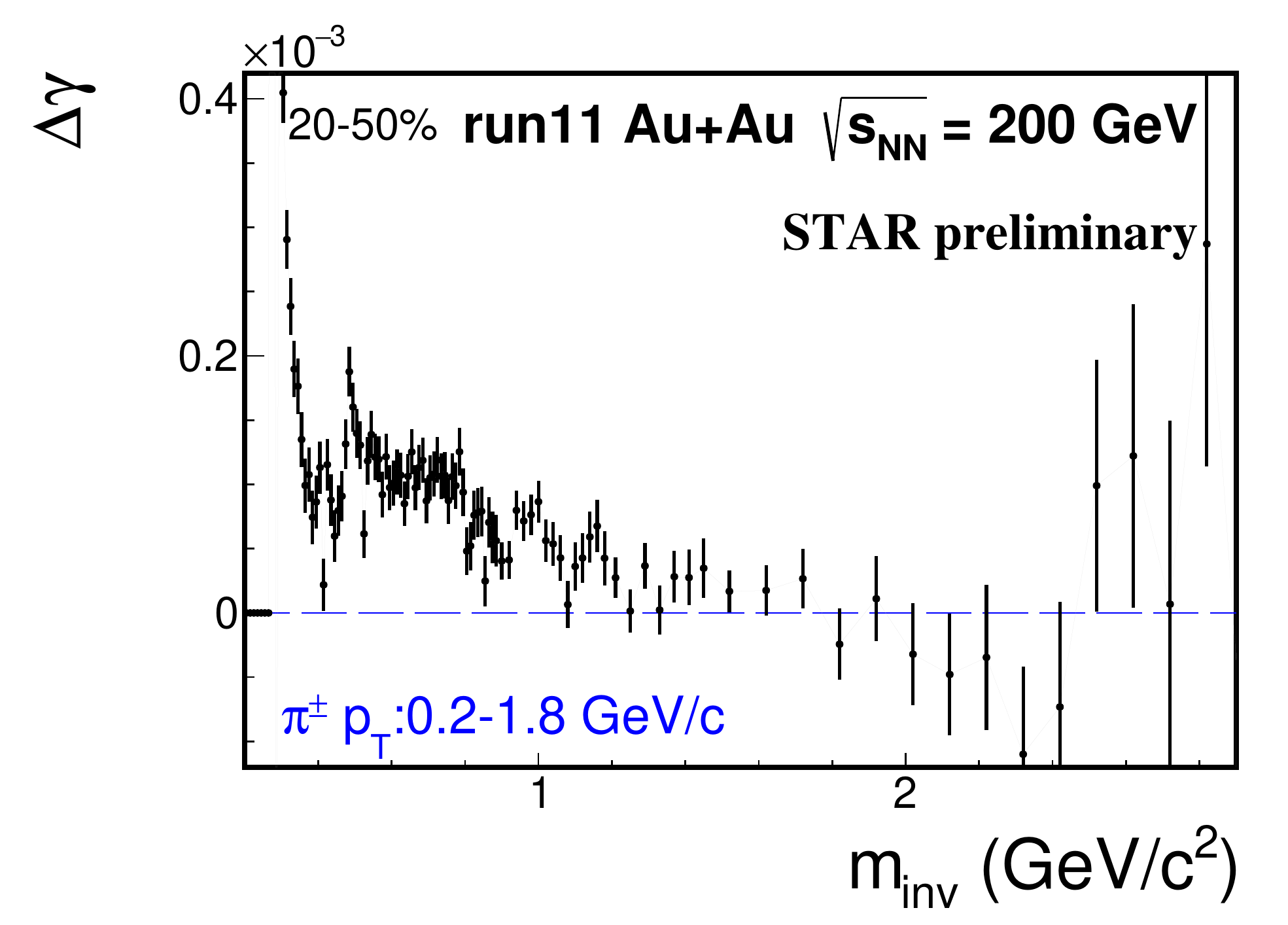}
  }
  \caption{The invariant mass ($\minv$) dependence of the relative excess of OS over SS pairs of charged pions, $r=(N_{\rm OS}-N_{\rm SS})/N_{\rm OS}$ (upper panel), and the azimuthal correlator difference, $\dg=\gOS-\gSS$ (lower panel) in 20-50\% \AuAu\ collisions at $\snn = 200$~GeV from Run-11 by STAR. Errors shown are statistical. Adapted from Refs.~\cite{Zhao:2017wck,Zhao:2018pnk,Zhao:2018blc}.} 
  \label{fig:mass}
\end{figure}

Figure~\ref{fig:mass} shows the preliminary results in mid-central \AuAu\ collisions by STAR~\cite{Zhao:2017wck,Zhao:2018pnk,Zhao:2018blc}.
The left panel shows the $\minv$ dependence of the relative OS and SS pion pair abundance difference, $r=(N_{\rm OS}-N_{\rm SS})/N_{\rm OS}$.
The pions are identified by the TPC and the time-of-flight (TOF) detector within pseudorapidity and $\pt$ ranges of $|\eta|<1$ and $0.2<\pt<1.8$~\gevc, respectively. The resonance peaks of $K_S$ and  $\rho$ are clearly seen. The large increase toward the low-$\minv$ kinematic limit is due to the acceptance edge effect, where the OS and SS pair acceptance difference of the detector amplifies~\cite{Zhao:2018blc,Adamczyk:2015lme}.
The right panel shows the $\dg$ measurement as a function of $\minv$. A clear peak at the $K_S$ mass is observed; a broad peak at the $\rho$ mass is observable.
The $\minv$ structures are similar in $r$ and $\dg$; the $\dg$ correlator traces the distribution of the resonances. 
\begin{figure}[!htb]
  \centerline{
    \includegraphics[width=0.45\hsize]{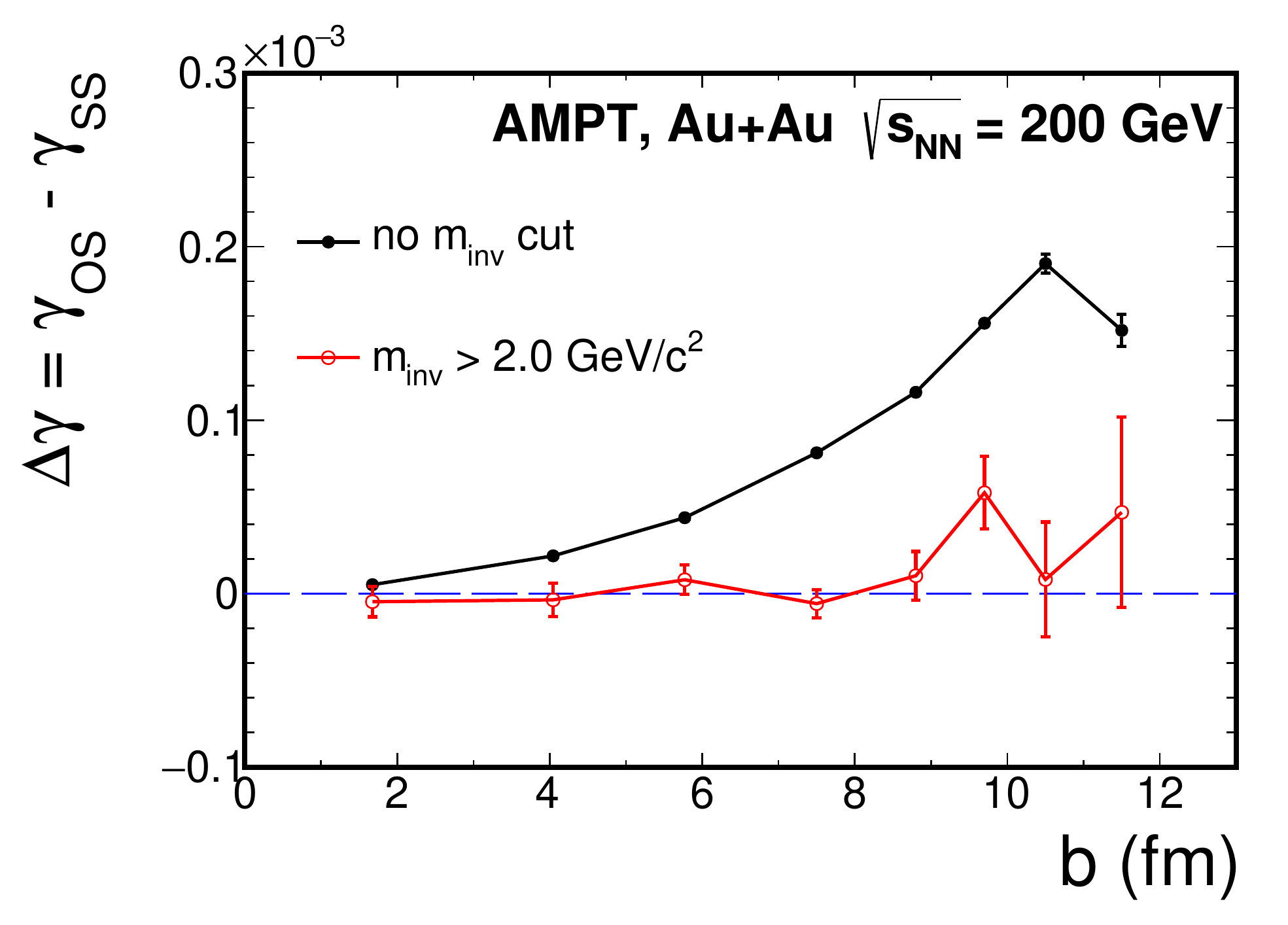}
    \hspace{0.05\hsize}
    \includegraphics[width=0.45\hsize]{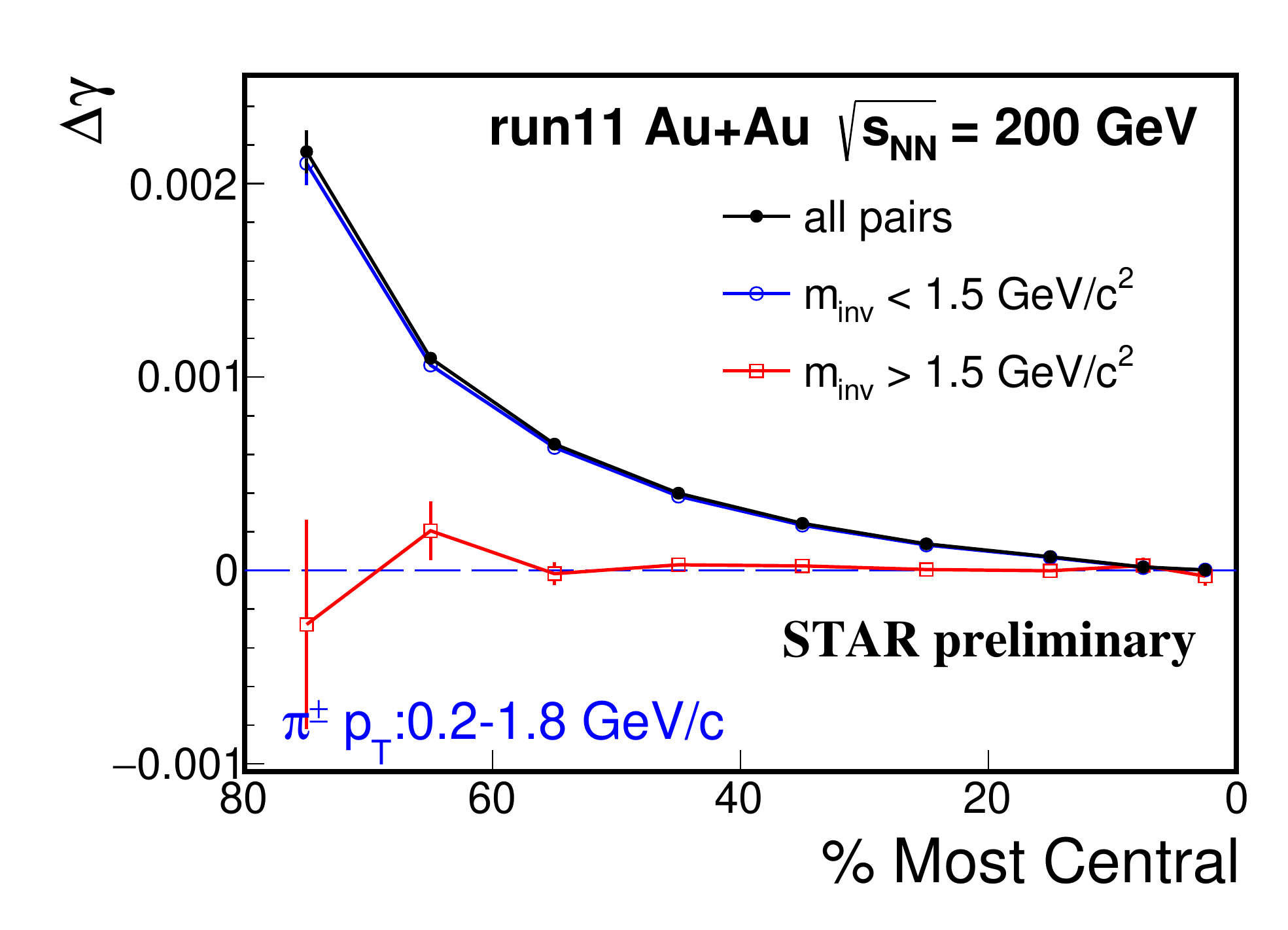}
  }
  \caption{The average $\dg$ at large pair mass, compared to the inclusive $\dginc$, in \AuAu\ collisions at $\snn=200$~GeV. Left panel: from AMPT simulation as function of the impact parameter ($b$)~\cite{Zhao:2017nfq,Li:2018oot}. Right panel: from Run-11 STAR data as function of centrality~\cite{Zhao:2017wck,Zhao:2018pnk,Zhao:2018blc}. Errors shown are statistical.}
  \label{fig:high}
\end{figure}

Most of the $\pi$-$\pi$ resonance contributions are located in the low $\minv$ region, below $\minv<1.5$~\gevcc~\cite{Agashe:2014kda,Adams:2003cc}; at higher $\minv$ the resonance contribution to the OS-SS difference can be neglected.
The easiest way to remove resonance contributions from $\dg$ is, therefore, to restrict the measurements to the large-$\minv$ region. AMPT model simulation shows that such a $\minv$ cut, although significantly reducing the pair statistics, can eliminate essentially all resonance decay backgrounds~\cite{Zhao:2017nfq,Li:2018oot}. This is shown in the left panel of Fig.~\ref{fig:high}. The AMPT $\dg$ at high $\minv$ is consistent with zero as expected because there is no CME in AMPT. This is also a good confirmation that a lower mass cut can eliminate all background contributions to $\dg$.
The right panel of Fig.~\ref{fig:high} shows the average $\dg$ from STAR data with a lower mass cut, $\minv>1.5$~\gevcc, in comparison to the inclusive $\dginc$ measurement~\cite{Zhao:2017wck,Zhao:2018pnk,Zhao:2018blc}. The high mass $\dg$ is drastically reduced from the inclusive data, by over an order of magnitude.
Preliminary STAR data combining Run-11 ($\sim$0.5 billion minimum-bias events taken in year 2011), Run-14 ($\sim$0.8 billion, year 2014), and Run-16 ($\sim$1.2 billion, year 2016) yield a $\dg$ at $\minv>1.5$~\gevcc\ of $(5\pm2\pm4)$\% of the inclusive $\dginc$ measurement in 20-50\% centrality Au+Au collisions at $\snn=200$~GeV~\cite{Zhao:2018blc}; the systematic uncertainty is currently estimated from the differences among the three runs~\cite{Zhao:2018blc}. The high mass $\dg$ is consistent with zero within two standard deviations.

\begin{figure*}[!htb]
  \centerline{\includegraphics[width=0.7\hsize]{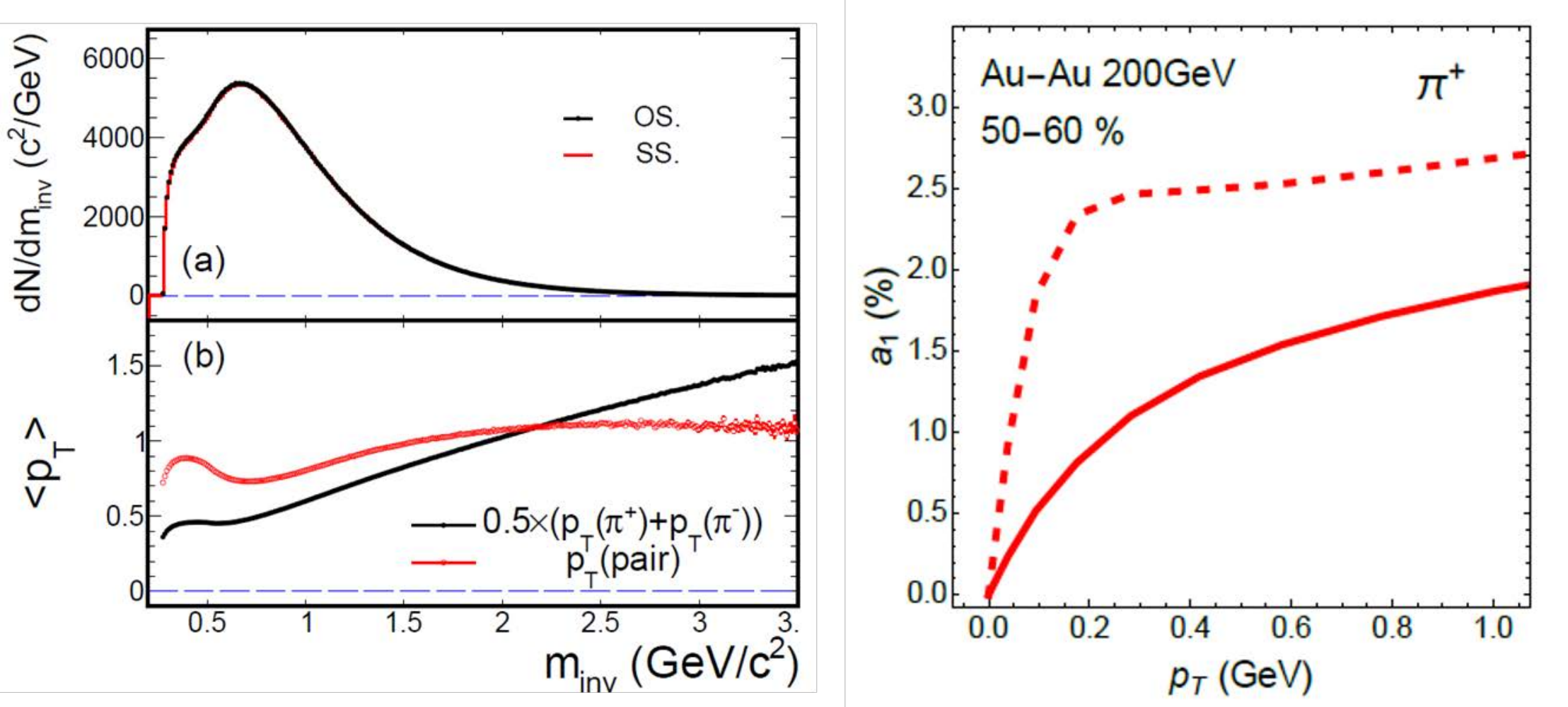}}
  \caption{(Color online) Upper left panel: typical $\minv$ distributions of pion pairs in relativistic heavy-ion collisions. Lower left panel: the $\mean{\pt}$ of single pions (black) and of pion pairs (red) as functions of $\minv$~\cite{Li:2018oot}. Right panel: the CME charge separation signal strength in directly produced pions (dashed) and in final-state pions (solid) as functions of $\pt$~\cite{Shi:2017cpu}.}
  \label{fig:pt}
\end{figure*}

It is generally expected that the CME is a low $\pt$ phenomenon and its contribution to high mass may be small~\cite{Kharzeev:2007jp,Abelev:2009ad}. However, as shown in Fig.~\ref{fig:pt} left panel, a $\minv$ cut of 1.5~\gevcc\ corresponds to $\pt\sim 1$~\gevc\ which is not particularly high. 
Moreover, a recent study~\cite{Shi:2017cpu} indicates that the CME signal is rather independent of $\pt$ at $\pt>0.2$ \gevc\ (Fig.~\ref{fig:pt} right panel). These studies suggest that the CME signal may persist to relatively high $\minv$.
Because of the vanishing background contributions, a positive measurement of $\dg$ at large $\minv$ would be a good indication of the existence of the CME.
It is worthwhile to note, however, that a null measurement of the CME at high $\minv$ does not necessarily mean that the CME at low $\minv$ is also zero.

One can take a step further to use the low $\minv$ data to extract the possible CME signal. In order to do so, resonance contributions must be excluded.
In a two-component model, the $\minv$ dependence of the $\dg$ can be expressed~\cite{Zhao:2017nfq,Li:2018oot} as
\be \dg(\minv) \approx r(\minv)R(\minv) + \dg_{CME}(\minv)\,. \label{eq:dgminv} \ee
The first term is resonance contributions, where the response function $R(\minv)$ should be a smooth function of $\minv$, while $r(\minv)$ contains resonance mass shapes. 
Consequently, the first term is not ``smooth'' but a peaked function of $\minv$.
The second term in Eq.~(\ref{eq:dgminv}) is the CME signal which should be a smooth function of $\minv$. 
The $\minv$ dependences of the CME signal and the background are distinctively different, and this can be exploited to extract CME signals at low $\minv$.
The feasibility of this method was investigated by a toy-MC simulation~\cite{Zhao:2017nfq} as well as in STAR data~\cite{Zhao:2017wck,Zhao:2018pnk}.
In principle, in order to extract the CME signal in the low-$\minv$ region, the $\minv$ dependence of the background contribution is needed. In the STAR analysis, a linear response function $R(\minv)$ was assumed, guided by AMPT simulations~\cite{Zhao:2017nfq}, and various forms of CME$(\minv)$ were studied~\cite{Zhao:2017wck,Zhao:2018pnk}.

\begin{figure}[!htb]
  \centerline{\includegraphics[width=0.55\hsize]{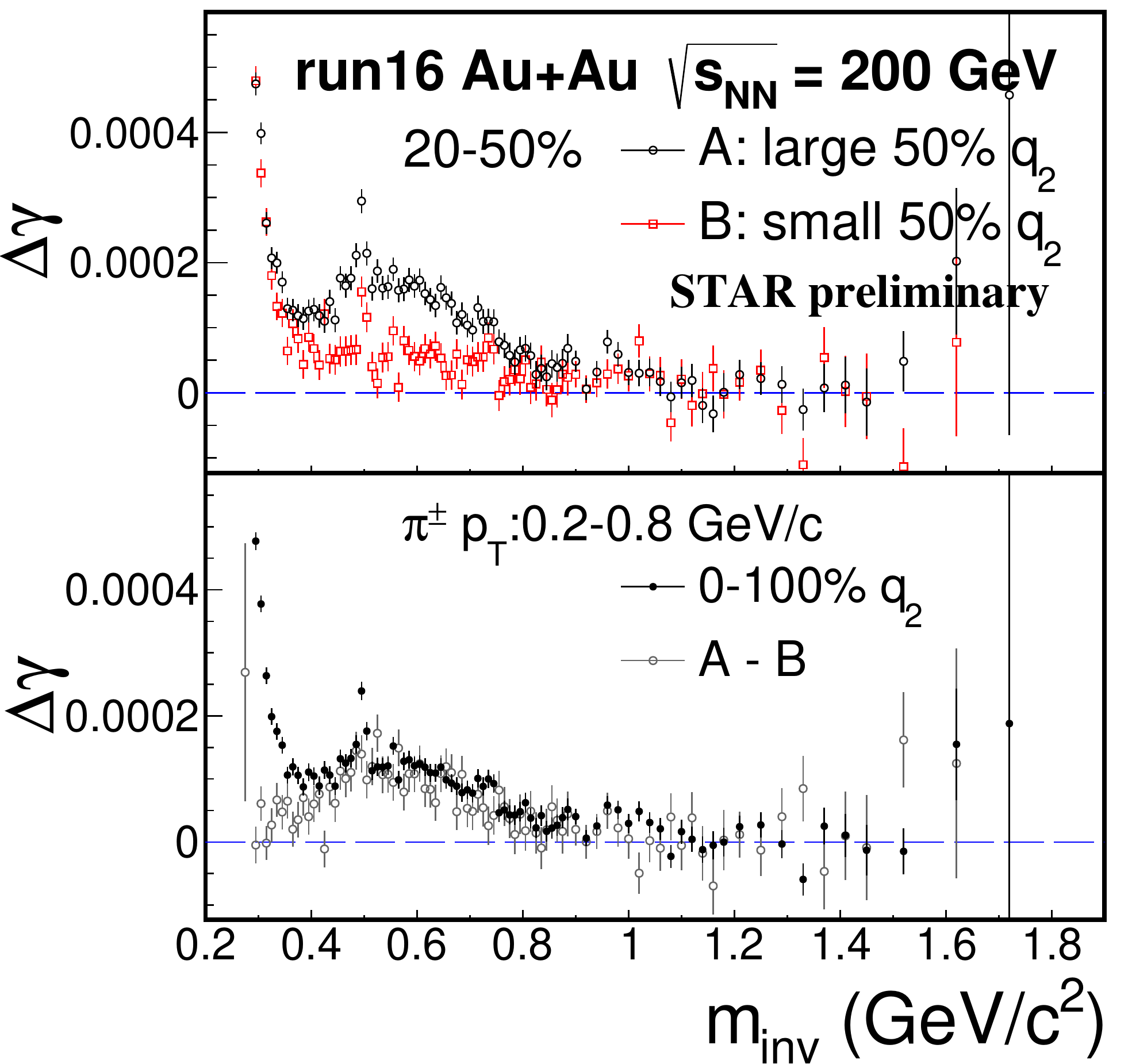}}
  \caption{The $\minv$ dependences of the $\dg$ in large and small $q_2$ events (upper panel), and the $\dg$ difference between large and small $q_2$ events together with the inclusive $\dginc$ (lower panel) in 20-50\% central \AuAu\ collisions at $\snn=200$~GeV from Run-16 by STAR. Errors shown are statistical. Adapted from Ref.~\cite{Zhao:2018blc}.}
  \label{fig:q2}
\end{figure}

One difficulty in the above method is that the exact functional form of $R(\minv)$ is presently unknown and requires rigorous modeling and experimental inputs.
To overcome this difficulty, STAR has recently improved the $\minv$ method, complemented by the ESE technique~\cite{Zhao:2018blc}. We call this the ``low $\minv$+ESE fit'' method. The events in each narrow centrality bin are divided into two classes according to the event-by-event $q_2$~\cite{Schukraft:2012ah}, calculated by Eqs.~(\ref{eq:Q}) and~(\ref{eq:q}) using the POIs. Since the magnetic fields are approximately equal within the narrow centrality bin while the backgrounds differ due to the different $q_2$ selections, the $\dg(\minv)$ difference between the two classes is a good measure of the background shape. 
Figure~\ref{fig:q2} shows the $\dg(\minv)$ distributions for such two $q_2$ classes ($\dg_A$ and $\dg_B$) in the upper panel, and the difference $\dg_A-\dg_B$ together with the inclusive $\dginc$ of all events in the lower panel in 20-50\% \AuAu\ collisions~\cite{Zhao:2018blc,Zhao:2018blc}. The $q_2$ selection is applied in narrower centrality bins, and then the data are combined over the range of 20-50\%. Note that the pion identification here was done using the TPC energy loss ($dE/dx$) information only, different from that in Figs.~\ref{fig:mass} and~\ref{fig:high}~\cite{Zhao:2018blc}.
\begin{figure}[!htb]
  \centerline{
    \includegraphics[width=0.45\hsize]{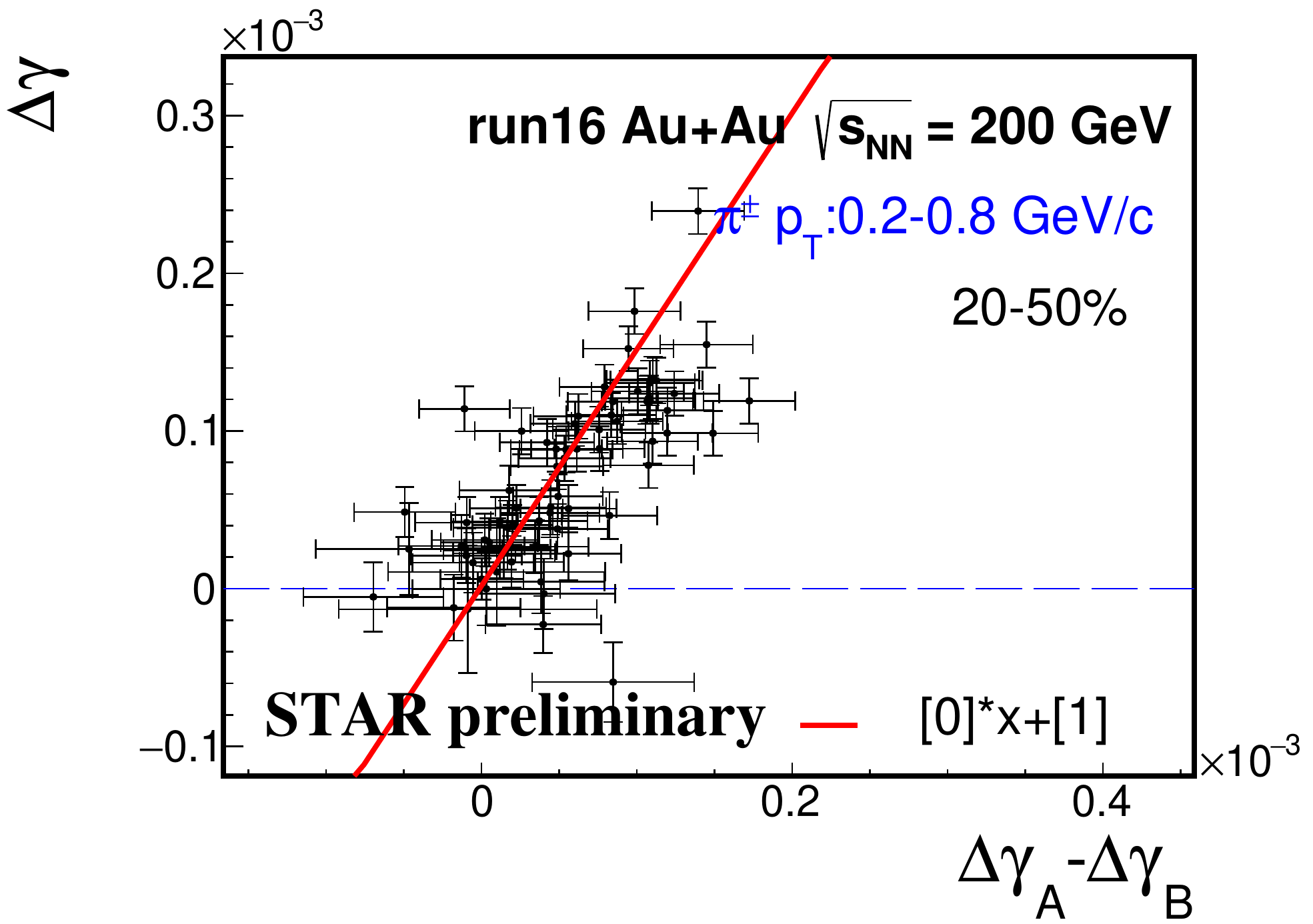}
    \hspace{0.05\hsize}
    \includegraphics[width=0.45\hsize]{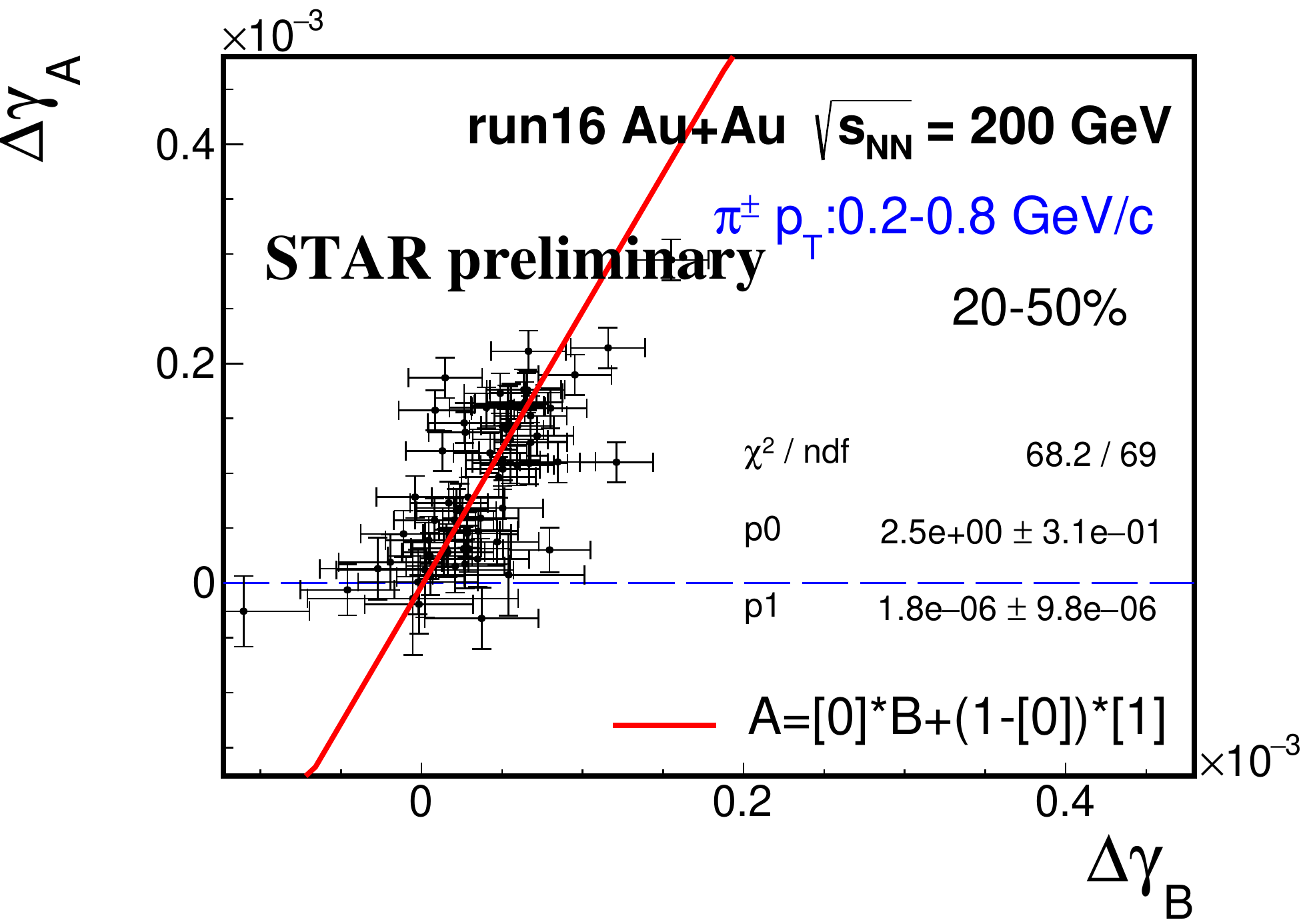}
  }
  \caption{The $\dg$ versus $\dg_A-\dg_B$ (left panel), and $\dg_A$ versus $\dg_B$ (right panel) in 20-50\% central \AuAu\ collisions at $\snn=200$~GeV from Run-16 by STAR. Each data point in the left (right) panel corresponds to one $\minv$ bin in the lower (upper) panel of Fig.~\ref{fig:q2}; only the $\minv>0.4$~\gevcc\ data points are included. Errors shown are statistical. Adapted from Ref.~\cite{Zhao:2018blc}.}
  \label{fig:fit}
\end{figure}

The overall $\dg$ contains both background and the possible CME. With the background shape given by $\dg_A-\dg_B$, the CME can be extracted from a two-component fit to the form:
\be\dg=b(\dg_A-\dg_B)+\dg_{\cme}\,.\label{eq:fitold}\ee
Figure~\ref{fig:fit} left panel shows $\dg$ as a function of $\dg_A-\dg_B$, where each data point corresponds to one $\minv$ bin in the lower panel of Fig~\ref{fig:q2}.
Only the $\minv>0.4$~\gevcc\ data points are included in Fig.~\ref{fig:fit} because the $\dg$ from the lower $\minv$ region is affected by edge effects of the STAR TPC acceptance~\cite{Zhao:2018blc,Adamczyk:2015lme}.
As seen in Fig.~\ref{fig:fit} left panel, there is a positive linear correlation between $\dg$ and $\dg_A-\dg_B$.
Since the same data are used in $\dg$ and $\dg_A-\dg_B$, their statistical errors are somewhat correlated.
To properly handle statistical errors, one can simply fit the independent measurements of $\dg_A$ versus $\dg_B$ by
\be
  \dg_A=k\dg_B+(1-k)\dg_{\cme}\,,
  \label{eq:fit}
\ee
where $k$ and $\dg_{\cme}$ are the fit parameters. Figure~\ref{fig:fit} right panel shows such a fit for the Run-16 \AuAu\ data~\cite{Zhao:2018blc}.
The straight line superimposed on the left panel of Fig.~\ref{fig:fit} is the same fit to Eq.~(\ref{eq:fit}), properly converted to Eq.~(\ref{eq:fitold}) by $b=(k+1)/2(k-1)$ assuming $\dg=(\dg_A+\dg_B)/2$.
The fit parameter $k$ reflects the relative background contribution in the large-$q_2$ (large-$v_2$) event class to that in the small-$q_2$ (small-$v_2$) event class, and since the background increases with $v_2$, the value of the $k$ parameter is larger than unity. The CME signal $\dg_{\cme}$ obtained from the fit is consistent with zero.

Note that in this fit model, unlike the ESE method described in Sect.~\ref{sec:ese}, the background is not required to be strictly proportional to $v_2$~\cite{Li:2018oot}. As long as the backgrounds are different for different $q_2$ event classes, one can extract the background shape as function of $\minv$. The slope fit parameter in Eq.~(\ref{eq:fit}) contains how good the linearity of the $\dg$ background versus $v_2$ is. The fit model, however, does assume the CME signal to be independent of $\minv$. The good fit quality seen in Fig.~\ref{fig:fit} indicates that this is a good assumption within the current statistical precision of the data. 

Combining Run-11, 14, and 16 data (total $\sim$2.5 billion minimum-bias events), STAR obtained the possible CME signal to be $(2\pm4\pm6)$\% of the inclusive $\dginc$ in 20-50\% centrality Au+Au collisions at $\snn=200$~GeV, where the systematic uncertainty is presently assessed from the differences among the three runs~\cite{Zhao:2018blc}.

\subsection{Harmonic-plane comparison method}\label{sec:plane}
The CME-induced charge separation is driven by the magnetic field, and is therefore the strongest along the magnetic field direction.
The magnetic field is mainly produced by spectator protons~\cite{Kharzeev:2007jp}, so its direction is on average perpendicular to the SP. 
The CME signal is therefore the largest if measured with respect to the SP~\cite{Xu:2017qfs}. 
The major background to the CME is related to the elliptic flow anisotropy $v_2$. The $v_2$ is generated by expansion of the participant geometry~\cite{Ollitrault:1992bk,Heinz:2013th}, and is therefore the largest when measured with respect to the PP~\cite{Alver:2006wh}.
The SP direction and the PP direction in heavy-ion collisions are correlated but different due to dynamical fluctuations of the nucleon positions in the colliding nuclei. 
These facts led to the novel idea to determine the CME signal (and flow background simultaneously) from $\dg$ measurements in the same collision event, one with respect to the SP (or RP) and the other with respect to the PP~\cite{Xu:2017qfs}. It is found that the spectator plane SP nearly coincides with the reaction plane RP in heavy-ion collisions except for very central collisions~\cite{Xu:2017qfs}, so we will simply use RP and SP interchangeably.
We call this method the ``RP-PP method.'' 

\begin{figure}[!htb]
  \centerline{
    \includegraphics[width=0.4\hsize]{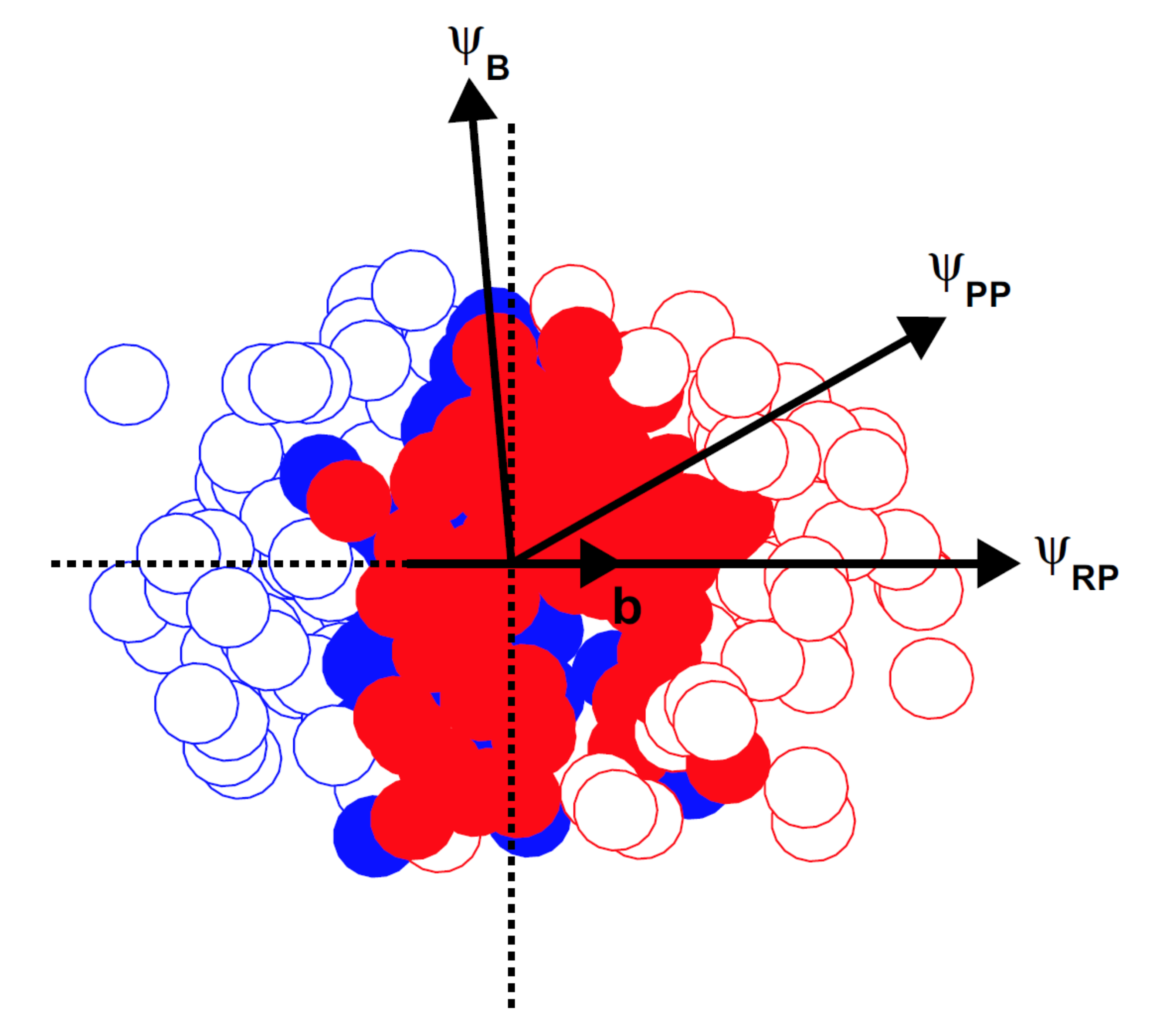}
    \hspace{0.05\hsize}
    \includegraphics[width=0.35\hsize]{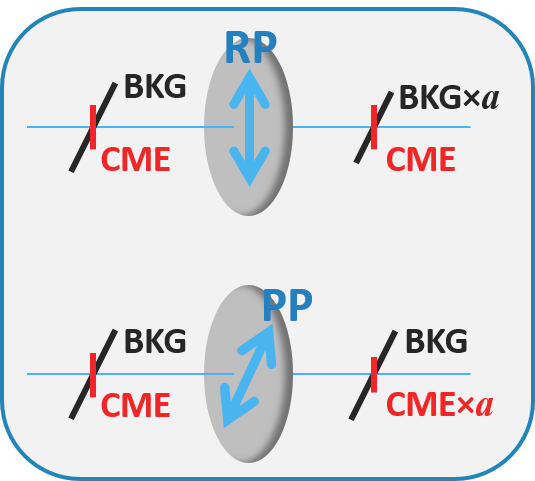}
  }
  \caption{Left: sketch of a heavy-ion collision projected onto the transverse plane (perpendicular to the beam direction). The $\psiRP$ is the azimuthal angle of the reaction plane (impact parameter, $b$) direction, $\psiPP$ the participant plane direction (of interacting nucleons, denoted by the solid circles), and $\psiB$ the magnetic field direction (mainly from spectator protons, denoted by the open circles together with spectator neutrons). Right: illustration of the ``CME-background filter.'' Present in a single collision are a CME signal ``along'' the RP and a background ``along'' the PP. The RP and PP are not the same but with an opening angle factor, $a=\mean{\cos2(\psiPP-\psiRP)}$. With the RP ``filter,'' the background is reduced by a factor of $a$ and the CME remains in entirety, whereas with the PP ``filter,'' the full background remains and the CME is reduced by the same factor of $a$.}
  \label{fig:plane}
\end{figure}

Due to fluctuations~\cite{Alver:2006wh}, the $\psiB$ and $\psiPP$ can be wildly different on event-by-event basis. The fluctuations of the participant nucleon positions cause $\psiPP$ to be different from $\psiRP$. The fluctuations of the positions of the spectator protons, mainly responsible for the the magnetic field, cause $\psi_{B}$ to be not always perpendicular to the $\psiRP$. In addition, the magnetic field directions and magnitudes are all different at different locations in the collision zone. The CME measurement would be an average effect of the magnetic fields along a particular plane where the measurement is made. The position fluctuations of the participant nucleons and spectator protons are independent (except that they are all in the nucleus), thus $\psiPP$ and $\psi_{B}$ fluctuate independently about $\psiRP$. See the cartoon in the left panel of Fig.~\ref{fig:plane}. 

\begin{figure*}[!htb]
  \centerline{\includegraphics[width=0.9\hsize]{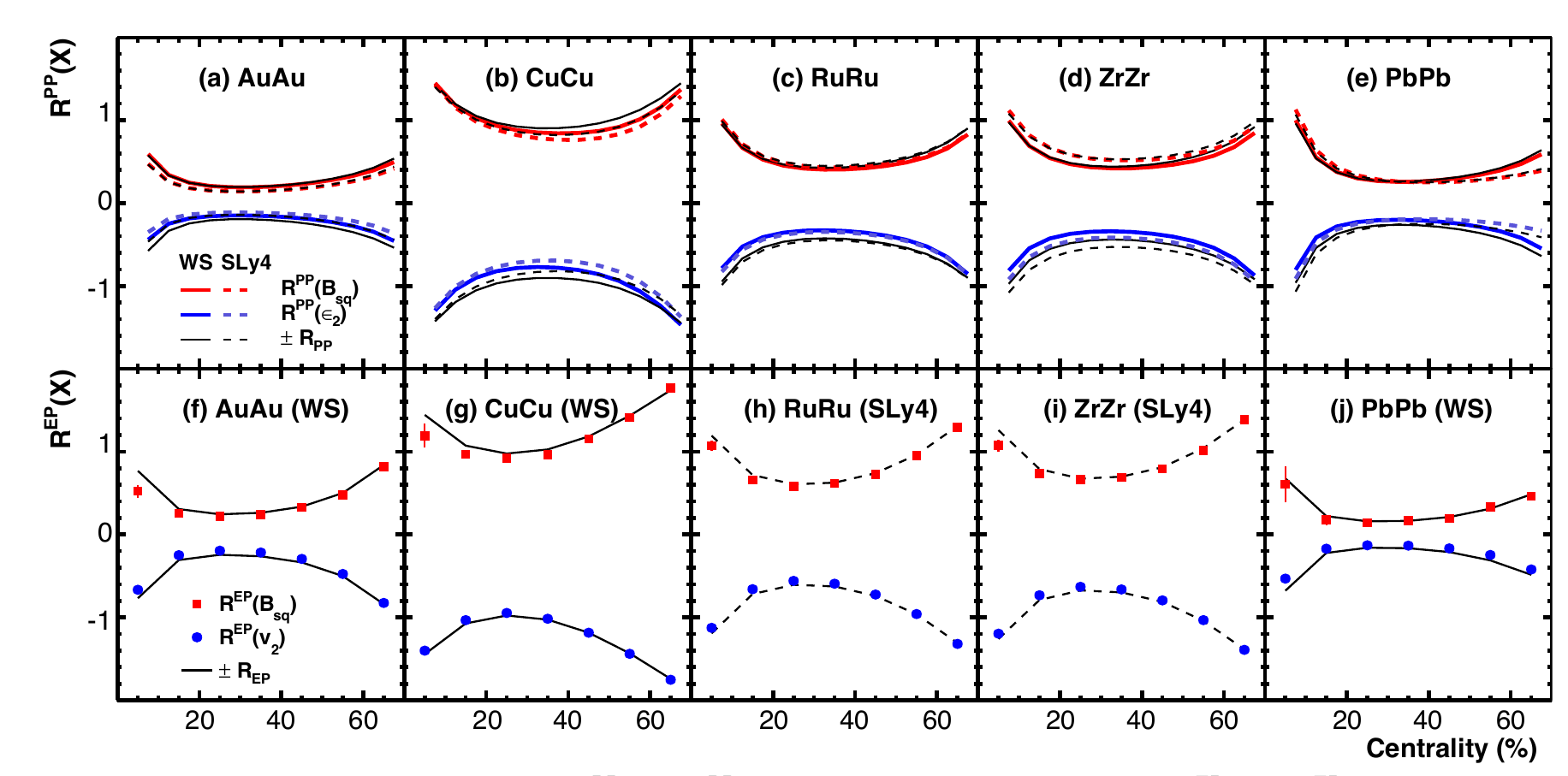}}
  \caption{(Color online) Relative differences $R_{\epsilon_2}$, $R_{\Bsq}$ from MC Glauber model (upper panels) and $R_{v_2}$, $R_{\Bsq}$ from AMPT (lower panels) for (a,f) \AuAu, (b,g) Cu+Cu, (c,h) Ru+Ru, and (d,i) Zr+Zr at RHIC, and (e,j) \PbPb\ at the LHC. Both the Woods-Saxon and density functional theory (DFT) calculated densities are shown for the MC Glauber calculations, while the used density profiles are noted for the AMPT results. Adapted from Ref.~\cite{Xu:2017qfs}.}
  \label{fig:RP2}
\end{figure*}

To put in mathematical terms, the eccentricity of the transverse overlap geometry is related to the PP. It yields the largest $v_2\{{\rm PP}\}$. Thus the $v_2$-induced background will be the largest in the $\dg$ measurement with respect to the PP, and will be weaker in the $\dg$ measurement with respect to the RP. The reduction factor is determined by the opening angle between the two planes and equals to
\be
a=\mean{\cos2(\psiPP-\psiRP)}\,. \label{eq:a}
\ee
The quantity relevant to the CME measured with respect to $\psi$ is the average magnetic field squared with correction from the event-by-event azimuthal fluctuations of the magnetic field orientation~\cite{Kharzeev:2007jp},
\be\Bsq\equiv\mean{(eB/\mpi^2)^2\cos2(\psi_{B}-\psi)}\,.\label{eq:Bsq}\ee
It is therefore strongest along the RP direction because the magnetic field is mainly generated by the spectator protons. The CME is smaller if measured with respect to the PP, reduced by the same factor $a$ of Eq.~(\ref{eq:a}). In other words, the relative difference in the eccentricities with respect to $\psiRP$ and $\psiPP$,
\be R_{\epsilon_2}\equiv2\cdot\frac{\epsilon_2\{\psiRP\}-\epsilon_2\{\psiPP\}}{\epsilon_2\{\psiRP\}+\epsilon_2\{\psiPP\}}\,,\label{eq:Recc}\ee
and that in the corresponding magnetic field strengths,
\be R_{\Bsq}\equiv2\cdot\frac{\Bsq\{\psiRP\}-\Bsq\{\psiPP\}}{\Bsq\{\psiRP\}+\Bsq\{\psiPP\}}\,,\label{eq:RBsq}\ee
are the opposite. Namely
\be R_{\Bsq}=-R_{\epsilon_2}=2(1-a)/(1+a)\,.\label{eq:R}\ee
This is verified by MC Glauber model calculations~\cite{Xu:2014ada,Zhu:2016puf} for various collision systems, shown in the upper panels of Fig.~\ref{fig:RP2}~\cite{Xu:2017qfs}. The AMPT simulations using the reconstructed EP, shown in the lower panels of Fig.~\ref{fig:RP2}, also confirm the conclusion~\cite{Xu:2017qfs}.
Because of fluctuations~\cite{Alver:2006wh}, the PP and RP do not coincide, so $a$ has a value always smaller than unity. In other words, the $\dg$ measurements with respect to the PP and the RP contain different amounts of the $v_2$ backgrounds and the CME signals. Thus, the two $\dg$ measurements can resolve two quantities, namely the $v_2$ background and the CME signal. This is illustrated pictorially by the cartoon in the right panel of Fig.~\ref{fig:plane}; the PP and RP serve as two different ``filters'' for the $v_2$ background and CME signal.

The $\psiRP$, $\psiPP$ and $\epsilon_2$ are, however, all theoretical concepts, and cannot be experimentally measured.
Usually the 1st-order harmonic EP from ZDC, which measures spectator neutrons~\cite{Reisdorf:1997fx,Herrmann:1999wu,Abelev:2013cva,Adamczyk:2016eux}, is a good proxy for $\psiRP$. As a proxy for $\psiPP$, the 2nd-order harmonic EP ($\psiEP$) reconstructed from final-state particles is used. Since $v_2$ is generally proportional to $\epsilon_2$, one can obtain the factor $a$ of Eq.~(\ref{eq:a}) by
\be a=v_2\{\psiRP\}/v_2\{\psiEP\}\,.\label{eq:a2}\ee

The $\dg$ variable contains the CME signal and the $v_2$-induced background:
\be\dg\{\psi\}=\dg_{\cme}(\Bsq\{\psi\})+\dg_{\bkg}(v_2\{\psi\})\,.\label{eq:dg2comp}\ee
Assuming the $\dg_{\cme}(\Bsq\{\psi\})$ is proportional to $\Bsq\{\psi\}$ and $\dg_{\bkg}(v_2\{\psi\})$ is proportional to $v_2\{\psi\}$, one can obtain the relative CME signal to background contribution by
\be r\equiv\frac{\dg_{\cme}(\Bsq\{\psiRP\})}{\dg_{\bkg}(v_2\{\psiEP\})}\approx \frac{R_{v_2}-R_{\dg}}{R_{v_2}+R_{\dg}}\,,\label{eq:r}\ee
where
\be
R_{v_2}\equiv2\cdot\frac{v_2\{\psiRP\}-v_2\{\psiPP\}}{v_2\{\psiRP\}+v_2\{\psiPP\}}\,,\hspace{0.3in}{\rm and}\hspace{0.3in}
R_{\dg}\equiv2\cdot\frac{\dg\{\psiRP\}-\dg\{\psiPP\}}{\dg\{\psiRP\}+\dg\{\psiPP\}}\,.
\label{eq:Rv2Rdg}
\ee
The CME signal fraction in the measurements with respect to $\psiEP$ is then
\be \fcme^{\rm EP}=\dg_{\cme}(\Bsq\{\psiEP\})/\dg\{\psiEP\}=r/(r+1/a)\,.\label{eq:fcme}\ee

We note that the experimentally measured harmonic planes are affected by systematics, such as nonflow correlations. However, our method does not require a precise determination of the RP and PP~\cite{Xu:2017qfs}. As long as there are two experimentally assessable planes onto which the projections of the magnetic field and the elliptic flow are the opposite, our method is robust and is not affected by the uncertainties in assessing the true RP and PP. The plane projection relationship is given by Eq.~(\ref{eq:a}) where the $\psiPP$ and $\psiRP$, in an experimental data analysis context, should be taken as the experimentally measured harmonic planes.

STAR has measured the $\dg$ with respect to the $\psi_1$ from the ZDCs, in addition to the measurements with respect to the $\psi_2$ from produced particles. The data are shown in Fig.~\ref{fig:star_y7} in Sect.~\ref{sec:STAR}. The results for $\psi_1$ and $\psi_2$ are equal within the large statistical uncertainties. 
Nevertheless, one may extract the CME signal and the background contribution from those data, as was done in Ref.~\cite{Xu:2017qfs}. The statistics of the data are too poor to be definite; the extracted CME fraction ranges essentially from 0\% to 100\%.

\begin{figure}[!htb]
  \centerline{ 
    \includegraphics[width=0.45\hsize]{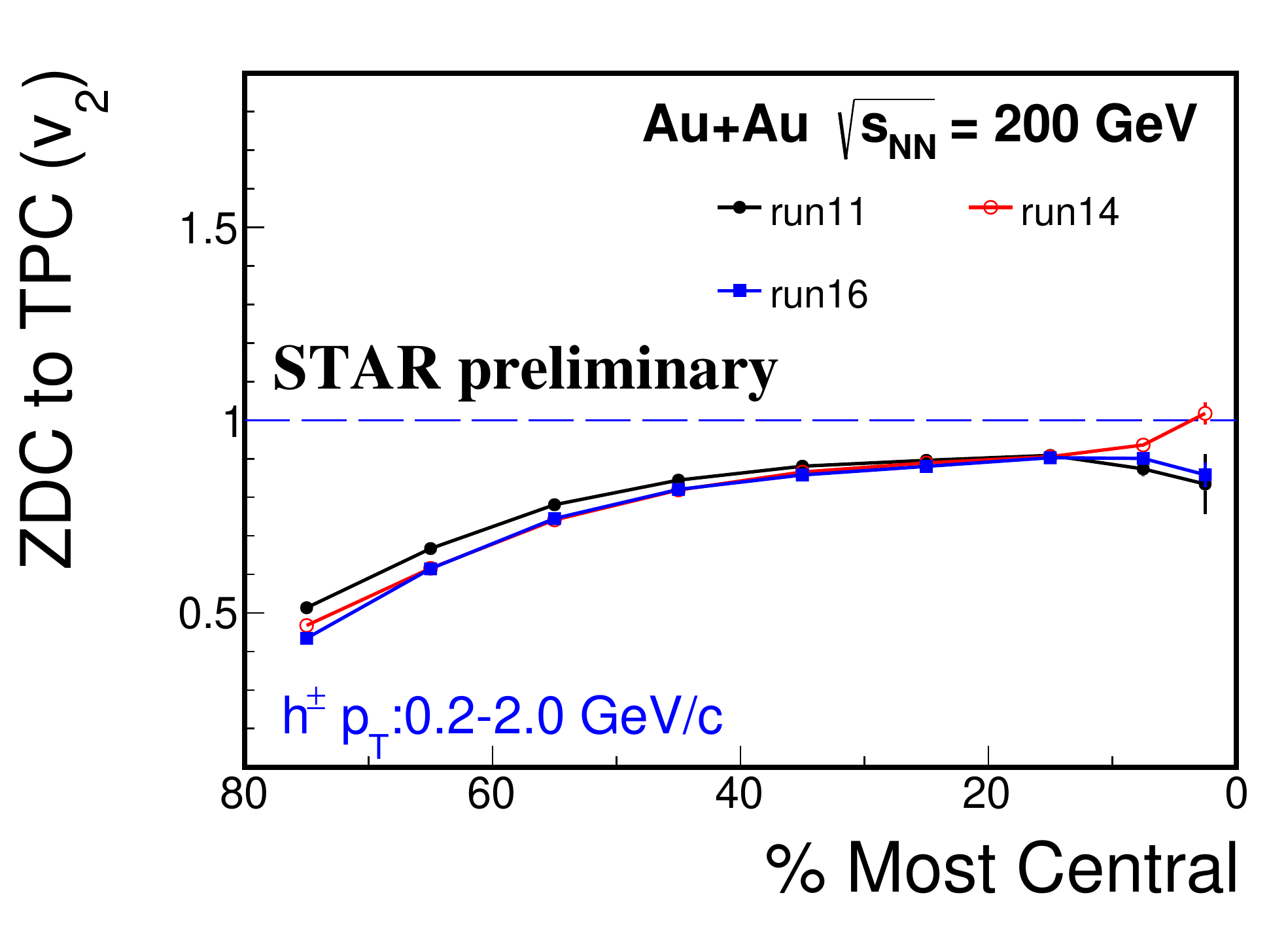}
    \hspace{0.05\hsize}
    \includegraphics[width=0.45\hsize]{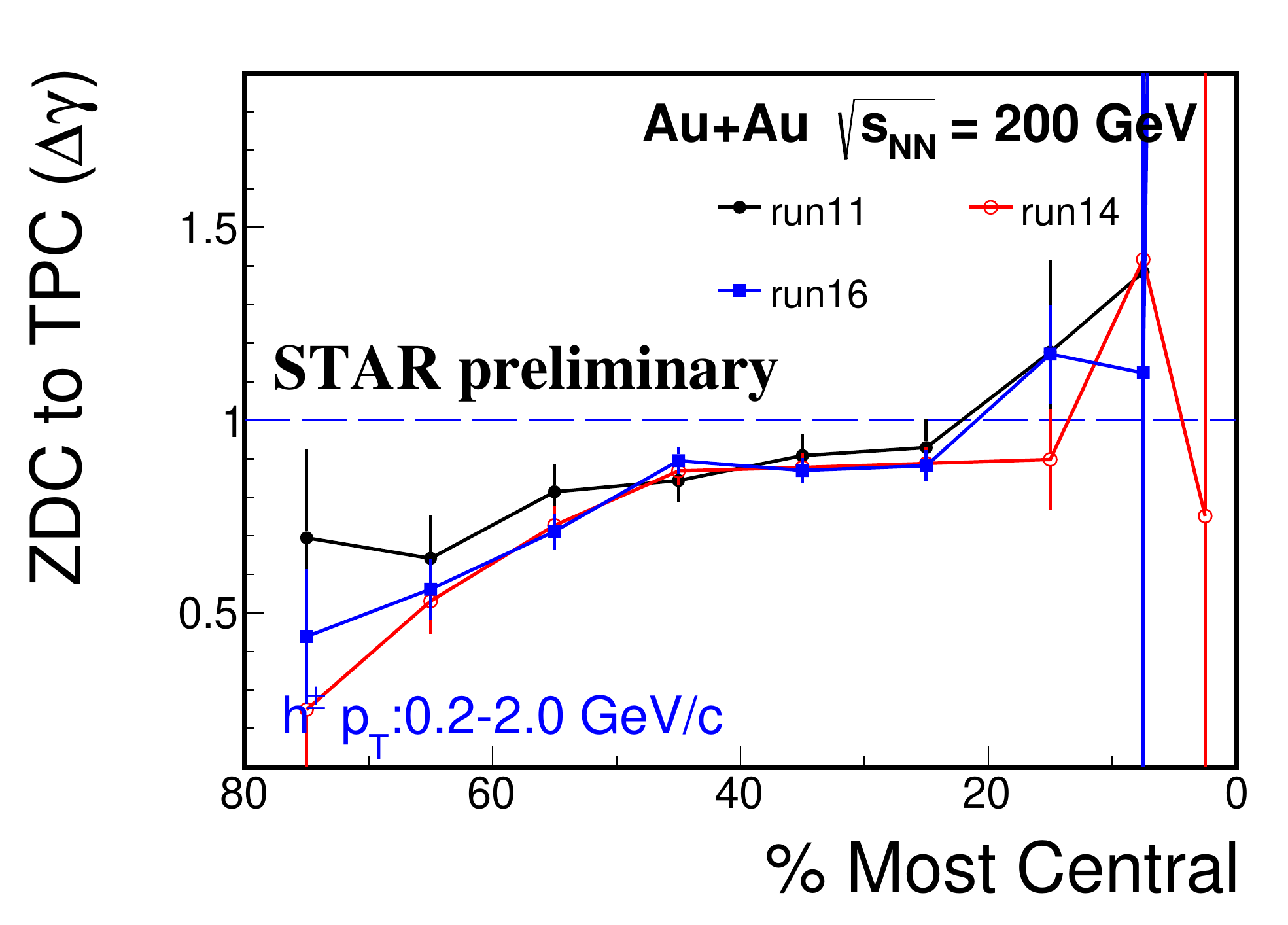}
  }
  \caption{The centrality dependences of the ratios of the $v_2$ (left panel) and $\dg$ (right panel) measured with respect to the ZDC event plane to those with respect to the TPC event plane. The sub-event method is used where the POIs come from half of the STAR TPC and the TPC EP is reconstructed from the other half. Adapted from Ref.~\cite{Zhao:2018blc}.}
  \label{fig:PPRP}
\end{figure}
\begin{figure}[!htb]
  \centerline{\includegraphics[width=0.5\hsize]{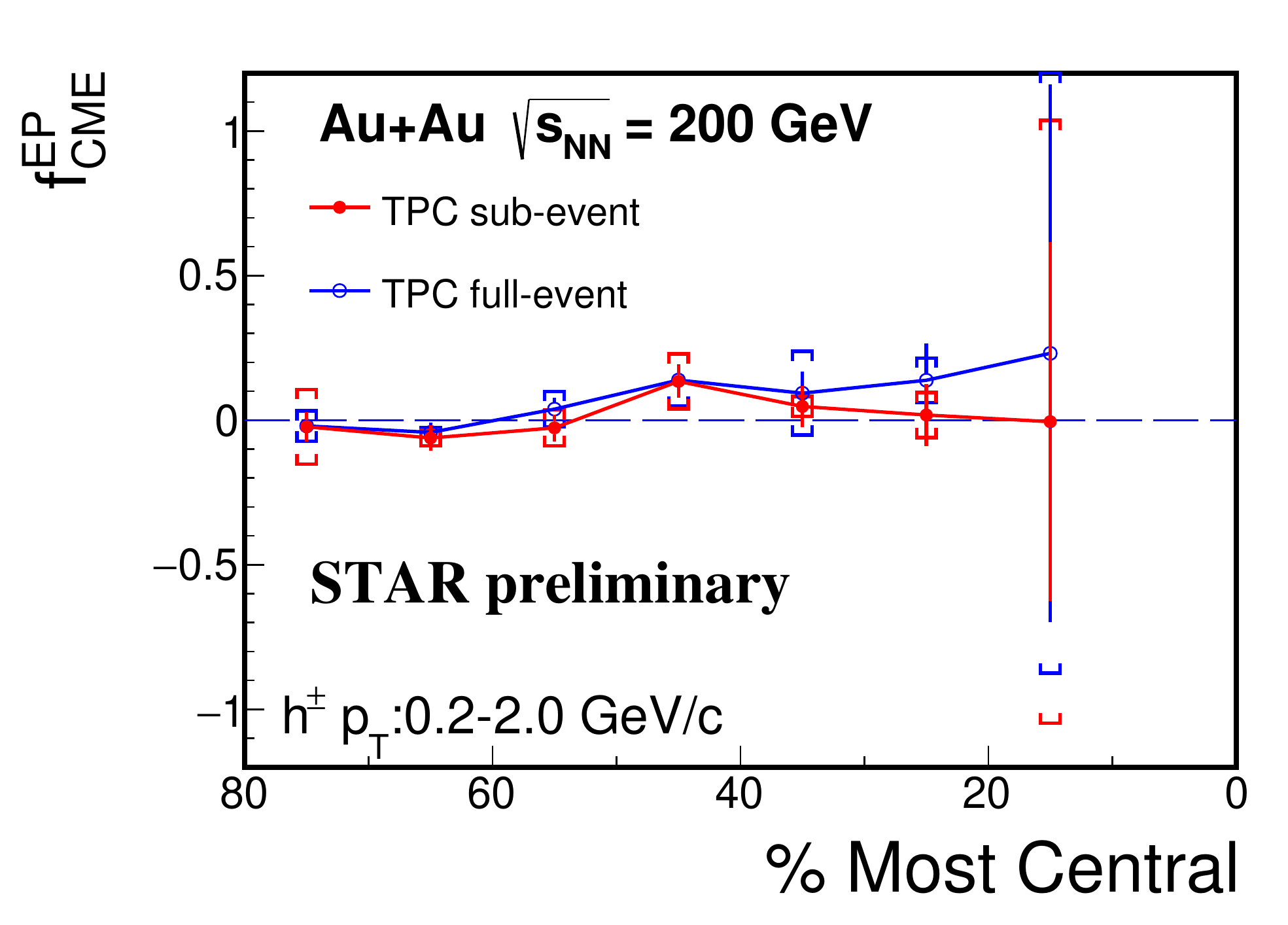}}
  \caption{The extracted fraction of potential CME signal, $\fcme^{\rm EP}$, as a function of collision centrality in 200~GeV \AuAu\ collisions by STAR, combining data from Run-11, Run-14, and Run-16. Error bars (horizontal caps) represent statistical (systematic) uncertainties. Adapted from Ref.~\cite{Zhao:2018blc}.}
  \label{fig:fCME}
\end{figure}

STAR has now accumulated two orders of magnitude more data compared to those published in Ref.~\cite{Adamczyk:2013hsi}. STAR has employed this novel RP-PP method to extract the CME signal from those high statistics data~\cite{Zhao:2018blc}.
Figure~\ref{fig:PPRP} left panel shows the ratio of $v_2$ measured with respect to the ZDC 1st-order harmonic plane to that with respect to the TPC 2nd-order harmonic EP, and the right panel shows the corresponding ratio of $\dg$~\cite{Zhao:2018blc}. The sub-event method is used where the $\gamma$ correlators are obtained by Eq.~(\ref{eq:gammaEP}) with the POIs ($\alpha$ and $\beta$) from one half of the TPC in pseudorapidity and the TPC EP from the other half. Figure~\ref{fig:fCME} shows the extracted CME fraction $\fcme^{\rm EP}$ by Eq.~(\ref{eq:fcme})~\cite{Zhao:2018blc}. The $\fcme^{\rm EP}$ from the full-event method is also shown in Fig.~\ref{fig:fCME}, where the $\gamma$ correlators are obtained by Eq.~(\ref{eq:gamma}) and all three particles are from anywhere in the TPC. 
Within errors, there is no measurable difference between sub-events and full events, though nonflow effects are expected to be larger in the latter. 
Combining Run-11, 14, and 16 data (total $\sim$2.5 billion minimum-bias events), the extracted CME fractions are $(9\pm4\pm7)$\% from TPC sub-events and $(12\pm4\pm11)$\% from TPC full events in 20-50\% centrality \AuAu\ collisions at 200~GeV~\cite{Zhao:2018blc}.

\subsection{Current best estimate and discussions}
\begin{figure}[!htb]
  \includegraphics[width=0.45\hsize]{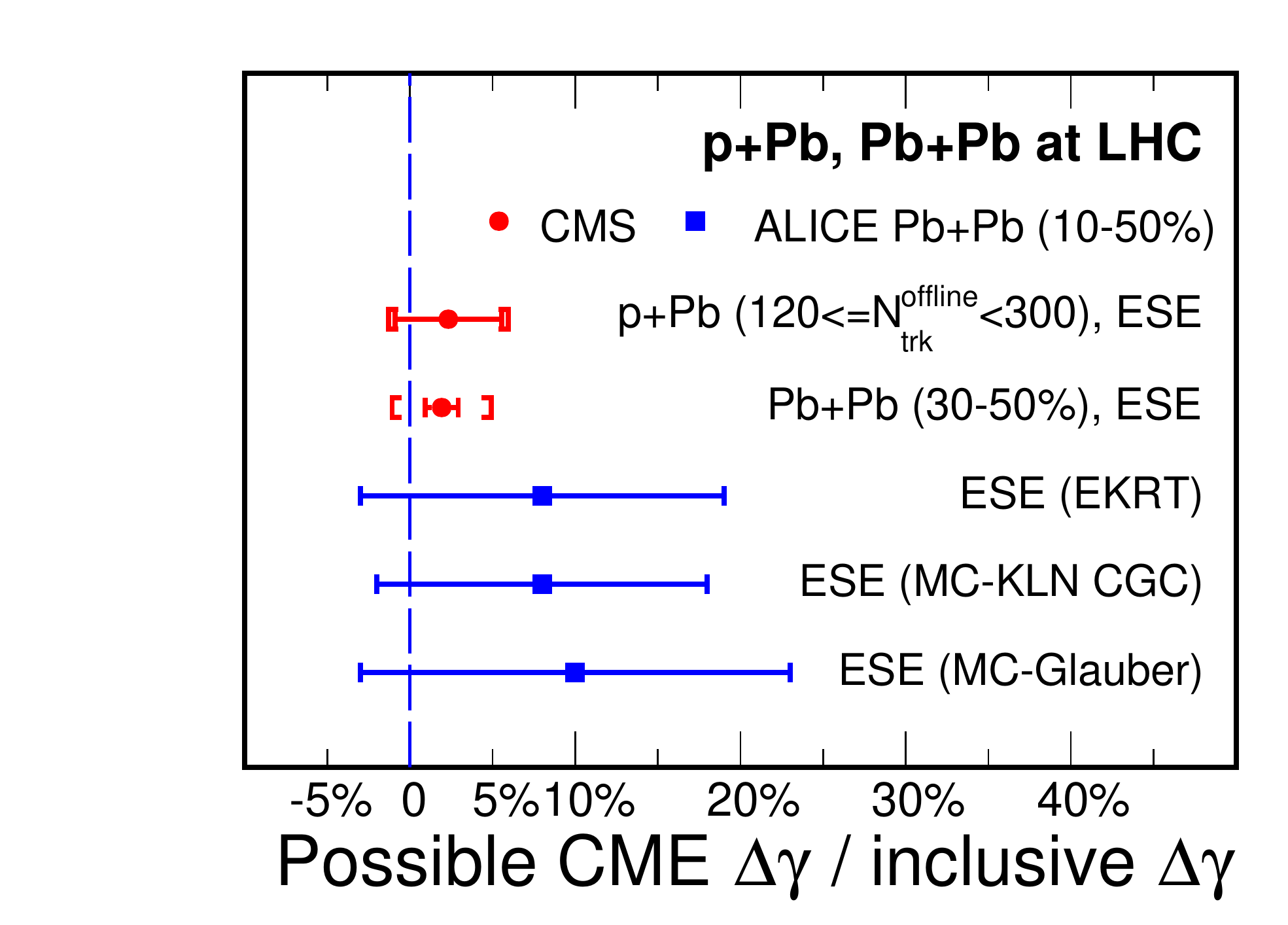}
  \hspace{0.03\hsize}
  \includegraphics[width=0.45\hsize]{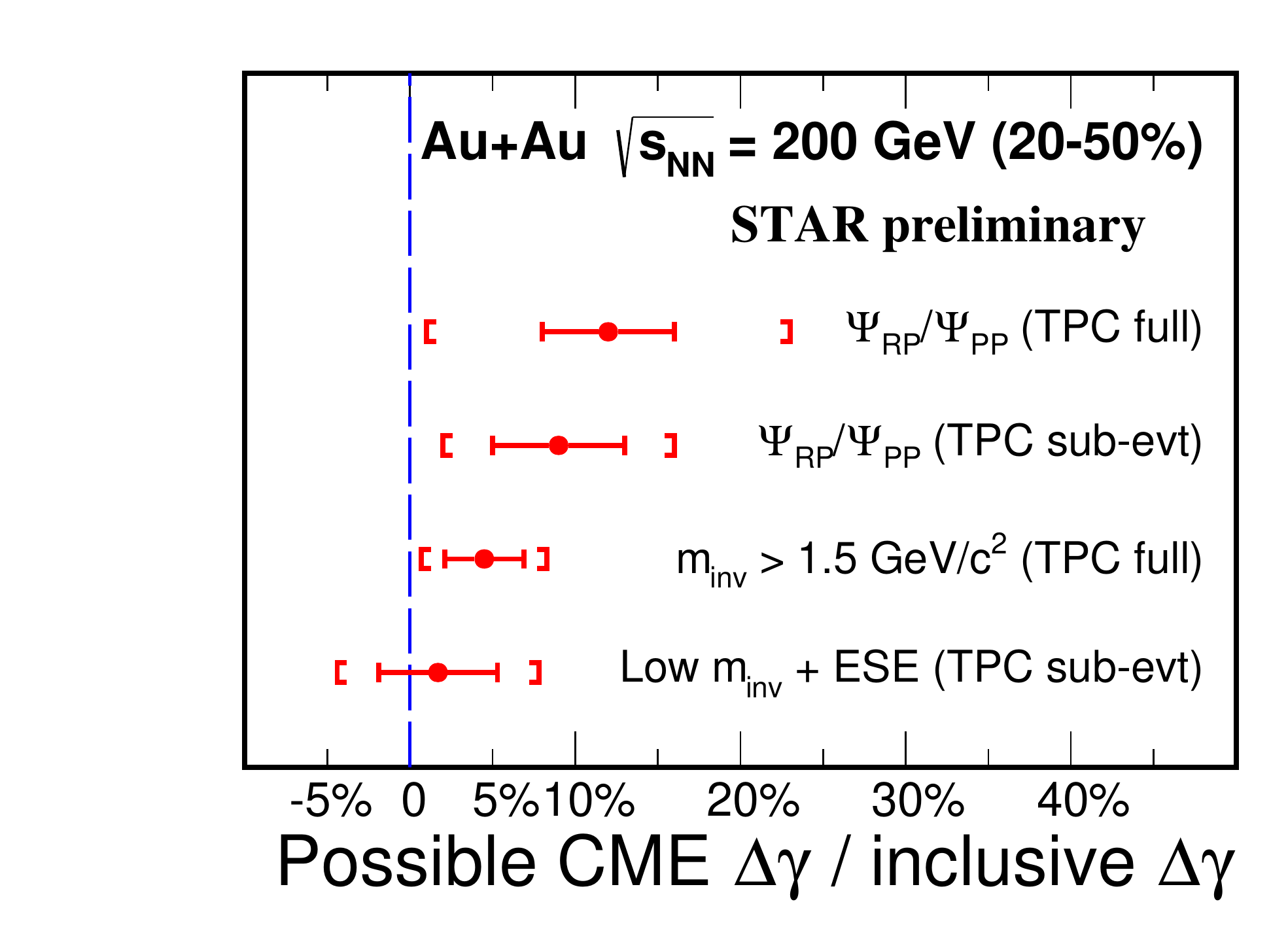}
  \caption{The possible CME signal, relative to the inclusive $\dginc$ measurement, extracted from \PbPb\ (and \pPb) collisions at the LHC (left panel) and from Au+Au collisions at $\snn=200$~GeV at RHIC~\cite{Zhao:2018blc} (right panel). The LHC results are obtained by the ESE method and include 10-50\% centrality (approximately $6.5\times10^6$) \PbPb\ collisions at $\snn=2.76$~TeV from ALICE~\cite{Acharya:2017fau}, 30-50\% centrality (approximately $6.0\times10^7$) \PbPb\ collisions at $\snn=5.02$~TeV and high multiplicity ($120\leq\Noff<300$, taken with an online high multiplicity trigger corresponding to a total $1.2\times10^9$) \pPb\ collisions at $\snn=8.16$~TeV from CMS~\cite{Sirunyan:2017quh}. The RHIC results are obtained by the $\minv$ method (both low $\minv$+ESE fit and high $\minv$ cut) and the RP-PP method (labeled ``$\psiRP/\psiPP$'') in 20-50\% centrality (approximately $7.5\times10^8$) \AuAu\ collisions at $\snn=200$~GeV by STAR~\cite{Zhao:2018blc}. Error bars (square brackets) represent statistical (systematic) uncertainties.}
  \label{fig:QMfrac}
\end{figure}

The three innovative analysis methods described in this section give, we believe so far, the best estimates of the possible CME signals in heavy-ion collisions without major flow background contaminations. 
At the LHC, the ALICE and CMS experiments used the ESE method to extract the CME signal by extrapolating the $\dg$ measurements to vanishing $v_2$. 
Figure~\ref{fig:QMfrac} left panel summarizes the current status of the possible CME signals in \PbPb\ collisions (and in \pPb\ collisions where no CME is expected to be observable) at the LHC. The current results are consistent with zero within the statistical and systematic uncertainty. 

At RHIC, the STAR experiment used the $\minv$ method by evaluating the $\dg$ in the high $\minv$ region and by fitting a two-component model complemented with the ESE technique to the low $\minv$ data. STAR has also used the RP-PP method by simultaneously solving the CME signal and the flow background from two $\dg$ measurements in the same event.
Figure~\ref{fig:QMfrac} right panel summarizes the current status of the possible CME signal at RHIC in 20-50\% central \AuAu\ collisions at $\snn=200$~GeV. The data show that the CME signal is small, on the order of a few percent of the inclusive $\dginc$ measurement, with relatively large errors~\cite{Zhao:2018blc}. It should be noted that these data points are from the same data using different analysis methods. 

Table~\ref{tab:QMfrac} summarizes the current numerical results on the fraction of the possible CME signal in the measured inclusive $\dginc$ observable with respect to the event plane (participant plane) at RHIC and the LHC.
\begin{table*}[!htb]
  \caption{The fraction of the possible CME signal (in percentage) in the measured inclusive $\dginc$ observable with respect to the event plane (participant plane) at the LHC by ALICE~\cite{Acharya:2017fau} and CMS~\cite{Sirunyan:2017quh} and at RHIC by STAR~\cite{Zhao:2018blc}. All results were obtained from charged hadrons except the $\minv$ method from STAR where charged pions were used, $0.2<\pt<1.8$~\gevc\ for the $\minv>1.5$~\gevcc\ result and $0.2<\pt<0.8$~\gevc\ for the $0.4<\minv<1.5$~\gevcc\ result~\cite{Zhao:2018blc}. The LHC inclusive $\dginc$ values are obtained from least-$\chi^2$ fitted averages of the ALICE~\cite{Abelev:2012pa} and CMS~\cite{Khachatryan:2016got} data. When two errors are quoted, the first is statistical uncertainty and the second is systematic uncertainty; otherwise the quoted error is statistical.}
  \label{tab:QMfrac}
  \begin{center}
  \begin{tabular}{lcl}
    \hline\hline
    \multicolumn{3}{c}{LHC/ALICE~\cite{Acharya:2017fau}, $|\eta|<0.8$, $0.2<\pt<5.0$~\gevc} \\\hline
    \multirow{3}{3in}{\PbPb~(10-50\%), $\snn=2.76$~TeV $\dginc=(7.96\pm0.45)\times10^{-5}$~\cite{Abelev:2012pa}}
    & $(10\pm13)$\% & [with $B(v_2)$ from MC-Glauber] \\
    & $( 8\pm10)$\% & [with $B(v_2)$ from MC-KLN CGC] \\
    & $( 8\pm11)$\% & [with $B(v_2)$ from EKRT] \\
    \hline\hline
    \multicolumn{3}{c}{LHC/CMS~\cite{Sirunyan:2017quh}, $|\eta|<2.4$, $|\deta|<1.6$, $0.3<\pt<3.0$~\gevc} \\\hline
    \begin{varwidth}{3.5in}\pPb~($120\leq\Noff<300$), $\snn=8.16$~TeV \hspace{0.2in} $\dginc=(6.38\pm0.11\pm0.41)\times10^{-4}$~\cite{Khachatryan:2016got}\end{varwidth}
    & $(2\pm3\pm4)$\% & \\
    \begin{varwidth}{2.7in}\PbPb~(30-50\%), $\snn=5.02$~TeV \hspace{0.2in} $\dginc=(1.47\pm0.01\pm0.28)\times10^{-4}$~\cite{Khachatryan:2016got}\end{varwidth}
    & $(2\pm 1\pm 3)$\% & \\
    \hline\hline
    \multicolumn{3}{c}{RHIC/STAR~\cite{Zhao:2018blc}, $|\eta|<1$, $0.15<\pt<2.0$~\gevc} \\\hline
    \multirow{4}{3in}{\AuAu~(20-50\%), $\snn=200$~GeV \hspace{0.2in} $\dginc=(1.82\pm0.03)\times10^{-4}$}
    & $(12\pm4\pm11)$\% & [RP-PP method, TPC full events] \\
    & $( 9\pm4\pm 7)$\% & [RP-PP method, TPC sub-events] \\
    & $( 5\pm2\pm 4)$\% & [$\minv>1.5$~\gevcc] \\
    & $( 2\pm4\pm 6)$\% & [$0.4<\minv<1.5$ fit method] \\
    \hline\hline
  \end{tabular}
  \end{center}
\end{table*}
      
As for all searches of physics importance, in order to claim discovery, one must remove all ambiguities so the signal is beyond $5\sigma$ of the total statistical and systematic uncertainty. The strategy is often to remove all possible background contaminations to leave the final signal absolutely clean, although part of the signal could inevitably be removed together with the background removal. Only when there is an unambiguous signal beyond any reasonable doubt, should one claim the discovery of the CME. On the other hand, if the final signal is consistent with zero, it does not necessarily mean that the CME is nonexistent, because of the same reason that the CME could be removed together with the background removal.

\section{Future perspective}\label{sec:future}

\subsection{Isobaric collisions}\label{sec:isobar}
\begin{figure}[!htb]
  \centerline{
    \includegraphics[width=0.4\hsize]{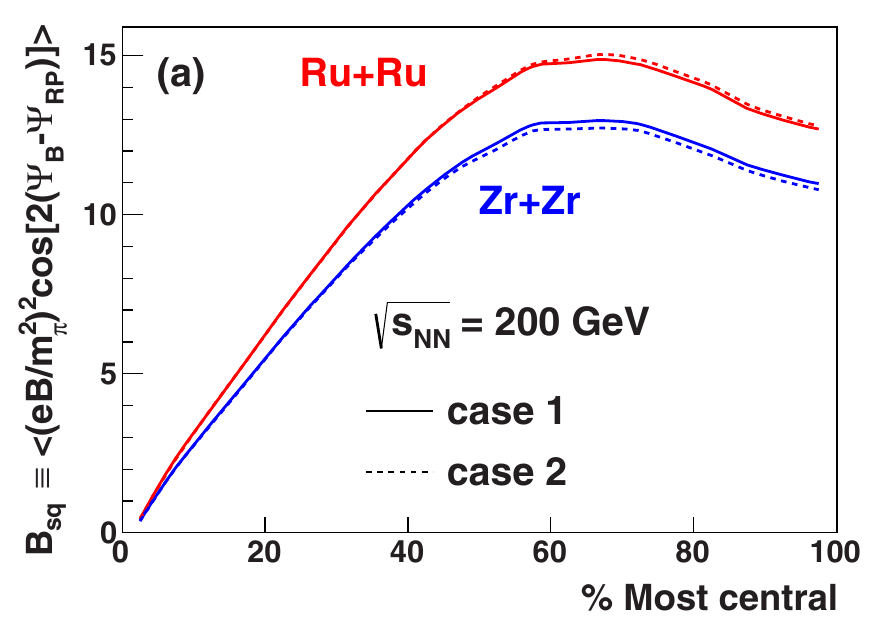}
    \hspace{0.05\hsize}
    \includegraphics[width=0.4\hsize]{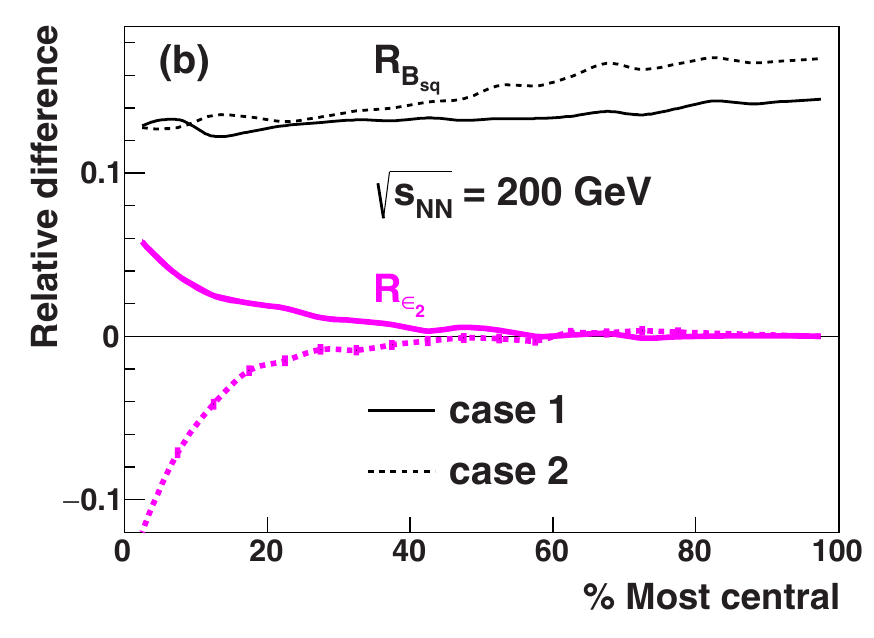}
  }
  \caption{(Color online) (a) Event-averaged initial magnetic field squared at the center of mass of the overlapping region, with correction from event-by-event fluctuations of the magnetic field azimuthal orientation, for Ru+Ru and Zr+Zr collisions at 200 GeV, and (b) their relative difference as functions of the collision centrality. Also shown in (b) is the relative difference in the initial eccentricity. The line styles correspond to two extreme cases of the isobaric nuclear deformation parameters. Adapted from Ref.~\cite{Deng:2016knn}.}
  \label{fig:isobar_B}
\end{figure}

The CME is related to the magnetic field while the background is produced by $v_2$-induced correlations. In order to gauge differently the magnetic field relative to the $v_2$, isobaric $\RuRu$ and $\ZrZr$ collisions have been proposed~\cite{Voloshin:2010ut,Skokov:2016yrj}. $\Ru$ and $\Zr$ have the same mass number but different charge (proton) number. One would thus expect the same $v_2$, which is insensitive to isospin, and 10\% difference in the magnetic field. MC Glauber calculations of the spatial eccentricity and the magnetic field strength in Ru+Ru and Zr+Zr collisions have been carried out~\cite{Deng:2016knn,Huang:2017azw}. The Woods-Saxon density distribution is used~\cite{Deng:2016knn,Huang:2017azw},
\be
\rho(r,\theta) = \frac{\rho_0}{1+\rm{exp}\{[r-R_0-\beta_2R_0Y_2^0(\theta)]/ a \}}\,, \label{eq:rho}
\ee
where $R_0$ is the charge radius parameter of the nucleus, $a\approx0.46$~fm is the surface diffuseness parameter, $Y_2^0$ is the spherical harmonic, and $\rho_0$ is the normalization factor. The charge radii of $R_0=5.085$~fm and 5.020~fm were used for $\Ru$ and $\Zr$, respectively, for both the proton and neutron densities. The deformity quadrupole parameter $\beta_2$ has large uncertainties; extreme cases were taken and yielded 2\% difference in $\epsilon_2$ between Ru+Ru and Zr+Zr collisions at 20\% centrality and significantly smaller towards higher (more peripheral) centralities~\cite{Deng:2016knn,Huang:2017azw}. The magnetic field strengths were calculated by using Lienard-Wiechert potentials with the HIJING model taking into account the event-by-event azimuthal fluctuations of the magnetic field orientation~\cite{Deng:2012pc}.
Figure~\ref{fig:isobar_B}(a) shows the calculated quantity $\Bsq\equiv\mean{(eB/\mpi^2)^2\cos2(\psi_{B}-\psiRP)}$, relevant to the CME with respect to $\psiRP$, at the initial encounter time of the nuclei in Ru+Ru and Zr+Zr collisions at 200 GeV. 
Figure~\ref{fig:isobar_B}(b) shows the relative difference in $\Bsq$ between Ru+Ru and Zr+Zr collisions,
\be R_{\Bsq}=2(\Bsq^{\rm Ru+Ru}-\Bsq^{\rm Zr+Zr})/(\Bsq^{\rm Ru+Ru}+\Bsq^{\rm Zr+Zr})\,.\label{eq:ReccIso}\ee
The difference is approximately 15\%. Figure~\ref{fig:isobar_B}(b) also shows the relative difference in the initial eccentricity,
\be R_{\epsilon_2}=2(\epsilon_2^{\rm Ru+Ru}-\epsilon_2^{\rm Zr+Zr})/(\epsilon_2^{\rm Ru+Ru}+\epsilon_2^{\rm Zr+Zr})\,.\label{eq:RBsqIso}\ee
The relative difference in $\epsilon_2$ is small in the 20-60\% centrality range, suggesting that the $v_2$-induced backgrounds will be almost equal for Ru+Ru and Zr+Zr collisions. Note that, although the same symbol $R$ is used, the $R$ in Eqs.~(\ref{eq:ReccIso}) and (\ref{eq:RBsqIso}) are different from those in Eqs.~(\ref{eq:Recc}), (\ref{eq:RBsq}), and (\ref{eq:Rv2Rdg}) which refer to the relative differences between measurements with respect to different harmonic planes in the same event.

A dedicated run of isobar collisions has been conducted at RHIC in 2018. The isobar run has accumulated total approximately 2.5 billion events each for Ru+Ru and Zr+Zr collisions in the STAR detector. One may estimate the $\dg$ magnitudes in Ru+Ru and Zr+Zr collisions based on the available $\dg$ measurements in \AuAu\ and Cu+Cu collisions at 200 GeV. If the CME signal is 5\% of the inclusive $\dginc$ measurement, as implied by the latest STAR results~\cite{Zhao:2018blc}, then the magnetic field and eccentricity differences between Ru+Ru and Zr+Zr collisions calculated in Refs.~\cite{Deng:2016knn,Huang:2017azw} would yield a 1-2$\sigma$ effect.

\begin{figure}[!htb]
  \centerline{
    \includegraphics[width=0.35\hsize]{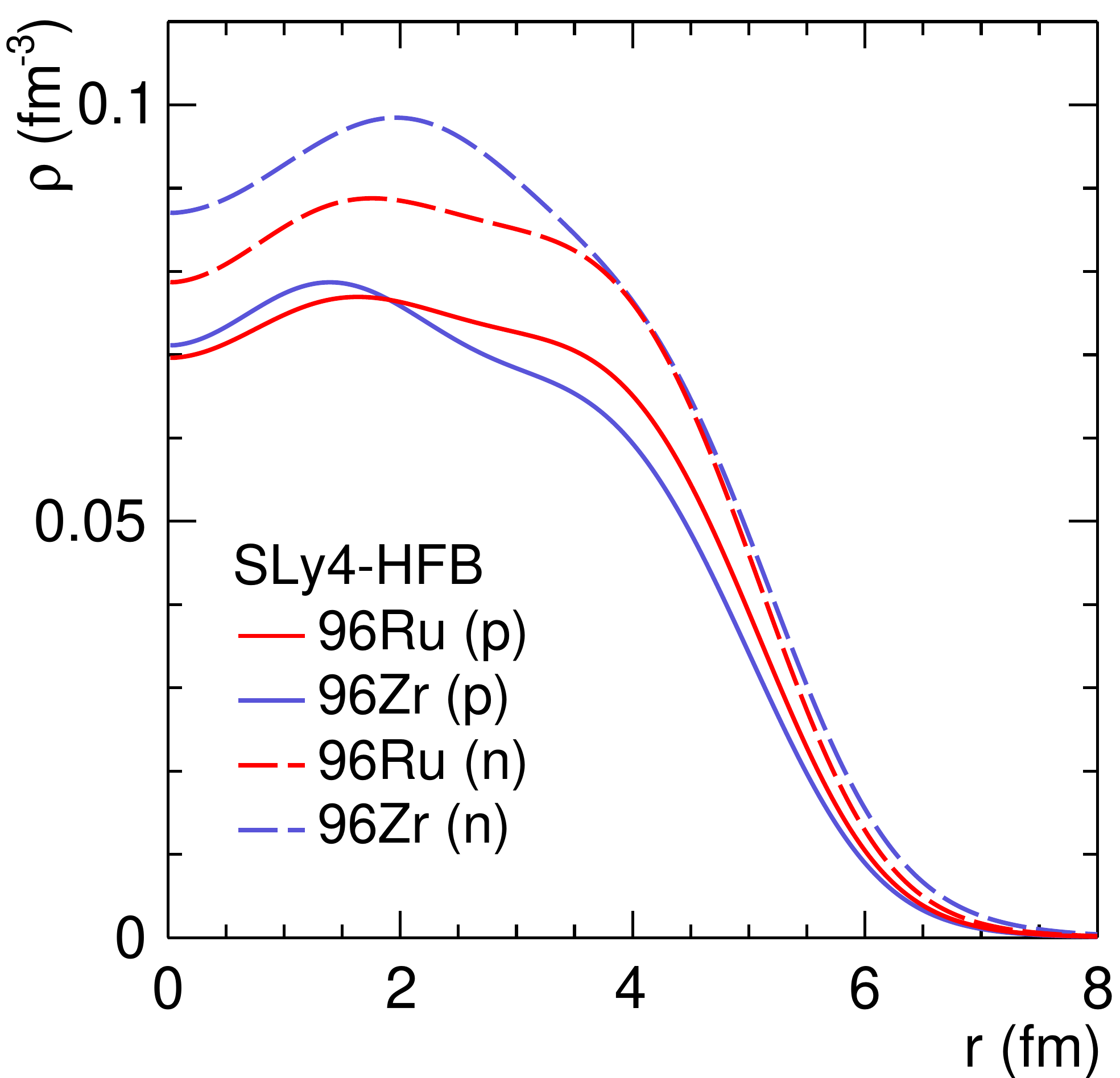}
    \hspace{0.05\hsize}
    \includegraphics[width=0.40\hsize]{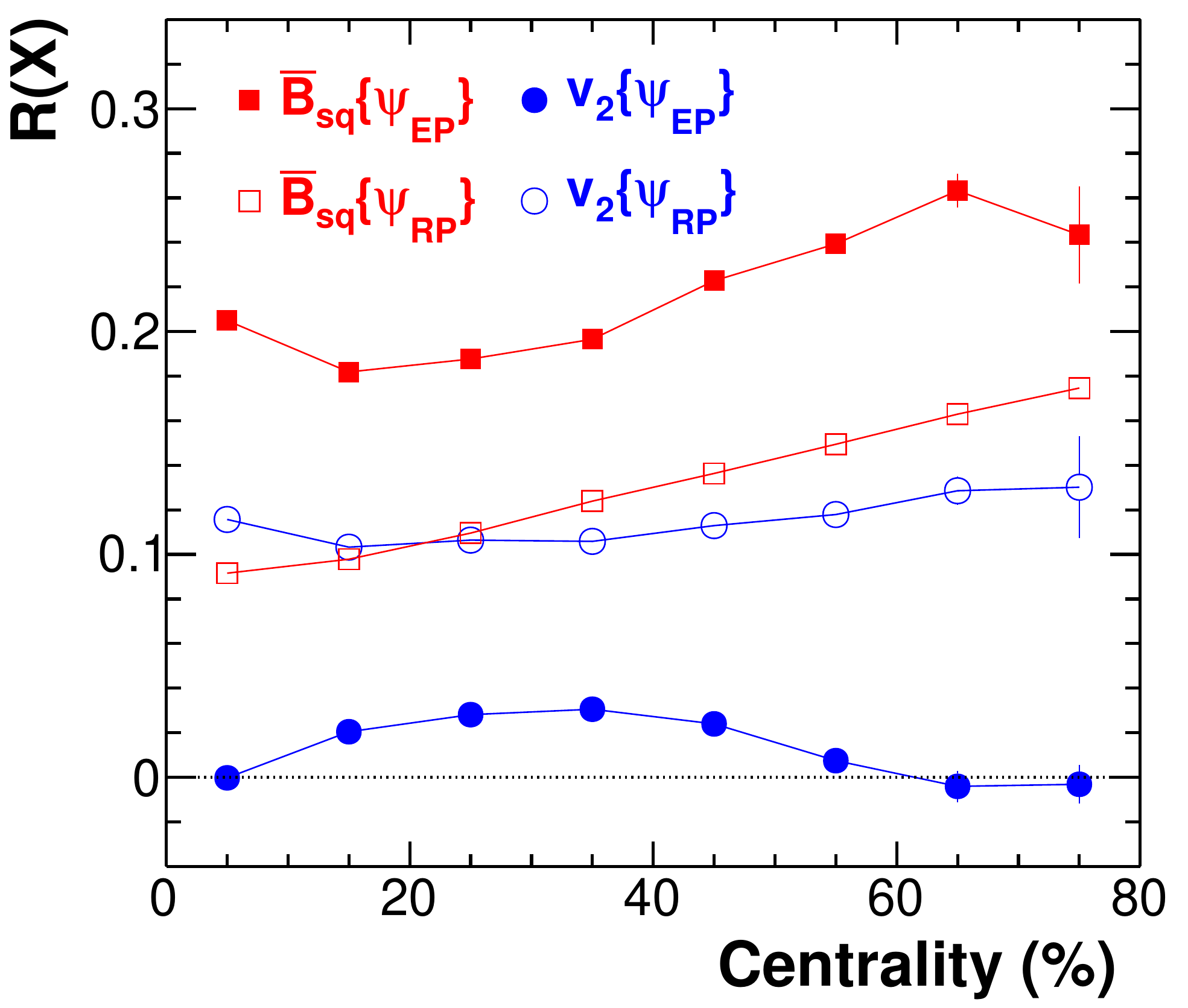}
  }
  \caption{(Color online) Left Panel: proton and neutron density distributions of the $\Ru$ and $\Zr$ nuclei, assumed spherical, calculated by DFT. Right Panel: relative differences between Ru+Ru and Zr+Zr collisions as functions of centrality in $v_2\{\psi\}$ and $\Bsq\{\psi\}$ with respect to $\psiRP$ and $\psiEP$ from AMPT simulations using the DFT densities from the left panel. Adapted from Ref.~\cite{Xu:2017zcn}.}
  \label{fig:isobar_rho}
\end{figure}

The above estimates assume Woods-Saxon densities, identical for proton and neutron distributions. Using the energy density functional theory (DFT) with the well-known SLy4 mean field~\cite{Chabanat:1997un} including pairing correlations (Hartree-Fock-Bogoliubov, HFB approach)~\cite{Wang:2016rqh,Bender:2003jk,ring1980nuclear}, the ground-state density distributions of $\Ru$\ and $\Zr$, assumed spherical, were calculated~\cite{Xu:2017zcn}. The results are shown in the left panel of Fig.~\ref{fig:isobar_rho}. They show that protons in Zr are more concentrated in the core, while protons in Ru, 10\% more than in Zr, are pushed more toward outer regions. The neutrons in Zr, four more than in Ru, are more concentrated in the core but also more populated on the nuclear skin. The right panel of Fig.~\ref{fig:isobar_rho} shows the relative differences in $v_2\{\psi\}$ and $\Bsq\{\psi\}$ between Ru+Ru and Zr+Zr collisions as functions of centrality from AMPT simulations with the densities calculated by the DFT method~\cite{Xu:2017zcn}. Results with respect to both $\psiRP$ and $\psiEP$ are depicted.
With respect to $\psiRP$, the differences in $v_2$ and $\Bsq$ are both on the order of 10\%. The isobaric collisions cannot distinguish the CME signal and the flow background. Most of the $\dg$ results have been analyzed with respect to the EP. The relative difference in $\Bsq$ with respect to the EP is the expected $\sim$20\%. However, the relative difference in $\epsilon_2$ and $v_2$ with respect to $\psiEP$ are not zero but as large as $\sim$3\%, over a wide range of centralities. This difference is much smaller than that in the magnetic field, but presents a significant background if the CME signal is small. For example, if the CME signal is 5\% and the background is 95\% in the measured $\dg$, then the $\dg$ difference between Ru+Ru and Zr+Zr collisions would be 4\%, only 1\% of which is due to the CME signal, the other 3\% is from background. This suggests that the premise of isobaric collisions for the CME search may not be as good as originally anticipated. 

\begin{figure}[!htb]
  \centerline{\includegraphics[width=0.45\hsize]{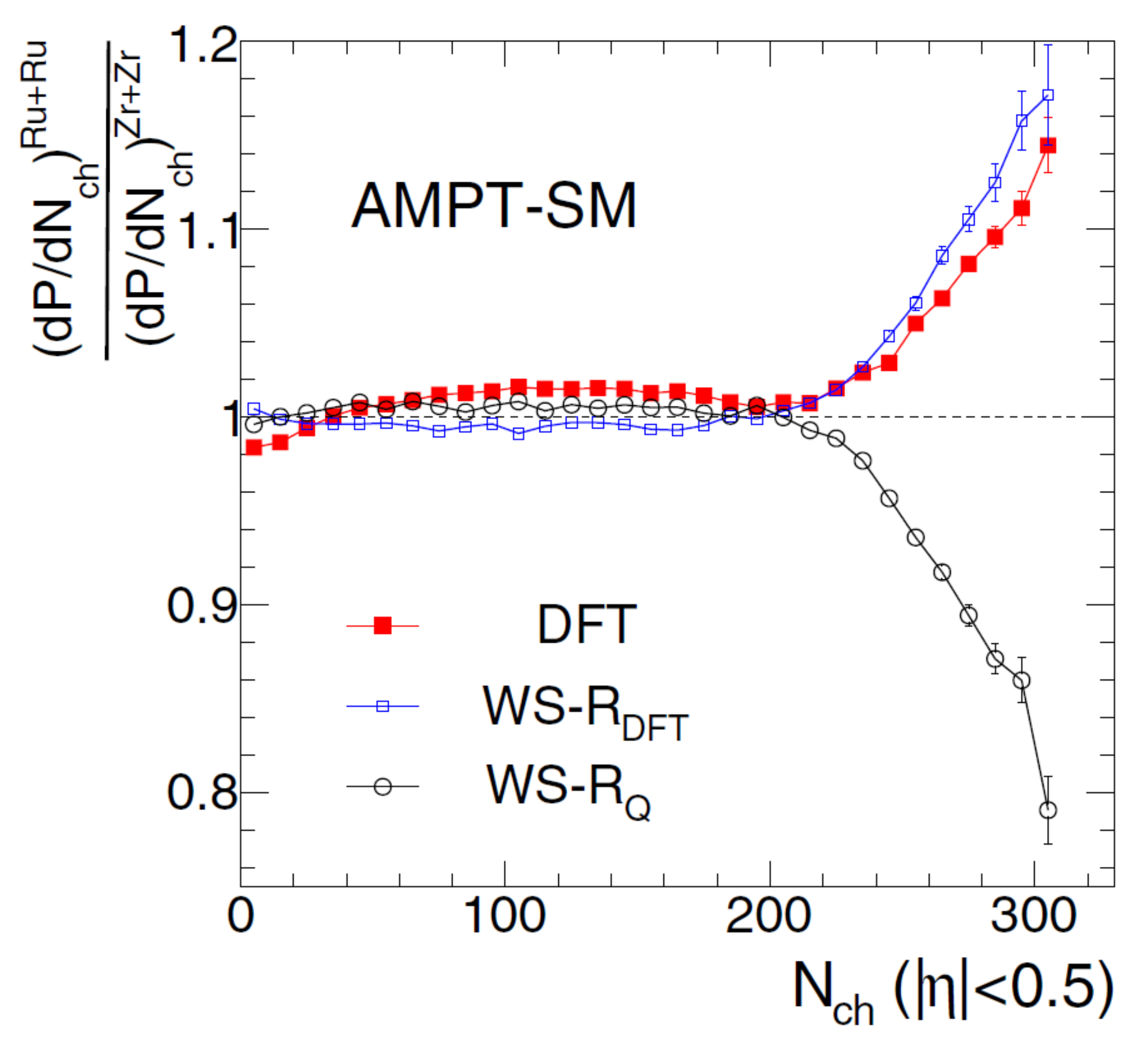}}
  \caption{Ratio of the charged particle multiplicity distributions in Ru+Ru and Zr+Zr collisions simulated by AMPT (string melting). Three types of nuclear density distributions are shown: DFT, WS with charge radii, and WS with DFT effective radii. Adapted from Ref.~\cite{Li:2018oec}.}
  \label{fig:isobarNch}
\end{figure}

Woods-Saxon nuclear density distribution is a simplistic parameterization. The nuclear distributions calculated from DFT are more trustworthy, but are not without theoretical uncertainties. How can the two types of density distributions be experimentally distinguished? The anisotropic flow measurements would be one way. If large anisotropic flow difference is observed, then it would be a clear evidence that the Woods-Saxon nuclear density distributions are incorrect. If the observed flow difference is small, then both the Woods-Saxon and DFT nuclear density distributions can accommodate it so the flow measurement would be insufficient to tell them apart. However, the multiplicity distribution difference between Ru+Ru and Zr+Zr collisions may still remain as a viable discriminator. The Woods-Saxon nuclear density distributions with the charge radius parameters give a larger multiplicity tail in Zr+Zr collisions~\cite{Deng:2016knn,Deng:2018dut,Li:2018oec}, while those with the effective nuclear radius parameters from DFT distributions yield the opposite~\cite{Li:2018oec}. As a result, the ratio of the multiplicity distributions in Ru+Ru to Zr+Zr reveals dramatic features at large multiplicities, as shown in Fig.~\ref{fig:isobarNch} by AMPT simulations~\cite{Li:2018oec}. Although the effective radius is the most important parameter, the nuclear density shapes can yield subtle difference in the intermediate multiplicity range of the ratio~\cite{Li:2018oec}. This is shown in Fig.~\ref{fig:isobarNch}; the results from Woods-Saxon and DFT densities are distinctly different in the intermediate multiplicity range of $\dNdeta=50$-200. This can be used to distinguish the different types of nuclear density distributions given the same effective radius, since the multiplicity distributions can be measured very precisely.

It is worthwhile to note that these recent studies~\cite{Xu:2017zcn,Li:2018oec} indicate that nucleus-nucleus collisions at relativistic energies may be used to probe nuclear structures which have been typically studied in low energy nuclear reactions. This is nontrivial and illustrates the rich connections between the different subfields of nuclear physics.

\subsection{Increasing \AuAu\ statistics}\label{sec:auau}
\AuAu\ collisions at RHIC are the one system that is most extensively studied in the search for the CME. Our understanding about the backgrounds has been improved tremendously over the past several years. The possible CME signal is apparently very small, but its existence in current experimental observations cannot be clearly ruled out. There are hints that traces of the CME signal may be hidden in the existing \AuAu\ data. The isobar data discussed in Sect.~\ref{sec:isobar} will help further our understanding of the background issue and the possible CME signal. But no matter what the outcome of the isobar data is, the search for the CME shall continue. More statistics should be accumulated for \AuAu\ collisions at RHIC and \PbPb\ collisions at the LHC.

\subsection{Zero-degree calorimeter upgrade}\label{sec:zdc}
Future detector upgrades should be considered to improve the sensitivities to the CME. 
One of the most promising methods to discover the CME is the RP-PP method by comparative measurements in the same nucleus-nucleus collision with respect to the PP and the SP. The limiting factor in the current measurements discussed in Sect.~\ref{sec:plane} is the poor resolution of the ZDC first-order harmonic plane. Improvement in the ZDC resolution will help tremendously in terms of the statistical precision of the CME measurement. This appears to us to be the highest priority in the search for the CME.

One promising proposal to qualitatively improve the ZDC resolution is to detect not only single neutrons, but also spectator protons and fragments. This can be achieved by measuring the deflections of charged fragments in the magnetic field of the collider elements~\cite{Tarafdar:2014oua}.

\subsection{New ideas and new observables}
The $\dg$ variable is the most widely used observable in experimental analysis searching for the CME, and arguably the easiest one to interpret. On the other hand, the CME is a parity violating effect, but the $\dg$ observable is essentially two-particle correlations and is intrinsically parity even, and therefore has inevitably large background contaminations. Sect.~\ref{sec:efforts} has presented several novel ideas to reduce backgrounds in the $\dg$ measurements. Nevertheless, more new ideas are called for to reduce backgrounds in measurements using the $\dg$ observable or its variants. Additional novel analysis techniques should be developed.

Parity-odd observables would be intrinsically more sensitive to parity-odd effects like the CME. However, since the topological charge signs are random, it may not be possible to identify a parity odd observable to search for the CME. It is, however, important to continue to look for new observables that are less background prone.

The major background is intrinsic two-particle correlations, mostly resonance decays. Three particle correlations might be one way to avoid most of the resonance decay contributions~\cite{Wang:2016iov}. Given the smallness of the CME signal, three-particle correlations may, on the other hand, prove prohibitively difficult to identify the CME.

The CME magnitude depends on the magnetic field strength, so does the polarization difference between $\Lambda$ and $\bar{\Lambda}$ hyperons. Event-by-event correlations between electric charge separation and $\Lambda$ polarization would be a strong evidence for the CME~\cite{Finch:2018ner}. Such an event-by-event analysis is certainly difficult, but the existing large statistics of \AuAu\ data may prove it viable~\cite{Finch:2018ner}.

\section{Summary}\label{sec:summary}
The chiral magnetic effect (CME) arises from local $\mathcal{P}$ and \CP\ violations caused by topological charge fluctuations in QCD. 
An observation of the CME would confirm several fundamental properties of QCD and could resolve the strong \CP\ problem responsible for the matter-antimatter asymmetry in today's universe. 
Relativistic heavy-ion collisions provide an ideal environment to search for the CME with the strong color gluon field and electromagnetic field. 
Charge-dependent azimuthal correlations with respect to the reaction plane RP (and participant plane PP) are sensitive to the CME.
Many observables have been proposed, studied, and used in data analysis, the most commonly used being the three-point azimuthal correlator $\dg$. All observables are contaminated by major physics backgrounds arising from the coupling of resonance/cluster decays and their elliptic flows $v_2$. 
Intensive theoretical and experimental efforts have been devoted to eliminate those backgrounds.

Experimental efforts include studies of both heavy-ion and small-system collisions, and significant progresses have been made.
This review provides a synopsis of the long-lasting development and sophistication of analysis methods in the search for the CME. Some earlier efforts to remove background contaminations are discussed. Candid assessments of their advantages and disadvantages are provided.
Emphasis and majority of the review are devoted to the more recently developed novel methods in eliminating background contaminations in the $\dg$ measurements. These methods include event-shape engineering (ESE), invariant-mass ($\minv$) dependence, and RP-PP comparative measurements.
The current estimates on the strength of the possible CME signal are on the order of a few percent of the inclusive $\dg$ values, consistent with zero with large uncertainties. The prospect of the recently taken isobaric collision data is discussed.

It is clear from this review that the experimental challenges in the CME search are daunting. There is no doubt that the physics of the CME is of paramount importance. The important physics warrants continued efforts despite of the experimental challenges.

\section*{Acknowledgments}
We thank Dr.~Xuguang Huang, Dr.~Wei Li, Dr.~Jinfeng Liao, Dr.~Zi-Wei Lin, Dr.~Feng Liu, Dr.~Berndt M\"uller, Dr.~J\"urgen Schukraft, Dr.~Zhoudunming Tu, Dr.~Nu Xu, and Dr.~Bill Zajc for valuable discussions. We thank the STAR, CMS, and ALICE collaborations for the beautiful data we have used extensively in this review and for their discussions. 
This work was supported in part by U.S.~Department of Energy (Grant No.~de-sc0012910) and National Natural Science Foundation of China (Grant No.~11847315).

\bibliographystyle{unsrt}
\bibliography{../../rc/ref}
\end{document}